Министерство образования и науки, молодежи и спорта Украины

Севастопольский национальный технический университет

**Юпиков Олег Александрович**

# РАЗРАБОТКА МОДЕЛИ И ОПТИМИЗАЦИЯ ПАРАМЕТРОВ ГИБРИДНОЙ АНТЕННОЙ СИСТЕМЫ ДЛЯ РАДИОТЕЛЕСКОПОВ С ШИРОКИМ ПОЛЕМ ОБЗОРА

Специальность 05.12.07 — Антенны и устройства микроволновой техники

Диссертация на соискание ученой степени кандидата технических наук

Научный руководитель:
кандидат технических наук, доцент
Щекатурин Андрей Алексеевич

Научный консультант:
кандидат технических наук, доцент
Ивашина Марианна Валерьевна

Севастополь 2012

# СОДЕРЖАНИЕ

# ПЕРЕЧЕНЬ УСЛОВНЫХ ОБОЗНАЧЕНИЙ

*I*-ДН — это зависимость напряженности электрического поля от угловых координат $\theta$ и $\varphi$, при возбуждении активного антенного элемента источником тока $I_o = 1$ А. При этом остальные элементы имеют разрыв в цепи питания

*SKA* — *Square Kilometer Array telescope*

*V*-ДН — это зависимость напряженности электрического поля от угловых координат $\theta$ и $\varphi$, при возбуждении активного антенного элемента источником напряжения с амплитудой $V_o = 1$ В и остальными элементами, порты которых закорочены

*WSRT* — *Westerbork Synthesis Radio Telescope*

АФР — Антенная Фокальная Решетка — антенная решетка, расположенная в фокусе зеркала

ДН — диаграмма направленности

КИП — коэффициент использования поверхности

КНД — коэффициент направленного действия

МВ — микроволновый

МПФ — микрополосковый фидер

МШУ — малошумящий усилитель

ФДН — формирователь ДН

ХН — характеристика направленности

ЭМ — электромагнитный

# ВВЕДЕНИЕ

За последние 50 лет антенных технологии значительно продвинулись в области радиоастрономии. Первый специальный выпуск трудов по антеннам и распространению радиоволн Института радиоинженеров (*Institute of Radio Engineers — IRE*) был издан в 1961 году и посвящен радиоастрономии [1]. Ллойд Беркнер, президент института в то время, отметил во введении: «На сегодняшний день характеристики антенн и фидеров ограничивают прогресс в радиоастрономии… эти [новые] разработки [в области антенн и фидеров] дали всей радио науке и технологиям неиссякаемый источник знаний, которые будут полезны для будущих разработок». Представление об уровне антенной техники для радиоастрономии того времени дают статьи Нэша Крауса (360-футовый телескоп Огайского государственного университета), доклады Свенсона и Ло (радиотелескоп университета Иллинойса) и другие публикации по интерферометрии и технологиям приема радиоволн [1].

Характеристика Беркнера радиоастрономии, как движущей силы прогресса в антеннах и микроволновой технике, была справедлива на протяжении 50 лет. Однако, существующие принципы построения радиотелескопов накладывают некоторые ограничения на такие характеристики антенной системы радиотелескопов, как чувствительность и ширина поля обзора. Если чувствительность может быть увеличена путем разработки и внедрения новых малошумящих усилителей и алгоритмов обработки сигналов, то для расширения поля обзора необходимо, кроме разработки алгоритмов обработки сигналов, пересмотреть концепцию построения самих антенных систем. Одной из таких новых концепций является использование в качестве облучателей зеркальных антенн плотных антенных решеток вместо традиционных рупоров, что позволяет одновременно формировать множество близко расположенных лучей [2].

Более подробно существующие решения и проблемы описаны в первом разделе.

В национальном институте радиоастрономии *ASTRON* (Нидерланды) были разработаны прототипы антенной решетки для использования ее в качестве облучателя зеркал Вестерборгского радиотелескопа-интерферометра (*Westerbork Synthesis Radio Telescope — WSRT*), а также система приема и обработки сигналов с антенных элементов решетки. Проект получил название *APERTIF* (*APERture Tile In Focus*). Цель проекта — заменить используемые рупорные облучатели WSRT облучателями в виде фокальных решеток с целью увеличения скорости обзора неба примерно в 20 раз [3]. К настоящему времени изготовлен прототип системы *APERTIF* для одной из 25-метровых рефлекторных антенн, который включает в себя фокальную решетку, МШУ, приемник и коррелятор с функцией обработки сигнала. Для полного анализа разработанного прототипа и его функционирования в составе рефлекторной системы необходимо решить следующие **актуальные технические задачи**:

— провести анализ чувствительности разработанной гибридной антенной системы с фокальной решеткой и выявить конструктивные параметры и физические эффекты, которые обеспечивают наибольший вклад в формирование чувствительности (такие как, например, потери на согласование в антенне и шумовые параметры МШУ);

— на основе проведенного анализа разработать более эффективные варианты конструкции антенной решетки, системы согласования и питания ее элементов;

— разработать методику расчета и измерения весовых коэффициентов элементов фокальной решетки, оптимальных по критерию максимальной чувствительности и с ограничением на другие параметры, такие как минимальное количество требуемых лучей в поле обзора и эффективность их калибровки с точки зрения минимума калибровочных параметров лучей.

Также существуют другие актуальные технические задачи, которые в настоящее время изучаются. А именно: а) разработка эффективной методики калибровки антенной системы с фокальной решеткой; б) разработка методики улучшения поляризационной дискриминации антенной системы с фокальной решеткой.

Чтобы решить указанные проблемы, необходимо разработать математическую и компьютерную модели системы, которые описывают гибридную антенную систему (состоящую из рефлектора, облучающей фокальной решетки, МШУ и формирователя диаграммы направленности ФДН) и которые позволяют провести численный анализ ее чувствительности и выявить ее определяющие факторы.

**Целью работы является** разработка модели гибридной антенной системы с фокальной решеткой, включающей также малошумящие усилители и формирователь диаграммы направленности (в дальнейшем — просто «система»), используя существующие модели отдельных ее частей, а также анализ прототипа *APERTIF* с помощью разработанной модели, включая оптимальные по нескольким критериям схемы возбуждения элементов решетки.

В соответствии с целью в диссертации решаются следующие **задачи**:

— разрабатывается математическая модель антенной системы с фокальной решеткой, МШУ и ФДН и ее компьютерная модель, реализованная с помощью существующего пакета *CAESAR* [4];

— рассчитываются основные параметры системы на основе созданной модели;

— выполняется оптимизация весовых коэффициентов решетки для заданных критериев оптимизации;

— осуществляется экспериментальная проверка адекватности модели системы.

**Объектом исследований** являются метод формирования многолучевого поля обзора радиотелескопа, а также процессы, происходящие при этом в системе.

**Предметом исследований** являются как отдельные части системы (микрополосковый фидер элемента Вивальди, антенная решетка из элементов Вивальди), так и вся система в комплексе и ее характеристики излучения системы (диаграмма направленности лучей, КИП), матрица сопротивлений между антенными элементами решетки, шумовые характеристики системы и ее чувствительность в поле обзора.

В работе использованы следующие методы исследования и модели:

— *метод характеристических базисных функций* (расширение метода моментов) [4…7] — для анализа распределения токов в решетке и последующего расчета характеристик излучения отдельных элементов решетки в присутствии остальных элементов;

— *эквивалентная модель системы, состоящей из антенной решетки и приемника* [8] — для анализа шумовых характеристик антенной решетки и, в частности, расчета шумовой температуры усилителей с учетом эффекта рассогласования с элементами антенной решетки и весовых коэффициентов ФДН;

— *метод компьютерного моделирования* в программном обеспечении MatLab — для построения компьютерной модели, а также проведения численного моделирования и исследования интересующих параметров системы;

— *метод обработки сигнала в антенных решетках* [9], который основан на построении целевой функции в виды отношения других функций записанных в виде квадратичных форм, например как отношение сигнал-шум.

— *методы экспериментальных исследований*, которые включают использование измерителя комплексных коэффициентов передачи для измерения матрицы сопротивлений между антенными элементами решетки, экспериментальный метод измерения шумовой корреляционной матрицы приемной системы с фокальной решеткой [10] с последующим расчетом чувствительности системы, а также метод измерения диаграммы направленности лучей с помощью сильного источника (звезды) на небесной сфере [11].

**Научная новизна** полученных результатов заключается в следующем:

1. Разработаны математическая и компьютерная модели антенной системы с фокальной решеткой, позволяющая выполнять численное моделирование электродинамических и шумовых характеристик системы.

2. Получены значения оптимальных весовых коэффициентов решетки, позволяющие получить заданную величину равномерности чувствительности в многолучевом поле обзора телескопа. Реализованный подход к оптимизации весовых коэффициентов элементов решетки обеспечивает чувствительность, которая ниже теоритического максимума не более чем на 10%, однако позволяет увеличить ее равномерность в поле обзора по сравнению с классическим ФДН, реализующим максимум чувствительности. Также, используя этот подход, формируются более симметричные диаграммы направленности, что позволяет проводить калибровку радиотелескопа более эффективно, используя простые математические модели лучей с малым количеством неизвестных [96].

**Практическое значение** полученных результатов:

1. Для разработанной модели антенной системы предложены методики расчета важнейших параметров радиотелескопов, таких как шумовая температура системы и ее составляющие, коэффициент использования поверхности (КИП) и его составляющие, распределение чувствительности в поле обзора, и др.

2. Разработанная модель системы была использована для расчета параметров Вестерборгского радиотелескопа, одна из антенн которого оборудована антенной решеткой из элементов Вивальди, результаты сравнены с измерениями, проведенными для этого прототипа системы. Полученное хорошее совпадение результатов подтверждает практическую значимость работы, так как модель может быть использована для предсказания характеристик системы, а также разработки усовершенствованных прототипов, один из которых был недавно разработан, и на данный момент находится в процессе изготовления [97].

3. Разработана инженерная программа для расчета и анализа характеристик зеркальных антенн с фокальными решетками, позволяющая также оптимизировать весовые коэффициенты решетки по любому описанному пользователем критерию. Эта программа используется в настоящее время в проекте *DVP* (*Dish Verification Program under the US SKA project*) для анализа и оптимизации геометрии рефлекторных систем (в частности, для сравнения симметричных и офсетных систем) в комбинации с облучателями в виде фокальных решеток.

4. Модель может быть использована для анализа динамического диапазона радиоастрономических карт (например, как это было сделано для *WSRT* радиотелескопа [98]), а также может служить основой для разработки новых эффективных техник калибровки радиотелескопов [96].

**Апробация результатов диссертации.** Результаты работы докладывались и обсуждались на следующих конференциях: 4[th] International *Conference on Ultrawideband and Ultrashort Impulse Signals UWBUSIS'*2008 (*Sevastopol*, 2008); 3-ої міжнародної конференції «Проблеми телекомунікацій ПТ-2009» (Київ, 2009); 5-й молодежной конференции «Современные проблемы радиотехники и телекоммуникаций РТ-2009» (Севастополь, 2009); 13-м молодежном форуме «Радиоэлектроника и молодежь в XXI веке» (Харьков, 2009); *Int. Symposium On Antennas and Propagation* (*Charleston*, *SC*, *USA*, 2009); 4[th] *and* 5[th] *European Conference On Antennas and Propagation*

*EuCAP* (*Barcelona*, 2010 *and Rome*, 2011); *International Conference on Electromagnetics in Advanced Applications ICEAA'2010* (*Sydney*, 2010)

**Публикации.** По результатам исследований опубликовано 13 научных трудов, в том числе 3 статьи в специальных изданиях из перечня ВАК Украины, 2 статьи в *IEEE Transactions on Antenna and Propagation* и 8 работ в материалах международных научно-технических конференций.

**Структура и объем диссертации.** Диссертация состоит из введения, четырех разделов, заключения, списка используемых источников и одного приложения.

**В первом разделе** приведен обзор некоторых существующих антенных систем для радиоастрономии; описаны структура, требования и проблемы разрабатываемого в настоящее время телескопа *Square Kilometer Array* (*SKA*); описан проект *APERTIF* и структура системы, модель которой разработана в следующих разделах, обоснован метод моделирования металлической структуры антенной решетки и рассмотрены несколько схем формирования ДН, наиболее интересных для радиоастрономических наблюдений.

**Во втором разделе** разработана теоретическая модель антенной системы с фокальной решеткой. В частности:

— рассмотрена электромагнитная модель антенной решетки из элементов Вивальди;

— разработана микроволновая модель микрополоскового фидера (МПФ) для элементов Вивальди и показан способ нахождения одного из ее неизвестных параметров (коэффициента трансформации);

— разработан алгоритм моделирования комплексной гибридной антенной системы и расчета ее параметров по трем критериям: а) обеспечение приема максимальной мощности падающей плоской волны, б) обеспечение максимальной чувствительности системы, в) нахождение компромисса между максимальной чувствительностью и ее равномерностью в широком поле обзора.

**В третьем разделе** приведены результаты моделирования отдельных частей общей модели системы. В частности:

— проведено моделирование МПФ элементов Вивальди, показано влияние его параметра — коэффициента трансформации — на $S$-матрицу решетки;

— приведено описание усовершенствований первого прототипа решетки системы *APERTIF*, в том числе способ улучшения ее шумовых характеристик и способ улучшения стабильности электродинамических параметров при механическом воздействии;

— приведены результаты расчета $S$-матрицы антенной решетки с МПФ (прототип *APERTIF*) и произведено ее сравнение с полученной экспериментально;

— произведена экспериментальная проверка точности моделирования ДН антенных элементов решетки путем расчета реального фокусного расстояния зеркала.

**В четвертом разделе** приведены результаты моделирования всей антенной системы на примере второго прототипа системы *APERTIF*. В частности, рассчитаны такие ее параметры, как

— первичные (до отражения от зеркала) и вторичные (после отражения от зеркала) ДН элементов решетки и всей решетки;

— весовые коэффициенты, оптимальные по трем критериям (максимальная принятая мощность, максимальная чувствительность, максимальная равномерность чувствительности при уменьшении чувствительности не более чем на 10%);

— коэффициенты эффективности системы при заданных весовых коэффициентах (КИП, коэффициенты эффективности амплитудного и фазового распределений в апертуре зеркала, коэффициент перехвата энергии облучателя, КПД);

— шумовая температура системы и ее составляющие: за счет приема шумов земли ($T_{sp}$) и неба ($T_{sky}$), за счет шумящих МШУ и взаимного влияния элементов решетки друг на друга ($T_{lna}$), за счет шумящих элементов системы после МШУ ($T_{sec}$), за счет потерь энергии в системе ($T_{rad}$);

— чувствительность системы для трех критериев оптимизации весовых коэффициентов;

— поляризационные свойства системы, которые для радиотелескопа можно определить в виде коэффициента ортогональности пар лучей, сканирующих в одном направлении при наблюдении источника сигнала со случайной поляризацией [99].

Для расчета весовых коэффициентов по критерию равномерной чувствительности была проведена оптимизация параметра $g_{const}$ (см. подраздел 2.3.4) данного критерия.

Также в четвертом разделе приведено сравнение чувствительности, полученной в результате моделирования, с измеренной чувствительностью.

# РАЗДЕЛ 1
# ОСОБЕННОСТИ ПОСТРОЕНИЯ И МОДЕЛИРОВАНИЯ АНТЕНН ДЛЯ РАДИОАСТРОНОМИИ

## 1.1. Развитие радиоастрономии в последние десятилетия

В течение уже длительного времени в радиоастрономии широко используются большие рефлекторные антенны, из которых самыми большими являются 100-метровый телескоп Грин-Бэнк (*Green Bank Telescope — GBT*) [12] (рис. 1.1) и 300-метровый телескоп в Аресибо (*Arecibo Telescope*) [13] (рис. 1.2), а также синтезированные решетки, такие как Вестерборгский радиотелескоп (*Westerbork Synthesis Radio Telescope — WSRT*) [14], *Very Large Array* (*VLA*) в Соккоро, штат Нью Мексико [15], глобальная сеть *VLBI* (*Very Large Baseline Interferometer*) [16]. Более новыми инструментами радиоастрономии являются антенная решетка *Atacama Large Millimeter Array* (*ALMA*) в Чили [17], *Giant Metrewave Radio Telescope* (*GMRT*) в Индии [18] (рис. 1.3) и *Allen Telescope Array* в Калифорнии [19]. Также разработано и исрользуется большое количество других инструментов общего и специального назначения. Ключевыми технологимяи, позволившими сделать шаг вперед, стали алгоритмы формирования изображений, полученных методом синтезированных апертур; стабильные системы синхронизации и записи для VLBI (рис. 1.4); цифровые спектрометры и корреляторы, построенные на современных интегральных схемах и программируемых логических интегральных схемах (ПЛИС); высокоэффективные облучатели, такие как гофрированные рупоры; методы цифровой обработки для детектирования пульсаров. Особую важность для высокочувствительных радиотелескопов представляют малошумящие приемники и интегральные СВЧ усилители, построенные с использованием дискретных криогенных транзисторных усилителей, в которых достигается шумовая температура, близкая к абсолютному нулю.

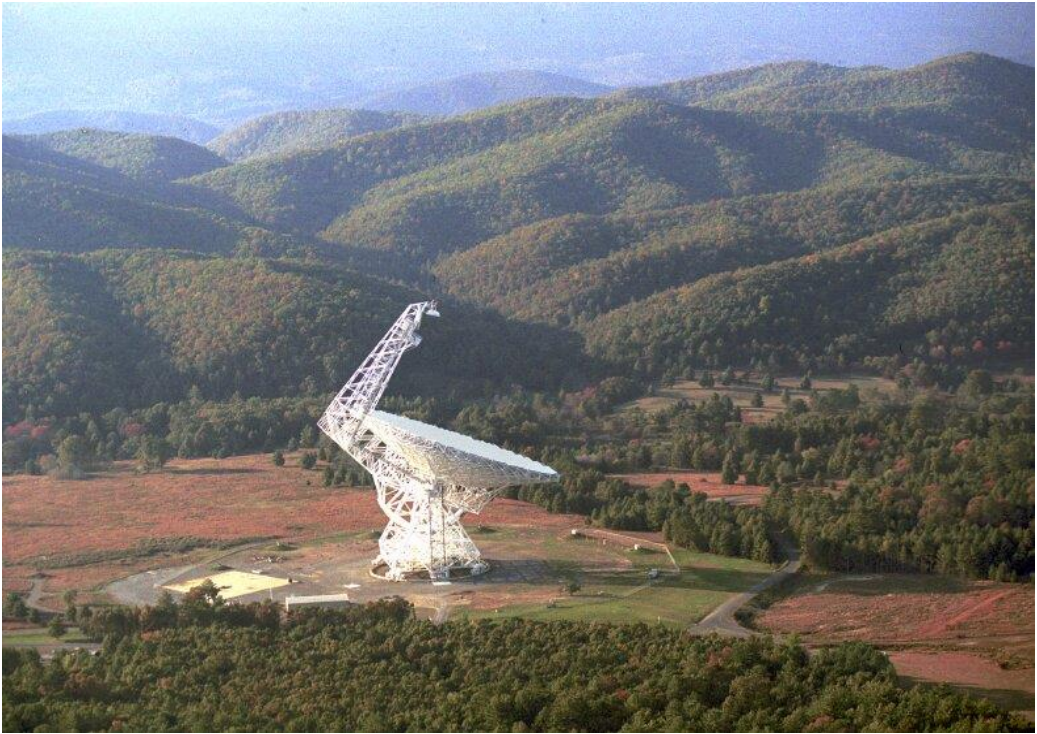

Рис. 1.1. *Green Bank Radio Telescope* (США)

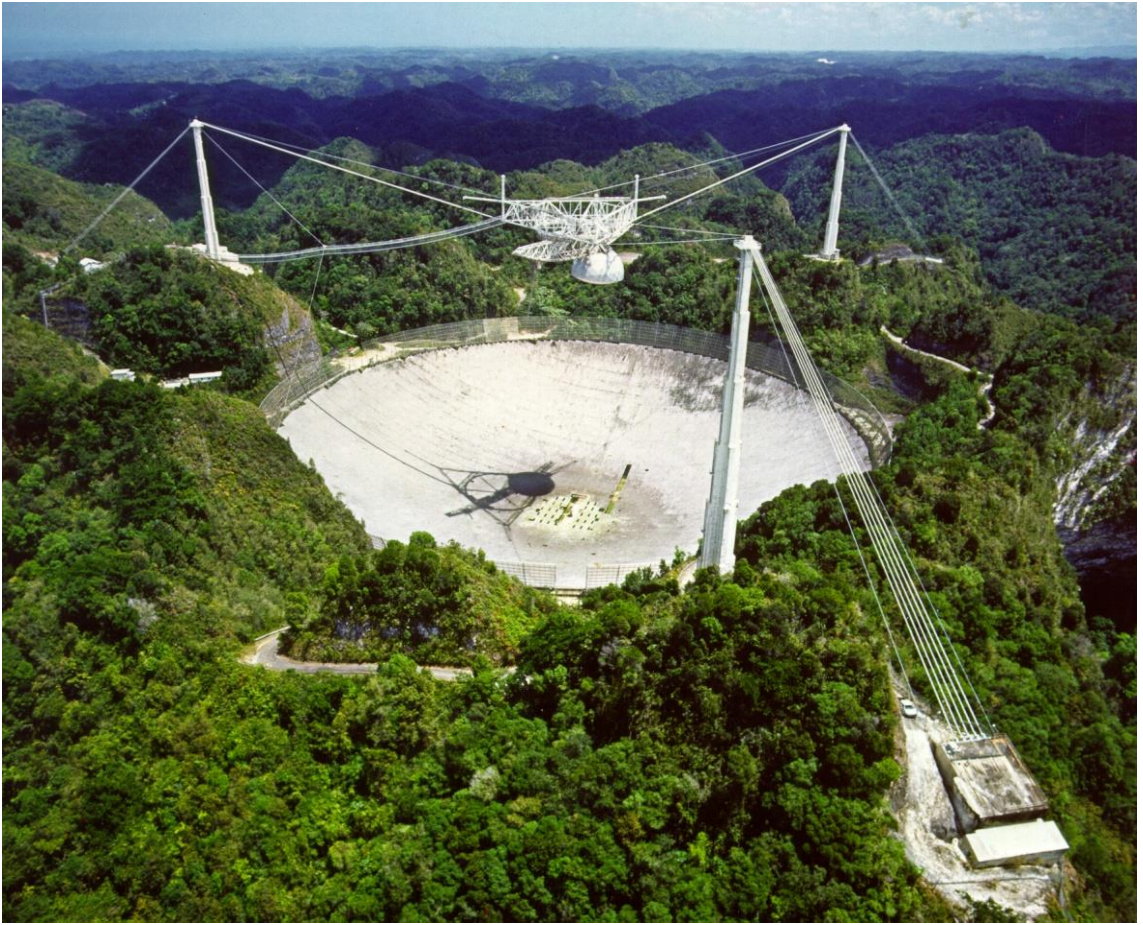

Рис. 1.2. *Arecibo Radio Telescope* (Нью Мексико)

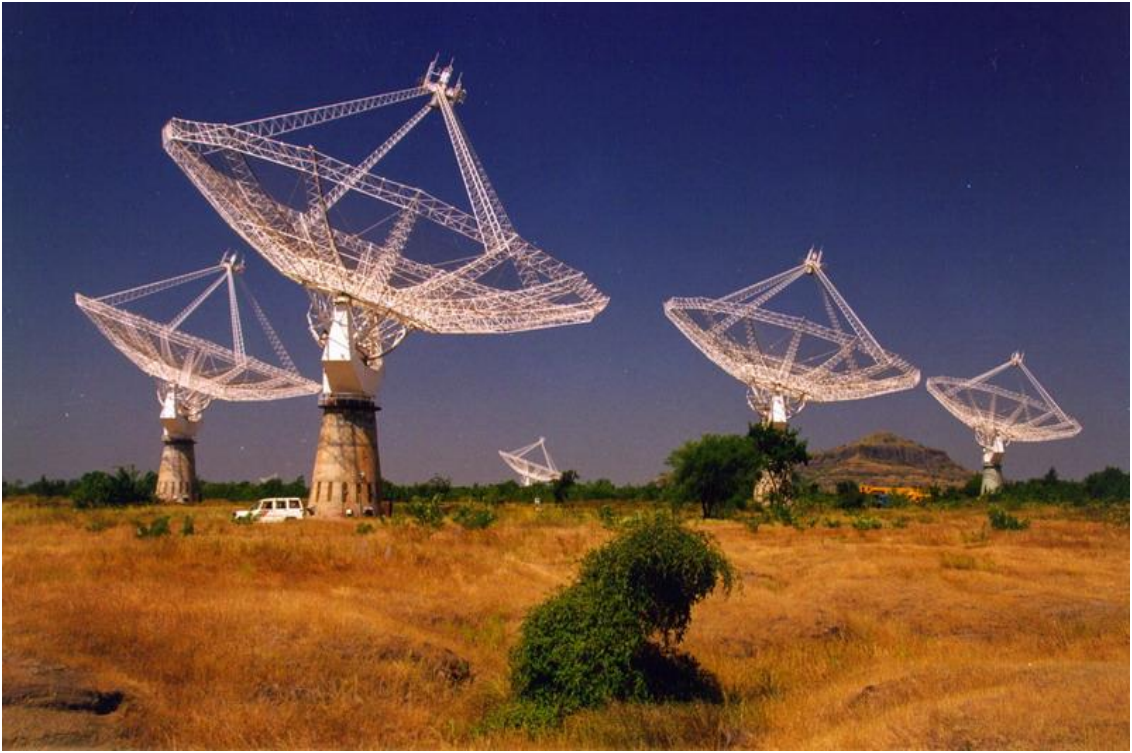

Рис. 1.3. *Giant Metrewave Radio Telescope* (*GMRT*) в Индии

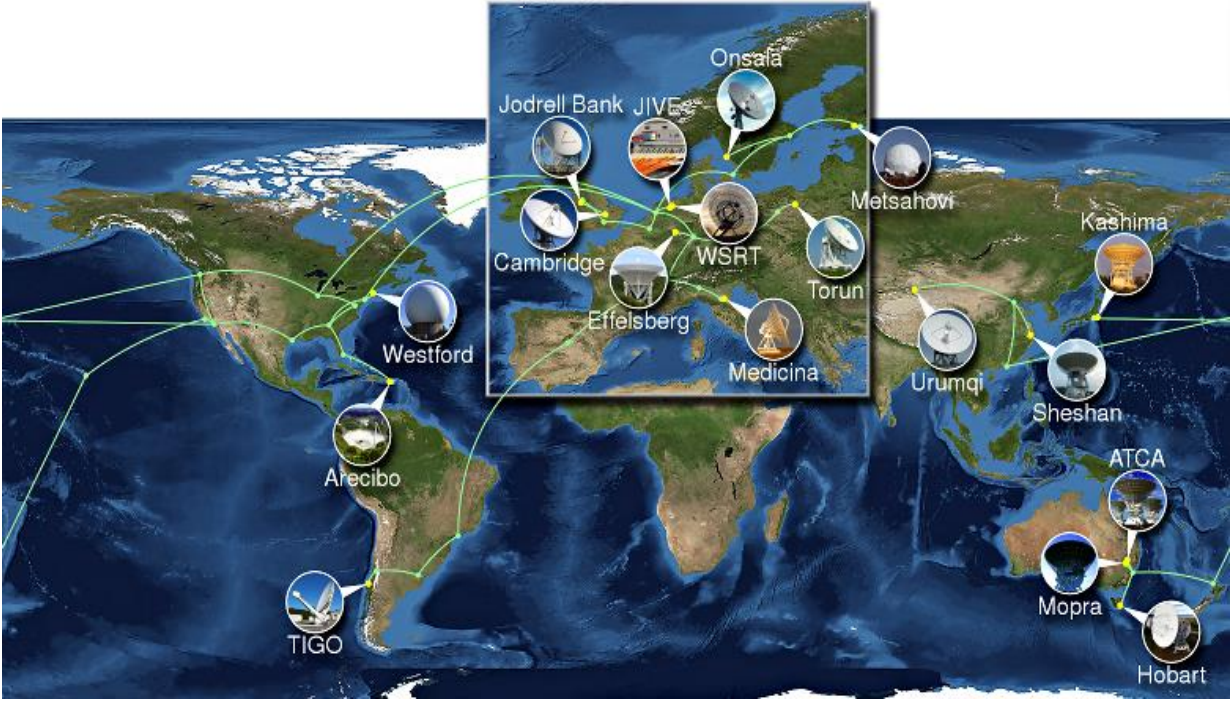

Рис. 1.4. Узлы сети *VLBI*

Радиоастрономия продолжает интенсивно развиваться. Так, в Китае идет подготовка для построения 500-метрового сферического телескопа *Five hundred meter Aperture Spherical radio Telescope* (*FAST*) [20]**,** который станет самым большим однорефлекторным телескопом в мире**.** Также на разных стадиях находится планирование и проведение усовершенствований существующих систем, таких как *Expanded Very Large Array* (*EVLA*), оборудование фидерами-решетками телескопов Аресибо, *WSRT* и *GBT*. Такие проекты, как *Precision Array to Probe the Epoch of Reionization* (*PAPER*), *Long Wavelength Array* (*LWA*), *European Low Frequency Array* (*LOFAR*), *MeerKAT* в Южной Африке, и *Australian Square Kilometre Array Pathfinder* (*ASKAP*), создают в настоящее время фундамент для радиотелескопа нового поколения, известного как *Square Kilometer Array* (*SKA*) [20]. Телескоп *SKA* является слишком сложным и дорогостоящим (примерно $2,9 млрд.), чтобы быть построенным какой-либо одной страной. Поэтому был сформирован консорциум из 19 стран, общими усилиями которых будет разработан и построен *SKA*. В октябре 2006 года консорциум выдвинул две страны-претендента на размещение телескопа: Австралию и Южную Африку.

## 1.2. Обзор принципов построения облучающих систем

### 1.2.1. Одиночные облучатели

Традиционно в качестве облучателей зеркал рефлекторных антенн используются рупора, как правило, гофрированные. Некоторые варианты рупоров показаны на рис. 1.5 [21, 22, 23].

Недостаток таких облучателей – не достаточно широкая полоса рабочих частот, которая не превышает 80% (рупор на рис. 1.5, а) даже для облучателей с большой апертурой. Для увеличения диапазона частот было разработано множество вариантов широкополосных облучателей, например, облучатель для *Goldstone Apple Valley Radio Telescope* [24] и *Eleven* антенна [25], показанных на рис. 1.6. Однако, такие облучатели на сегодняшний день еще не нашли применения в радиоастрономических системах и находятся на стадии изучения и испытаний.

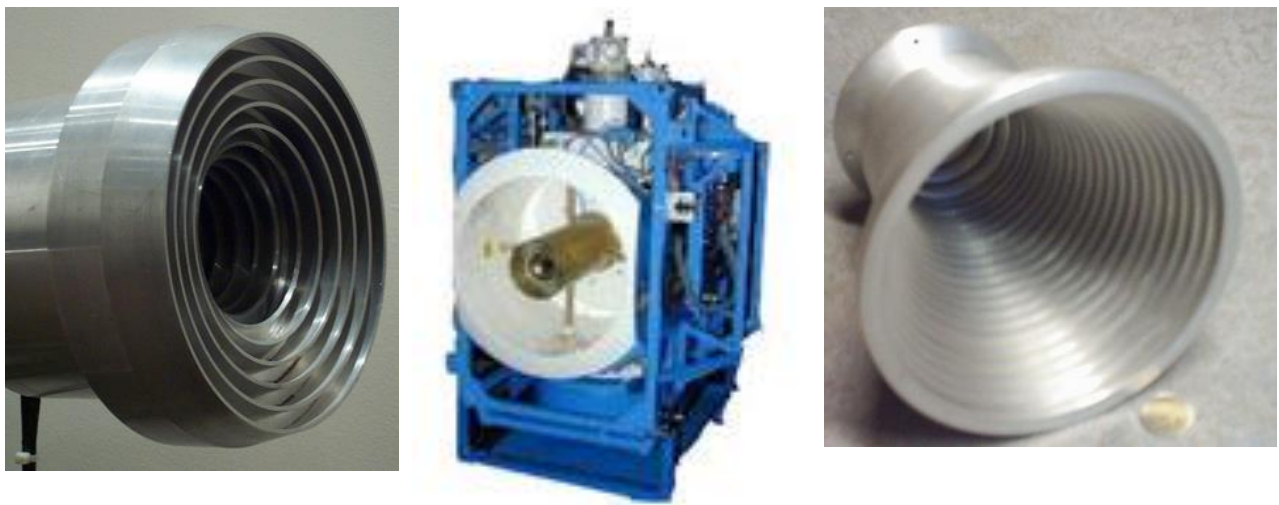

а) б) в)

Рис. 1.5. Традиционные рупорные облучатели:

а) Широкополосный рупор, предложенный *B.M.Thomas* и др.;

б) Двухдиапазонный рупор, используемый в *WSRT*; в) Широкополосный рупор, предложенный *J.Teniente* и др.

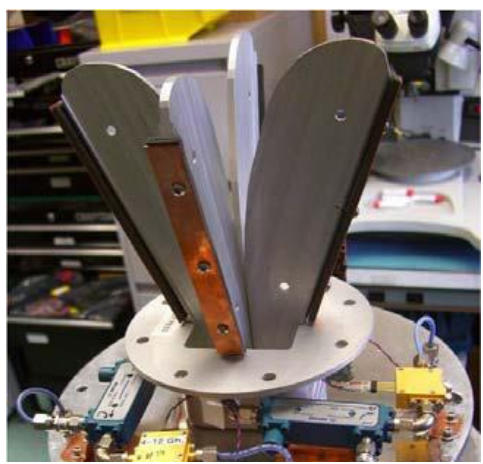 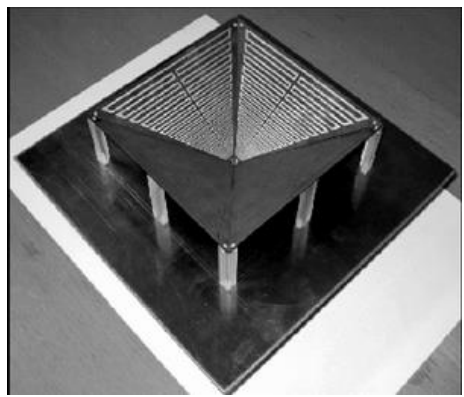

а) б)

Рис. 1.6. Широкополосные однолучевые облучатели:

а) Облучатель для *Goldstone Apple Valley Radio Telescope*, ($\Delta f/f = 3,5$);

б) *Eleven* антенна, ($\Delta f/f > 10$)

Однако, несмотря на относительную простоту систем с описанными облучателями, основная проблема современных радиотелескопов-интерферометров с такими облучателями — малая скорость обзора неба. Чтобы создать как можно больше баз при интерферометрических наблюдениях, один цикл измерений длится 12 часов (пока Земля совершит пол оборота вокруг своей оси). За этот цикл просматривается только участок неба площадью, равной площади поля обзора антенны. Так, например, у Вестерборгского радиотелескопа ширина ДН на частоте 1,42 ГГц равна примерно 0,5 градуса. Несложно подсчитать, что при такой ширине ДН построение карты полной небесной сферы (площадью $4\pi$ кв. радиан) займет около 140 лет. На более высоких частотах это время будет еще больше.

### 1.2.2. Кластеры рупоров

Одним из способов увеличения скорости обзора неба является применение в качестве облучателей групп, или кластеров, рупоров в конфигурации «один рупор — один луч». Это позволяет одновременно проводить наблюдения нескольких участков неба. Были предложены различные варианты. Некоторые из них показаны на рис. 1.7 [26, 27].

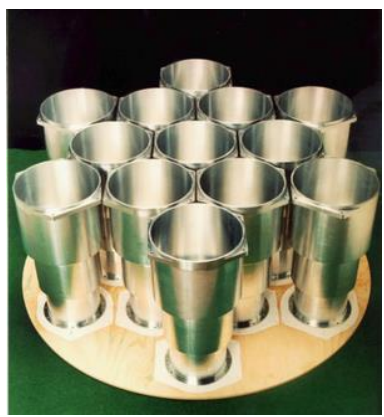      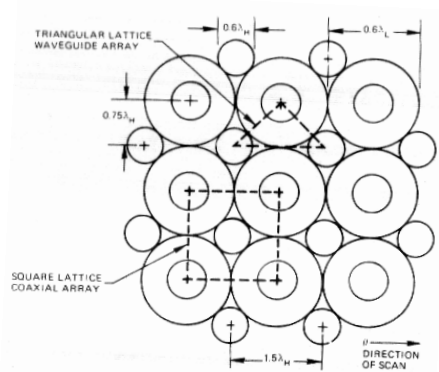

а)                          б)

Рис. 1.7. Рупорные кластеры:

а) облучатель для *Parkes Radio Telescope*, ($\Delta f/f = 0{,}16$);  б) концепция двухполосного кластерного облучателя

Однако, хотя использование кластеров из рупорных облучателей в конфигурации «один рупор – один луч» дает выигрыш в скорости обзора неба, оно имеет существенные недостатки: низкая чувствительность в точках пересечения соседних лучей и узкая рабочая полоса частот (для рупоров в составе кластеров она не превышает 30%) [26]. Остановимся более подробно на анализе чувствительности *Parkes Radio Telescope*.

Чувствительность интерферометра вычисляется по формуле [28]

$$S_{tot} = \frac{A_{eff}}{T_{sys}} \frac{\sqrt{\Delta f \cdot t \cdot n}}{K_S},$$ (1.1)

где $A_{eff}$ — эффективная площадь зеркала; $T_{sys}$ — общая шумовая температура системы; $\Delta f$ — полоса частот, в которой происходит наблюдение; $t$ — время интегрирования в корреляторе; $n$ — количество усредненных записей; $K_S$ — коэффициент чувствительности, который зависит от типа приемной системы и коррелятора.

Рассмотрим гексагональное расположение рупоров, как показано на рис. 1.7, а. Три соседних рупора такого облучателя будут создавать в небе три луча, показанных на рис. 1.8. На рисунке $\theta_b$ — это ширина главного лепестка (аберрацией пренебрегаем) и $\theta_s$ — угловое расстояние между максимумами соседних лучей.

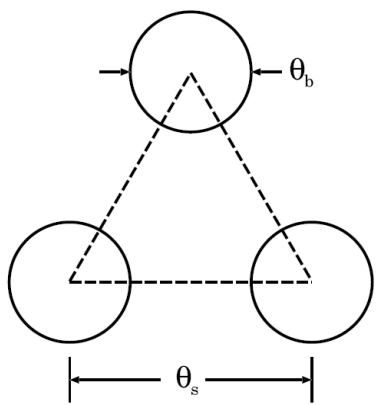

Рис. 1.8. Три луча гексагонального рупорного кластера

Площадь треугольной ячейки на рис. 1.8, $A_c$, при малых углах $\theta$ приблизительно равна

$$A_c \approx \frac{1}{2}\theta_s\left(\frac{\sqrt{3}}{2}\theta_s\right) = \frac{\sqrt{3}}{4}\theta_s^2 \quad [\text{кв. градусов}]. \qquad (1.2)$$

Площадь участка неба, покрываемого одним лучом, $A_b$, равна

$$A_b = \frac{\pi}{4}\theta_s^2. \qquad (1.3)$$

Тогда коэффициент заполнения поля обзора можно определить как

$$\eta_b = \frac{A_b}{A_c} = \frac{\pi}{\sqrt{3}}\left(\frac{\theta_b}{\theta_s}\right)^2 \qquad (1.4)$$

Для *Parkes Radio Telescope* с облучателем, показанным на рис. 1.7,а отношение ширины луча к расстоянию между лучами равно примерно 1/2, поэтому коэффициент заполнения поля обзора $\eta_b \approx 0,46$. Это означает, что телескоп необходимо позиционировать как минимум три раза, чтобы покрыть всю площадь многолучевого поля обзора. При этом из (1.1) следует, что чувствительность уменьшается в $\sqrt{3}$ раз по сравнению с некой системой, с помощью которой было бы возможно покрыть все поле обзора за одно позиционирование зеркала (при одинаковом общем времени интегрирования в поле обзора), т.к. для последней системы время интегрирования $t$ может быть в 3 раза больше. Кроме того, чувствительность зависит и от полосы частот. Приведенный выше облучатель (рис. 1.7,а) является узкополосным (отношение полосы частот к средней рабочей частоте $\Delta f / f = 0,16$), что также накладывает ограничения на максимальную чувствительность. Для более широкополосных рупоров, например рупора на рис. 1.5,а, коэффициент заполнения поля обзора меньше 0,2 [26].

### 1.2.3. Плотные фокальные решетки

Покрыть все поле обзора за одно позиционирование зеркала можно с использованием «плотных» антенных фокальных решеток (АФР), которые позволяют формировать хорошо перекрываемые лучи (уровень в точках пересечения соседних лучей уменьшается не более чем на 3 дБ). Под понятием «плотная» антенная решетка понимается решетка с расстоянием между ее элементами на верхней частоте рабочего диапазона менее $\lambda$ для рефлекторов с отношением фокусного расстояния к диаметру $F/D>1$ и менее $0,6\lambda$ для рефлекторов с $F/D<0,5$ [2, 29]. Это необходимо, чтобы удовлетворить теореме Котельникова при выборке поля в фокальной области.

Кроме хорошего перекрытия лучей, фокальные решетки имеют еще ряд преимуществ по сравнению с одиночными облучателями и кластерами [26]:

1) для параболических рефлекторов характерна сильная аберрация (особенно коматическая), и, поэтому, чем дальше ведется сканирование путем смещения рупорного облучателя, тем меньше становится КИП, и, следовательно, чувствительность. Причем эффект усиливается с увеличением глубины зеркала. При использовании плотных АФР есть возможность скомпенсировать это искажение фокального поля путем корректировки весовых коэффициентов решетки [30];

2) при использовании рупорных облучателей невозможно уменьшить эффект рассеяния поля от опор, держащих облучатель, и эффект блокировки падающей волны облучателем, что увеличивает шумовую температуру системы. В облучателях с АФР это возможно путем оптимального формирования лучей (выбора весовых коэффициентов) (см., например, экспериментально найденное распределение поля в апертуре зеркала в [11] для рупорного облучателя и АФР);

3) параметры лучей системы (ширина главного лепестка, КНД), формируемых рупорными облучателями, имеют ярко выраженную частотную зависимость. В АФР этот эффект может быть частично компенсирован, так как весовые коэффициенты также могут быть функцией частоты [31];

4) в отличие от рупоров, элементы фазированной плотной решетки могут быть широкополосными, что дополнительно дает выигрыш в чувствительности (см. (1.1));

5) В АФР возможно множество оптимальных схем возбуждения элементов решетки (по различным критериям);

6) С помощью АФР возможно скомпенсировать деформацию поверхности зеркала из-за гравитационного влияния или ошибок изготовления.

АФР имеют и недостатки. Среди них — высокая стоимость системы за счет ее усложнения (сама решетка, многоканальный приемник, более сложный коррелятор интерферометра), больших размеров решетки (по сравнению с традиционными облучателями), а так же более высокая (на 10...30%) шумовая температура $T_{sys}$ из-за сложности оптимального согласования по шумам всех элементов решетки (которые имеют разные значения активного импеданса) с малошумящими усилителями, работающими в неохлажденном режиме. Реализация охлажденных систем теоретически возможна, но с практической точки зрения реализуема на более высоких частотах или для фокальных решеток с небольшим количеством лучей. Но, как показано в [26], скорость обзора неба все равно может быть выше, чем для систем с кластерами рупоров.

АФР исследуются уже давно, начиная с 70-х годов. Теоретически рассматривались такие вопросы, как оптимальное расстояние между элементами решетки и оптимальное отношение $F/D$ зеркала [2], КИП, составляющие шумовой температуры системы, общие выражения для максимизации чувствительности системы с АФР [8], взаимное влияние элементов решетки друг на друга и вклад в шумовую температуру системы за счет этого [10], влияние искажения поверхности зеркала на параметры системы и другие вопросы.

В [32] приводятся первые экспериментальные результаты для АФР, состоящей из 144 элементов Вивальди (имеющих сильное взаимное влияние друг на друга), из которых активными были лишь 13 элементов. Но использован был простой ФДН, формирующий только осевой луч. В статье было экспериментально показано, что использование фокальных решеток дает выигрыш в КИП по сравнению с традиционными облучателями. Также в статье приводится метод анализа взаимной связи элементов решетки, и влияние этой связи на рассогласование элементов решетки с последующими цепями.

В [11] приведены результаты измерений ДН отдельных элементов решетки (первичные ДН) и ДН элеметвов решетки после отражения от зеркала (вторичные ДН), распределения чувствительности в поле обзора, приведено сравнение поля в апертуре зеркала для рупорного облучателя и АФР, а также сравнение зависимостей КИП от частоты, приведены параметры Стокса для одного из лучей.

В настоящей работе приведенные выше результаты теоретических исследований объединяются в единую модель рефлекторной антенной системы с фокальной решеткой, и с помощью этой модели проводится анализ прототипа многолучевой системы APERTIF.

### 1.3. Современное состояние разработок радиотелескопов

1.3.1. Радиотелескоп *Square Kilometer Array* и его характеристики

*Square Kilometer Array* (*SKA*) — это радиотелескоп нового поколения, который начнет функционировать в 2020 году. *SKA* станет революционным инструментом, у которого общая поверхность апертуры будет в 30 раз больше, а полоса частот в 10 раз шире, чем у самого большого традиционного радиотелескопа на сегодняшний день. Увеличение апертуры телескопа приведет к увеличению его чувствительности, что позволит принимать сигналы от более удаленных и более слабых небесных объектов. Одна из целей *SKA* — сделать возможным прием сигналов от объектов ранней Вселенной, которые являются наиболее удаленными. Естественно, эти сигналы очень слабые, и поэтому от инструмента требуется высокая чувствительность, и, следовательно, *SKA* должен иметь площадь не менее 1 км$^2$.

Чтобы обеспечить апертуру в 1 км$^2$ при относительно невысокой стоимости, *SKA* будет существенно отличаться от существующих телескопов, построенных по принципу одиночной антенны, и будет построен с использованием антенных решеток различных технологий. Институты, принимающие участие в разработке *SKA*, в настоящее время разрабатывают и строят прототипные системы; ключевые результаты этих разработок будут использоваться в *SKA*. Множество различных технологических решений будет отобрано и внедрено в проект. В качестве антенн предполагается использовать как планарные апертурные решетки из антенных элементов различных типов, так и зеркальные антенны, в том числе и гибридные зеркальные антенны. *Технология гибридных антенн должна обеспечивать многолучевость, то есть одновременное наблюдение множества* *источников в пределах большого поля обзора.* В *SKA* угол обзора будет от 20° до 1° в диапазоне частот от 1 до 10 ГГц.

Вся собирающая площадь *SKA* будет распределена между множеством станций, количество которых будет достигать нескольких сотен. Каждая станция будет иметь диаметр (100…200) м. Для сравнения, самый большой телескоп на сегодняшний день в Аресибо имеет диаметр 305 м, за ним идут телескоп *GBT* (100×110 м) и телескоп Эффельсберг в Германии (100 м).

Первая предложенная разработка *SKA* содержит: а) планарные апертурные решетки (АР), б) небольшие механически сканирующие зеркала с антенными фокальными решетками (АФР) и в) зеркала с широкополосными однолучевыми фидерами (ШФ). Перекрытие всего частотного диапазона будет происходить путем разбиения на три полосы частот: нижние частоты (антенны — АР), средние частоты (антенны — зеркальные антенны с АФР) и верхние частоты (антенны — зеркальные антенны с ШФ).

Если смотреть на *SKA* с высоты птичьего полета, расположение антенных станций будет иметь форму многозаходной спирали (рис. 1.9). Большое число зеркальных антенн и АР будут размещаться близко друг к другу в центральной области диаметром (15…20) км. Около половины собирающей поверхности телескопа будет находиться в этой области. Другая половина будет распределена по станциям, расположенным на лучах спирали. Самые дальние станции будут удалены на тысячи километров, что позволит получить высокую разрешающую способность с использованием интерферометрии со сверхдлинными базами (*Very Large Baseline Interferometer — VLBI*).

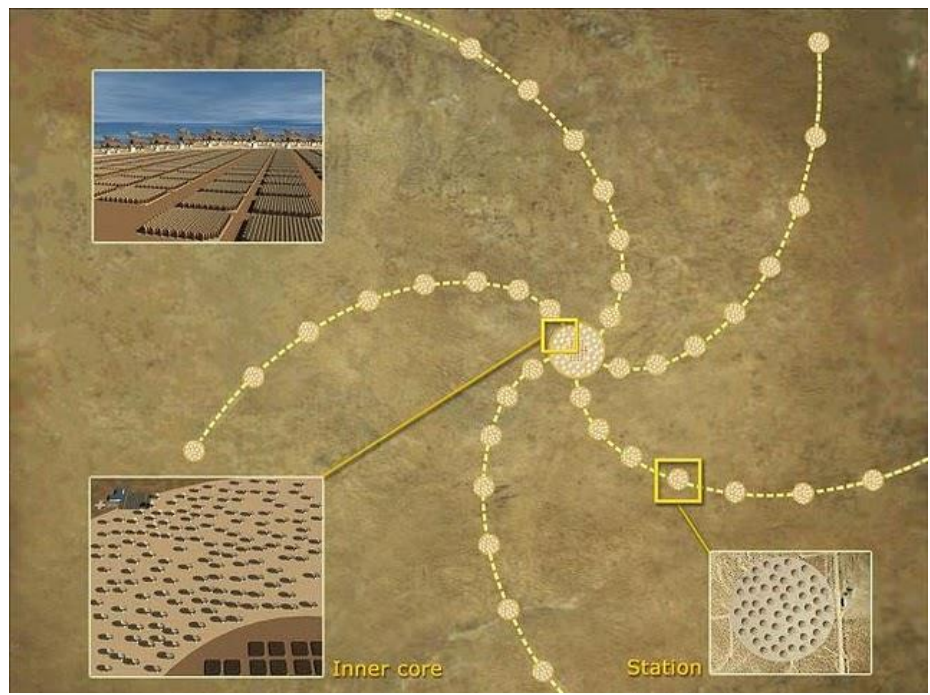

Рис. 1.9. Конфигурация *SKA*

На рис. 1.10…1.12 показаны отдельные части предложенного варианта телескопа SKA.

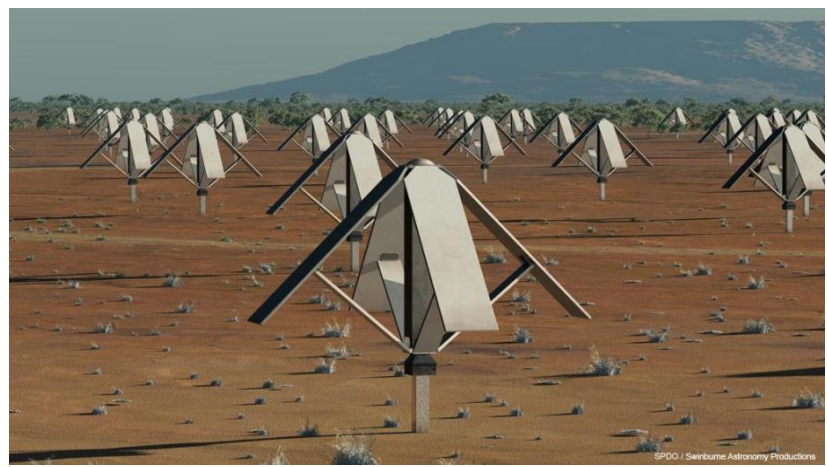

Рис. 1.10. Апертурная решетка из двуполяризованных дипольных антенн, используемых на нижних частотах рабочего диапазона

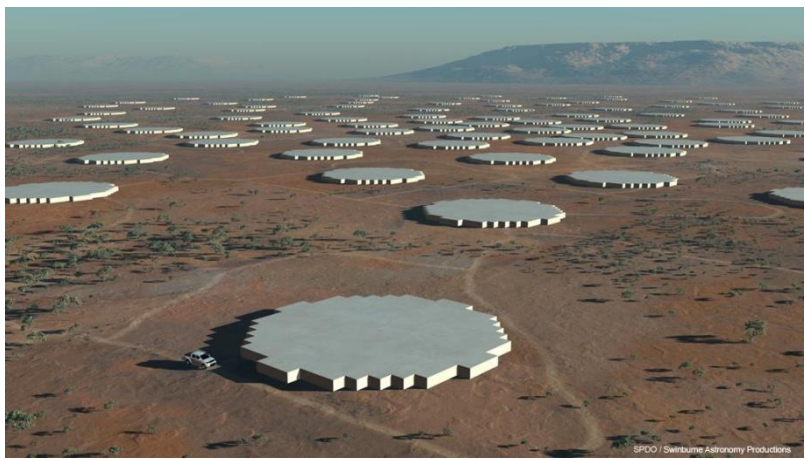

Рис. 1.11. Панели из апертурных решеток, используемых на средних частотах

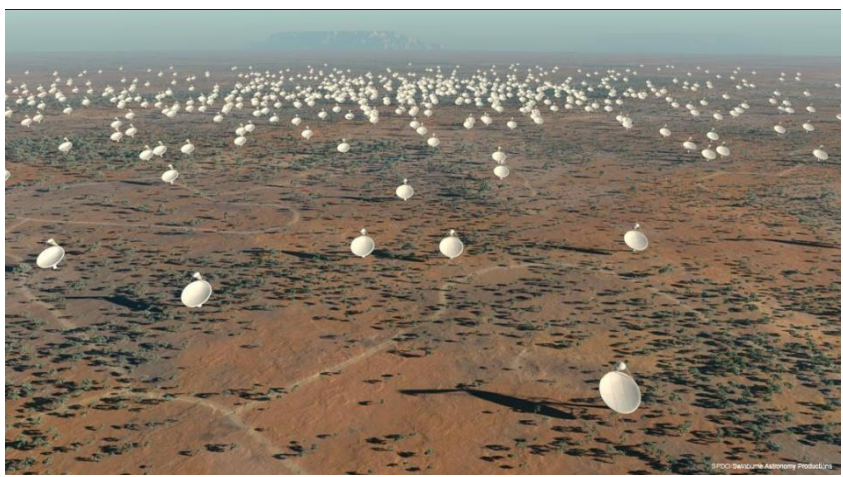

Рис. 1.12 — Зеркальные антенны с фокальными решетками и широкополосными однолучевыми фидерами, используемые в верхнем диапазоне частот

Характеристики радиотелескопа $SKA$.

Один из ключевых показателей качества любого радиотелескопа — это его чувствительность. Чувствительность характеризует способность телескопа измерять параметры излучения слабых источников и определяется как отношение эффективной площади антенной системы $A_{eff}$ к шумовой температуре всей системы $T_{sys}$. Чувствительность также зависит от полосы пропускания приемника, поэтому полоса пропускания является важным параметром для широкополосных систем. Так как космические объекты — это источники очень слабого излучения, то антенны радиотелескопов должны большую эффективную площадь, а приемный тракт — иметь минимальную шумовую температуру. Поэтому *SKA* имеет достаточно жесткие требования к антенной системе с приемником [33]:

— высокая чувствительность (до 12000 м$^2$/К);

— широкий диапазон рабочих частот телескопа (от 70 МГц до 35 ГГц);

— широкая полоса частот, в которой одновременно ведется прием (300…400) МГц;

— динамический диапазон антенной системы — не менее 60 дБ;

— шумовая температура системы ниже 20 К в нижнем диапазоне частот;

— стойкость к радиочастотной интерференции;

— высокая стабильность параметров системы;

— широкое поле обзора (на 1…2 порядка шире, чем у существующих радиотелескопов).

Системы, построенные после 1961 года, удовлетворяют большинству этих требований (в частности, требованию высокой эффективности и низкой шумовой температуры приемника), но достижение высокой стабильности параметров и высокой чувствительности в широком поле обзора и широкой полосе частот до сих пор представляют трудности для современных разработчиков антенных систем.

1.3.2. Система *APERTIF* — первый радиотелескоп с фокальными решетками

*APERTIF* (*APERture Tile In Focus*) — это зеркальная антенная система с АФР, разработку которой ведет институт *ASTRON* (*The Netherlands Institute for Radio Astronomy*), рис. 1.13. Цель проекта — заменить традиционные рупорные облучатели Вестерборгского радиотелескопа-интерферометра (*Westerbork Synthesis Radio Telescope — WSRT*) на фокальные решетки, тем самым увеличив скорость обзора неба примерно в 20 раз [3]. Планируется, что проект будет завершен в 2012…2013 годах.

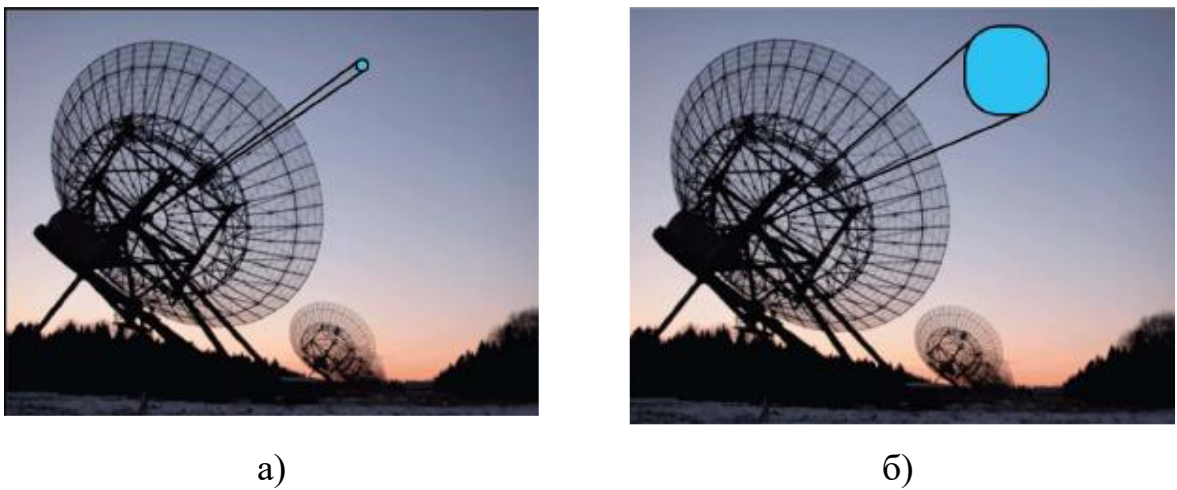

а)                                              б)

Рис. 1.13 — Схематическое изображение поля обзора Вестерборгского

радиотелескопа, расположенного в Нидерландах:

а) *WSRT* в настоящее время; б) *WSRT* в 2013 году

**Обзор системы *APERTIF*.**

*APERTIF* будет функционировать в диапазоне частот от 1000 до 1750 МГц, иметь полосу пропускания 300 МГц, шумовую температуру системы 55 К и коэффициент использования поверхности (КИП) 75% [11]. Увеличение поля обзора предполагается осуществить за счет одновременного формирования 37 лучей. Поле обзора имеет гексогональную форму с угловой площадью примерно 8 кв. градусов (рис. 1.14).

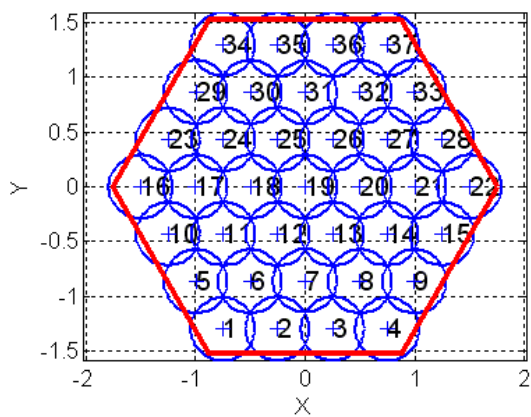

Рис. 1.14 — Расположение лучей в поле обзора

Расположение лучей в поле обзора на рисунке показаны в *True-View* координатной системе [34], соотношения для координат которой и координат сферической системы координат могут быть записаны следующим образом (третья координата *R* не используется, так как рассматривается дальняя зона и поэтому нас интересуют только направления):

$$\theta = \sqrt{X^2 + Y^2} \qquad (1.5)$$

$$\varphi = \arctan\left(\frac{Y}{X}\right). \qquad (1.6)$$

Представление лучей в координатной системе *True-View* удобно тем, что при малых углах $\theta$ (до 3 градусов) легко рассчитываются направления лучей путем равномерного заполнения ими гексогонального поля обзора на плоскости.

Рупорные облучатели, которыми в настоящее время оборудован *WSRT*, имеют шумовую температуру 30 К, КИП 55% и полосу пропускания 160 МГц. АФР несколько уменьшит чувствительность системы для однолучевых наблюдений, но это уменьшение компенсируется 37-кратным увеличением поля обзора. Таким образом, общий выигрыш в скорости обзора неба по сравнению с нынешней системой будет примерно в 20 раз.

На рис. 1.15 показана укрупненная структурная схема системы *APERTIF*.

Каждый из 14 рефлекторов интерферометра оборудован АФР и системой калибровки. Сигнал, принятый каждым элементом решетки и усиленный малошумящими усилителями (МШУ), поступает в приемник, где фильтруется, переводится на промежуточную частоту и оцифровывается. Далее он обрабатывается и поступает в центральный коррелятор, после которого поток данных сохраняется и в дальнейшем обрабатывается.

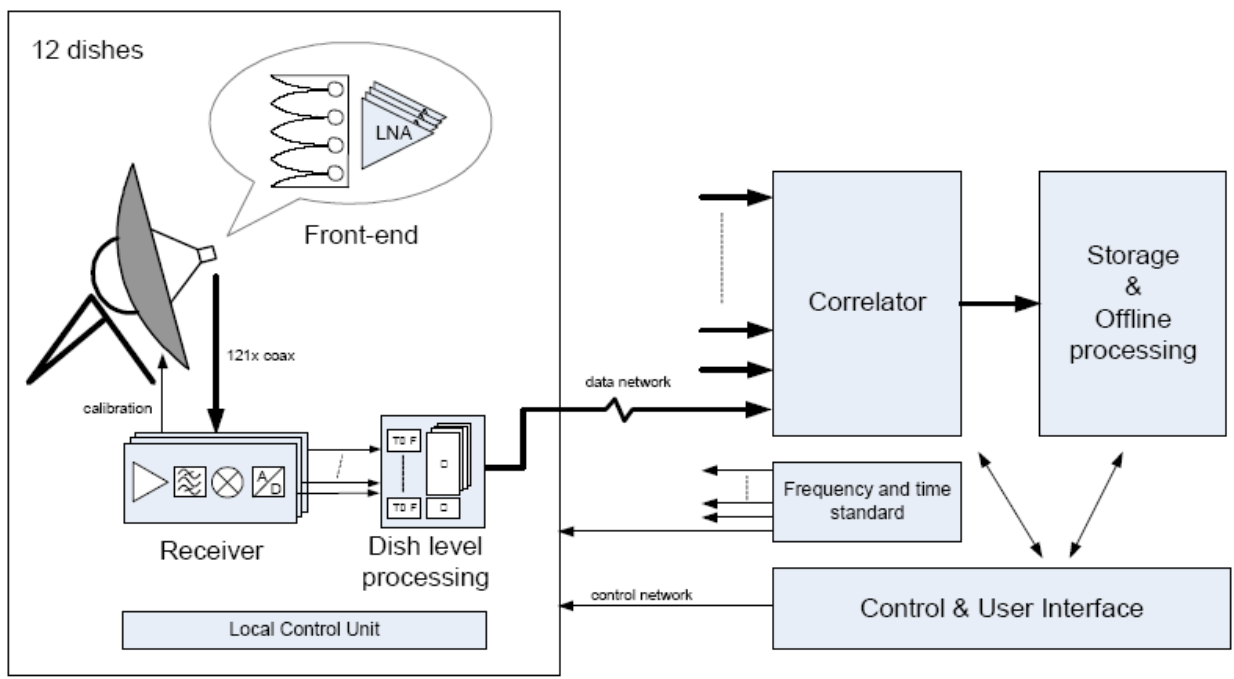

Рис. 1.15. Структурная схема системы *APERTIF*

Для качественного синтеза изображений из сохраненных данных необходимо точно знать ДН каждого из лучей. В этом отношении АФР создают дополнительные сложности в обработке сигналов. В традиционных рупорных фидерах конфигурации «один рупор — один луч» изменение усиления в приемнике ведет только к изменениям в принятой мощности сигнала, в то время как форма луча остается неизменной. Но в случае АФР каждый ее элемент участвует в создании каждого луча, и, так как каждый элемент решетки имеет свой приемный канал, изменение усиления в одном из них приведет не только к изменению принятой мощности, но и к изменению формы ДН лучей.

Для примера, типичная девиация коэффициента усиления рупорных облучателей *WSRT* в точках половинной мощности составляет около 1%, то есть *форма* ДН практически неизменна во времени, а меняется только общее усиление. В системе с АФР ожидаемые изменения усиления в каналах приема приведет к бо́льшим вариациям формы ДН, и, следовательно, требуется специальная схема калибровки.

Антенна *WSRT* в настоящее время представляет собой рефлектор диметром 25 м и фокусным расстоянием 8,25 м, в фокусе которого установлен рупорный облучатель. Проект *APERTIF* предполагает замену этого рупорного облучателя антенной решеткой с целью увеличения скорости обзора неба.

Антенная решетка состоит из 144 элементов Вивальди, с расстоянием между элементами 10 см, что составляет половину длины волны на частоте 1,5 ГГц. Каждый элемент решетки подключен к собственному МШУ через микрополосковое устройство питания (микрополосковый фидер — МПФ), выполненное на диэлектрической подложке с малым тангенсом угла потерь. МШУ работают в неохлажденном режиме, имеют коэффициент усиления 50 дБ и коэффициент шума примерно 0,4 дБ. Сигналы с выходов МШУ по коаксиальным кабелям передаются в блок приема и обработки сигналов, который физически находится в здании рядом с антенной. Для обеспечения временной стабильности луча в антенной системе существует подсистема калибровки в реальном времени.

Назначение приемника в антенной системе — преобразовать высокочастотный сигнал с выходов МШУ в цифровой поток данных. Принимая во внимание сложную внешнюю радиочастотную обстановку и сложность оцифровки и обработки сигналов в диапазоне частот выше 1 ГГц, приемник является супергетеродинным. Спектр принятого сигнала переносится в диапазон частот (450…750) МГц, а его оцифровка проводится 8-разрядным АЦП с частотой дискретизации 800 МГц. Таким образом, спектр сигнала находится во второй зоне Найквиста, и дискретизация является гармонической. Для сохранения высокого динамического диапазона оцифрованного сигнала к антиалиасинговому полосовому фильтру предъявляются повышенные требования. Скорость цифрового потока данных такой системы составляет 6,4 Гбит/с на каждый элемент решетки, или 774 Гбит/с на одну зеркальную антенну с АФР.

Лучи формируются полностью цифровым путем, используя сигналы с выходов каждого из элементов решетки. Последовательность цифровой обработки показана на рис. 1.16. Так как характеристики излучения антенных элементов и параметры аналоговых цепей приемника имеют частотную зависимость, то вся полоса частот принятого сигнала разбивается на поддиапазоны шириной до 1 МГц еще до формирователя. Далее сигналы каждого поддиапазона умножаются на собственные (для каждого поддиапазона) весовые коэффициенты и складываются.

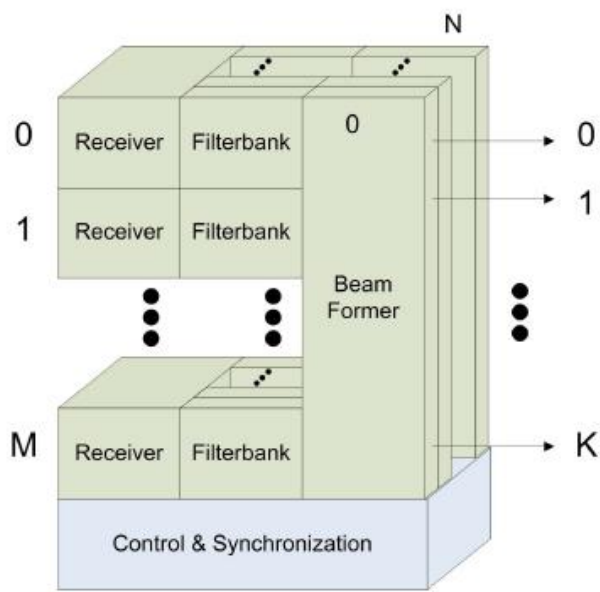

Рис. 1.16. Последовательность цифровой обработки сигналов (до коррелятора)

Для каждого поддиапазона и каждого луча существует свой набор весовых коэффициентов. Формирователь ДН (*beam former*) формирует 37 лучей, каждый из которых состоит из двух лучей с ортогональной поляризацией. Общее поле обзора, покрываемое этими 37 лучами, составляет 8 кв. градусов. Требуемая вычислительная мощность полосовых фильтров и формирователя оценивается в 7792 GMAC/s (*giga multiply and accumulate per* *second* — миллиардов умножений с накоплением в секунду) на одно зеркало, и почти 100 TMAC/s на 12 зеркал интерферометра. Выходной поток данных формирователя (178 Гбит/с) передается по 20 оптоволоконным кабелям с пропускной способностью 10 Гбит/с в центральный процессор для последующей обработки. Чтобы обеспечить такие скорости обработки, фильтры и формирователь будут выполнены на *FPGA* (*field-programmable gate array*).

Для калибровки характеристик каналов приема на зеркале установлен излучатель шума.

Центральный процессор производит обработку сигналов от всех зеркал интерферометра: осуществляет фильтрацию, рассчитывает корреляционную матрицу сигналов и синтезирует изображение.

**Прототип системы.**

К настоящему времени в рамках проекта *APERTIF* построена АФР из 144 элементов Вивальди и установлена в фокальной плоскости одного из зеркал *WSRT*. Это позволяет экспериментально определять характеристики системы, такие как ДН элементов решетки после отражения от зеркала (вторичные ДН элементов), сигнальную и шумовую корреляционные матрицы на выходах каналов приема, что позволяет вычислить реальную чувствительность системы. Чтобы минимизировать потери в проводниках и диэлектриках (и тем самым уменьшить шумовую температуру системы), микрополосковый фидер элемента решетки (МПФ) и МШУ выполнены на одной печатной плате, выполненной из диэлектрика с малым тангенсом угла потерь (рис. 1.17). Длина линии, соединяющей МПФ с МШУ, также минимизирована. При этом, как будет показано далее, КПД антенной решетки с МПФ равен 98%, что при температуре окружения 290 К эквивалентно вкладу 6 К в общую шумовую температуру системы. МШУ выполнены на дискретных элементах с малошумящим *HEMT* (*High Electron Mobility Transistor*) транзистором *ATF*-54143 на входе. В прототипе к приемникам подключены только 52 из 144 элементов решетки. Остальные элементы подключены к согласованной нагрузке.

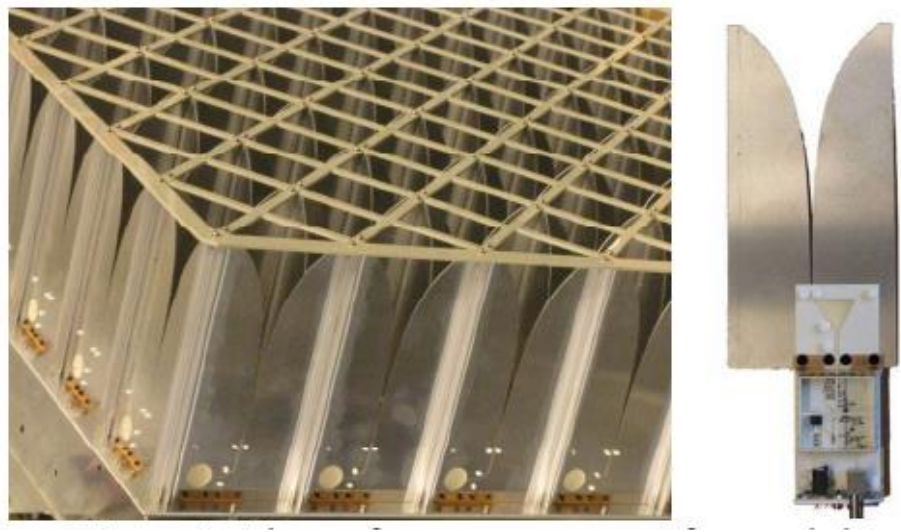

Рис. 1.17. Фотография прототипа антенной решетки и ее одного антенного элемента

В этом прототипе оцифрованный сигнал с каждого элемента решетки сохраняется и впоследствии обрабатывается в автономном режиме (выполняются фильтрация, формирование лучей и т.д.), как было описано выше. Также сигнал каждого из сформированных лучей может быть использован в корреляторе совместно с сигналами от трех других зеркал *WSRT*, которые оборудованы рупорными облучателями, что позволяет испытать новую систему в составе интерферометра.

## 1.4. Проблемы проектирования и практического использования радиотелескопов на базе антенных решеток

Точный анализ антенных систем является сложной проблемой, включающей в себя моделирование различных компонентов системы (таких как антенна, приемник, блок цифровой обработки сигнала) с дальнейшим объединением их в общую модель системы. Традиционно компоненты системы моделируются отдельно, а их взаимное влияние либо не учитывается вовсе, либо учитывается частично [29, 35…41]. Однако взаимное влияние компонентов следует учитывать, так как оно становится значительным при увеличении поля обзора (за счет использования многолучевости) и полосы частот, что является характерным для разрабатываемых радиотелескопов.

В течение последних (5…6) лет интенсивно исследовались несколько гибридных антенных систем, основанных на фазированных антенных решетках [29, 39…46]. Однако акцент делался в основном на экспериментальные исследования и демонстрации характеристик систем [29, 39], а также на разработку отдельных компонентов системы, таких как высокоэффективные решетки [9, 39…42], широкополосные однолучевые фидеры [9, 13, 21, 22], малошумящие усилители (МШУ) [47, 48], легкие и недорогие рефлекторы [20] и т.д., в то время как оптимизация всей системы в целом не производилась.

Одним из результатов исследований последних трех лет является понимание того, что важными параметрами для оптимизации системы являются взаимное влияние элементов решетки, шумы приемника и способ формирования лучей [45, 49…51]. Учет сильной взаимной связи между элементами фокальной решетки является важным для получения высокой чувствительности в большом поле обзора [51], так как она приводит к дополнительным потерям принятой мощности и сильной корреляции между полезными и шумовыми сигналами, которые распространяются через систему. Далее эти сигналы умножаются на комплексные весовые коэффициенты и складываются для получения требуемых лучей. Таким образом, оптимизация систем с фокальными решетками не является проблемой оптимизации независимо антенны и МШУ, а является проблемой комплексной системы, состоящей из антенны, МШУ и ФДН.

Для учета этих важных эффектов взаимодействия, были разработаны несколько подходов к моделированию системы численными методами [10, 52…55]. Они включают в себя анализ антенных решеток со множеством близко расположенных элементов [9, 46], совместный анализ электрически большого рефлектора и опорных структур [55], анализ активных цепей многоканального приемника и формирователя множественных лучей [8, 56…58]. Эти методы являются мощными инструментами для определения характеристик системы, но они слишком сложны для проведения ее оптимизации. Для этой цели необходимо разработать новые параметрические методы моделирования системы, в которые будут включены важные для практики критерии оптимизации, и которые будут построены на доминантных факторах, влияющих на систему. Разработка такой модели в настоящее время является одной из задач института *ASTRON* (Нидерланды), *Chalmers University of Technology* (Швеция) и *Brigham Young University* (США). Первая группа критериев оптимизации, описывающих основные составляющие шумовой температуры активных антенных решеток, рассмотрена в [8, 32, 56]. В настоящее время исследуются критерии, влияющие на свойства радиотелескопов с антенной решеткой как поляриметра (параметры Стокса и другие показатели качества поляризационной дискриминации) [59].

Другая серьезная проблема использования новых антенных систем для радиотелескопов — это отсутствие калибровочной модели и методики калибровки инструмента, которые необходимы для обработки данных и синтеза изображений. Калибровка подразумевает под собой подстройку параметров системы, как на аппаратном уровне, так и на программном. Она необходима для устранения инструментальных ошибок и эффектов, вызываемых окружающей средой [19, 58, 60…62]. Точность калибровочной модели имеет сильное влияние на динамический диапазон принимаемых сигналов, и, как следствие, на качество синтезируемых изображений. На практике динамический диапазон ограничен как уровнем шумов системы, так и ограничениями способов программной обработки сигналов и так называемой «очистки» изображений. Обработка сигналов всегда приводит к увеличению уровня шумов в синтезируемом изображении из-за накопления ошибок в процессе вычислений и исключения из данных систематических ошибок (в процессе калибровки) [63]. Чем больше калибровочных параметров (например, изменение коэффициентов усиления каналов) или функций, описывающих формы ДН лучей (такие как, например, гауссов луч), необходимо исключить из данных, тем меньше информации остается для построения изображения интересующего объекта [61, 64].

Традиционные методики калибровки [64…66] построены на эмпирических или упрощенных аналитических калибровочных моделях существующих сегодня антенных систем: механически сканирующих рефлекторов с одиночным фидером или кластером фидеров в конфигурации «один рупор — один луч», а также небольших апертурных решеток с одним или несколькими одновременно формируемыми лучами. Эти калибровочные модели включают измеренные параметры системы, антенна которой направляется на яркий небесный источник [65]. Построенная таким образом модель используется для «очистки» изображения во всем поле обзора.

Для новых зеркальных систем с АФР, которые имеют много большее поле обзора за счет множества одновременно формируемых лучей, существующие методики калибровки не подходят. Количество неизвестных параметров системы на 1…2 порядка больше, чем количество используемых при калибровке источников в небе. Поэтому, если это вообще возможно, методика калибровки каждого луча займет слишком много времени. Однако, лучи АФР не независимы, так как сигналы, принятые с разных направлений, коррелированны из-за существования взаимосвязи между элементами решетки.

Таким образом, для новых радиотелескопов с АФР требуется разработка новых калибровочных методик [45, 58, 62, 67], способных обеспечить необходимый результат для систем повышенной сложности, в состав которых входят АФР, а также при обработке данных со значительно большего поля обзора.

## 1.5. Обоснование выбора метода моделирования металлической структуры антенной решетки.

Существует несколько методов моделирования металлических структур. Первый, и наиболее известный, из них — это метод моментов. Этот метод дает точное решение, но расчет геометрически сложных и больших структур (с размером в несколько длин волн) требует решения системы из большого числа линейных алгебраических уравнений, что связано с трудностями, связанными с ограничениями на объем компьютерной памяти и производительность компьютеров.

Также существуют методы, основанные на анализе бесконечных антенных решеток с корректировкой краевых эффектов, например [68…70]. Однако, многие их них ограничены определенным типом возбуждения. Также эти методы дают хорошие результаты только лишь для больших решеток (характерный размер антенны более $10\lambda$, где $\lambda$ — рабочая длина волны), в которых центральные элементы ведут себя как элементы бесконечной решетки, и на которые практически не влияют краевые эффекты. Ввиду конструктивных и стоимостных ограничений размеры фокальной решетки не могут быть настолько большими, чтобы можно было использовать подход бесконечных решеток. В то же время, такие решетки достаточно велики, и применение чистого метода моментов становится слишком громоздким. Это относится, например, к решетке из элементов Вивальди, применяемой в проекте *APERTIF*, размер которой составляет от 4 до 6 длин волн в рабочем диапазоне частот.

Существует подход к моделированию больших конечных антенных решеток, основанный на уменьшении количества уравнений в системе линейных алгебраических уравнений метода моментов. Это может быть сделано путем введения базисных функций в виде полиномов высокого порядка [71].

Однако еще большего уменьшения количества неизвестных можно добиться путем введения так называемых макробазисных функций, описывающих макродомены структуры. В нашем случае макродоменами являются отдельные элементы антенной решетки. Этот метод и был применен в данной работе для моделирования металлической структуры антенной решетки. Он также получил название «Метод характеристических базисных функций» (ХБФ). Подробно метод ХБФ описан в [5…7].

### 1.6. Обзор методов оптимизации весовых коэффициентов антенной решетки

**Метод согласования по полю.**

Метод согласования по полю (*CFM – Conjugate Field Matching*) — самый простой способ вычисления весовых коэффициентов, при котором антенная система принимает максимум падающей на нее энергии, то есть ведется согласованный прием.

Весовые коэффициенты **w** в этом методе равны комплексно-сопряженной величине сигнала на выходе соответствующих элементов решетки при приеме сигнала от источника в заданном направлении, то есть

$$\mathbf{w} = \left[ \mathbf{e}\left( \theta_0, \varphi_0 \right) \right]^*, \tag{1.7}$$

где $\mathbf{w} = \left[ w_1, w_2, \ldots, w_M \right]^T$ — оптимальные по *CFM* критерию весовые коэффициенты; $\left( \theta_0, \varphi_0 \right)$ — направление требуемого максимума ДН системы.

Как будет показано в разделе 4, применение весовых коэффициентов, рассчитанных *CFM* методом, не дает максимальной чувствительности.

**Метод максимизации чувствительности.**

Метод расчета весовых коэффициентов, обеспечивающий достижение максимальной чувствительности известен уже давно (см., например, [72] и [73]). Заключается он в следующем.

Рассмотрим систему, показанную на рис. 1.18.

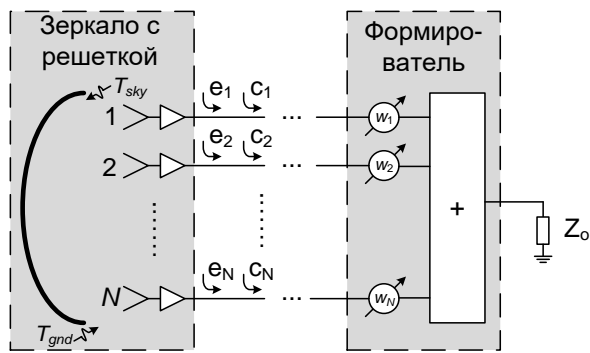

Рис 1.18 Структура антенной системы и матрицы, используемые для расчета весовых коэффициентов

Антенная система состоит из следующих частей: 1) зеркало, элементы решетки с устройством их питания и МШУ и 2) формирователь луча (управляемые аттенюаторы с фазовращателями и сумматор, как правило, цифровые). С выхода каждого элемента решетки выходят как сигнал $e_m$, так и шум $c_m$.

Чувствительность системы $S$ связана с отношением сигнал/шум на ее выходе формулой [31]

$$S = \frac{2k_B}{I}\left(\frac{C}{\text{Ш}}\right), \tag{1.8}$$

где $I$ — спектральная плотность потока мощности, создаваемая источником, $\left[\dfrac{W}{m^2 \cdot Hz}\right]$.

Таким образом, чувствительность будет максимальна, когда будет максимально отношение С/Ш при постоянной плотности потока мощности источника.

Введем обозначение $\rho$ для отношения С/Ш. Тогда отношение мощности сигнала к мощности шума на выходе формирователя можно записать как функцию весовых коэффициентов следующим образом:

$$\rho\left(\mathbf{w}\right) = \frac{\left(\mathbf{w}^H\mathbf{e}\right)^H \mathbf{w}^H\mathbf{e}}{\left(\mathbf{w}^H\mathbf{c}\right)^H \mathbf{w}^H\mathbf{c}} = \frac{\mathbf{w}^H\mathbf{e}\left(\mathbf{w}^H\mathbf{e}\right)^H}{\mathbf{w}^H\mathbf{c}\left(\mathbf{w}^H\mathbf{c}\right)^H} = \frac{\mathbf{w}^H\mathbf{e}\mathbf{e}^H\mathbf{w}}{\mathbf{w}^H\mathbf{c}\mathbf{c}^H\mathbf{w}} = \frac{\mathbf{w}^H\mathbf{P}\mathbf{w}}{\mathbf{w}^H\mathbf{C}\mathbf{w}}, \qquad (1.9)$$

где $\mathbf{P} = \mathbf{e}\mathbf{e}^H$ и $\mathbf{C} = \mathbf{c}\mathbf{c}^H$ — сигнальная и шумовая корреляционные матрицы соответственно, элементы которых равны $P_{mn} = e_m e_n^*$ и $C_{mn} = \overline{c_m c_n^*}$.

Найти максимум функции (1.9) можно, приравняв нулю все частные производные этой функции:

$$\frac{\partial \rho\left(\mathbf{w}\right)}{\partial w_m} = 0 \quad \text{и} \quad \frac{\partial \rho\left(\mathbf{w}\right)}{\partial w_m^*} = 0. \qquad (1.10)$$

Произведя необходимые преобразования, описанные в [74], получим, что для того, чтобы найти оптимальные весовые коэффициенты $\mathbf{w}_{opt}$, необходимо найти собственные значения матрицы $\mathbf{P}$, решив уравнение

$$\mathbf{P}\mathbf{w} = \rho\mathbf{C}\mathbf{w}. \qquad (1.11)$$

При этом максимальное собственное значение $\rho_{max}$ будет максимально достижимым отношением С/Ш, а соответствующий ему собственный вектор — оптимальными по критерию максимальной чувствительности весовыми коэффициентами $\mathbf{w}_{opt}$.

Также в [74] доказывается, что собственный вектор, соответствующий максимальному собственному значению для задачи (1.11), равен $\mathbf{w} = \mathbf{C}^{-1}\mathbf{e}$.

Таким образом, оптимальные по критерию максимальной чувствительности весовые коэффициенты рассчитываются в соответствии с следующим выражением:

$$\mathbf{w} = \mathbf{C}^{-1}\mathbf{e}(\theta_0, \varphi_0). \tag{1.12}$$

Расчеты, проведенные по данной методике, представлены в разделах 3 и 4.

**Метод максимизации чувствительности с ограничениями по направлениям.**

В радиоастрономии основной метод наблюдения и построения карт неба — это метод интерферометрии. Традиционные интерферометры являются однолучевыми, и наблюдение ведется только в пределах главного лепестка ДН антенных систем.

Целью же настоящей работы является разработка многолучевой системы, то есть усовершенствование интерферометра так, чтобы наблюдение можно было проводить одновременно в большой области, перекрываемой несколькими лучами антенн.

В случае использования многолучевой антенной системы перекрытие наблюдаемой области лучами должно осуществляться так, чтобы чувствительность в точке пересечения соседних лучей уменьшалась незначительно (рис. 2.19). Для некоторых радиоастрономических наблюдений распределение чувствительности для всего поля обзора должно быть как можно ближе к равномерному.

Рассмотренный ранее метод максимизации чувствительности оптимизирует чувствительность *отдельных* лучей, и поэтому не позволяет регулировать ее неравномерность в области многих лучей.

Существует другой способ, в котором расчет весовых коэффициентов **w** производится таким образом, чтобы в направлениях пересечения лучей их ДН имели бы заданный уровень, и, следовательно, общая чувствительность будет более равномерной. Общий уровень чувствительности при этом уменьшается, но можно найти компромисс между уровнем чувствительности и ее неравномерностью, что будет сделано в разделе 4.

Такой метод расчета весовых коэффициентов будем называть методом максимизации чувствительности с ограничениями по направлениям. В литературе метод известен как LCMV — *linear constrained minimum variance*, и математически записывается следующим образом [75]:

$$\mathbf{w}^{H} = \mathbf{g}^{H} \left[ \mathbf{G}^{H}\mathbf{C}^{-1}\mathbf{G} \right]^{-1} \mathbf{G}^{H}\mathbf{C}^{-1}, \tag{1.13}$$

где **G** — матрица размером $M \times N_{dir}$, содержащая сигнальные вектора с $N_{dir}$ направлений ограничений ($M$ — количество элементов в решетке); **g** — вектор размерностью $N_{dir} \times 1$, элементы которого устанавливают уровень ДН луча в каждом из $N_{dir}$ направлений.

Расчеты весовых коэффициентов по этому методу, оптимизация **g** для получения максимальной гладкости чувствительности, а также анализ системы с применением рассчитанных весовых коэффициентов приведены в разделе 4.

**1.7. Выводы по разделу**

1.7.1. С целью выявления требований к современным радиотелескопам, проведен обзор разрабатываемых в настоящее время антенных систем *SKA* и *APERTIF*. Показана необходимость увеличения скорости обзора неба.

1.7.2. Произведен обзор облучателей, используемых в настоящее время в радиотелескопах, указаны их достоинства и недостатки. Показана целесообразность использования плотных антенных решеток в качестве облучателей. Это позволяет формировать широкое многолучевое непрерывное поле обзора радиотелескопа, однако вызывает ряд проблем, связанных с сильным взаимным влиянием между антенными элементами решетки.

1.7.3. Подробно описана система APERTIF, модель которой разработана в данной работе.

1.7.4. Обоснован выбор метода моделирования антенной решетки из элементов Вивальди, используемой в качестве облучателя. Выбранный метод (метод характеристических базисных функций) позволяет проводить точное электромагнитное моделирование антенных решеток средних размеров (4…6 длин волн).

1.7.5. Приведен обзор нескольких, наиболее перспективных для радиоастрономии, методов расчета оптимальных весовых коэффициентов для элементов решетки.

1.7.6. Исходя из описанных в данном разделе проблем и трудностей, существует потребность в создании простой и эффективной модели гибридной антенной системы, что позволит рассчитывать характеристики системы, а также проводить ее параметрическую оптимизацию.

**РАЗДЕЛ 2**

**РАЗРАБОТКА ТЕОРЕТИЧЕСКОЙ МОДЕЛИ АНТЕННОЙ СИСТЕМЫ С ОПТИМАЛЬНЫМИ ВЕСОВЫМИ КОЭФФИЦИЕНТАМИ**

Одной из целей работы является построение модели антенной системы, с помощью которой возможно рассчитать интересующие параметры этой системы.

Моделирование состоит из следующих этапов:

1) электромагнитное моделирование металлической части антенной решетки с целью нахождения ДН отдельных элементов решетки, а также матрицы рассеяния **S** между входами антенных элементов решетки;

2) микроволновое моделирование системы с учетом МПФ, питающих антенные элементы решетки, и МШУ с целью нахождения уровней сигнала с выхода каждого элемента решетки и шумовой корреляционной матрицы;

3) используя полученные результаты, расчет оптимальных по нескольким критериям весовых коэффициентов и параметров антенной системы.

Этап 1 представлен в подразделе 2.1, модель МПФ для этапа 2 — в подразделе 2.2, анализ системы — в подразделе 2.3.

### 2.1. Электромагнитная модель антенной решетки

Рассмотрим антенную решетку вне зеркала. Матрица рассеяния **S** и ДН элементов решетки сильно зависят от типа и параметров цепи питания элементов решетки, которая оказывает влияние на распределение токов в металлической структуре антенны. Модель цепи питания будет более подробно описана в подразделе 2.2, а учет ее влияния на матрицу рассеяния и ДН элементов — в подразделе 2.3 данного раздела. Учет влияния проведен путем микроволнового моделирования цепи, в которой решетка заменяется
*N*-полюсником с известной матрицей рассеяния **S**.

Таким образом, первое, что необходимо сделать — это найти матрицу **S**
металлической структуры решетки (решетки без МПФ), а также ДН
отдельных ее элементов в присутствии других элементов.

Для численного электромагнитного (ЭМ) моделирования удобно
представить источники возбуждения элементов решетки (порты) как
четырехугольник, состоящий из двух ячеек (рис. 2.1). При этом источник
возбуждения каждого антенного элемента имеет внутреннее сопротивление,
равное нулю, то есть является источником напряжения.

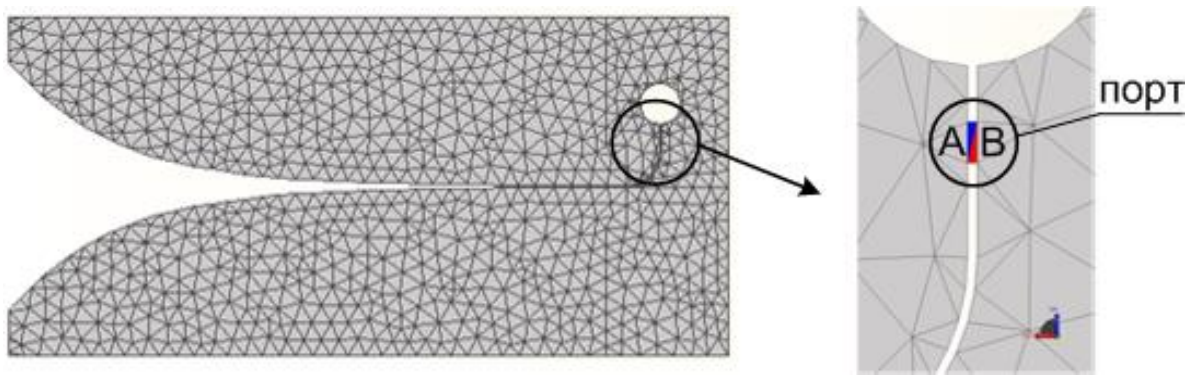

Рис.2.1 — Схема возбуждения антенного элемента Вивальди в ЭМ
модели

Каждый антенный элемент последовательно возбуждается напряжением с
амплитудой $V = 1$ В, в то время как порты остальных элементов
закорачиваются. При этом методом ХБФ рассчитывается распределение тока
во всех элементах решетки. По найденному распределению тока
рассчитывается поле в дальней зоне возбужденного элемента в присутствии
остальных элементов решетки.

Для дальнейших расчетов необходимы ДН антенных элементов,
возбуждаемых источником тока.

Ввиду различия способов возбуждения в ЭМ модели (источники напряжения и источники тока), ДН элементов для них будут отличаться. Поэтому введем следующие понятия:

1) *V*-ДН — это зависимость напряженности электрического поля от угловых координат $\theta$ и $\varphi$, при возбуждении активного антенного элемента решетки источником напряжения с амплитудой $V_o = 1$ В. При этом входы остальных элементов закорочены;

2) *I*-ДН — это зависимость напряженности электрического поля от угловых координат $\theta$ и $\varphi$, при возбуждении активного антенного элемента источником тока $I_o = 1$ А. При этом остальные элементы имеют разрыв в цепи питания.

Взаимосвязь между *V*-ДН и *I*-ДН показана в конце данного подраздела.

Таким образом, в результате ЭМ моделирования получены *V*-ДН всех элементов решетки.

Распределение тока, полученное при возбуждении каждого элемента по отдельности, позволяет также рассчитать проводимость $Y_{ab}^{ant}$ между портами элементов $a$ и $b$:

$$Y_{ab}^{ant} = -\frac{1}{V_a V_b} \iint\limits_{S_a} \boldsymbol{E}^s \left( \boldsymbol{J}_S^b \right) \cdot \boldsymbol{J}_S^a \, \mathrm{d}S \,, \qquad (2.1)$$

где $V_a$ и $V_b$ — напряжения возбуждения элементов $a$ и $b$; $\boldsymbol{J}_S^b$ и $\boldsymbol{J}_S^a$ — поверхностный ток во всей решетке при возбуждении элемента $b$ и $a$ соответственно.

Вывод формулы для расчета $Y_{ab}^{ant}$ приведен в [4]. Матрица проводимости может быть пересчитана в матрицу сопротивлений $\mathbf{Z_{ant}}$, которая необходима для дальнейшего микроволнового моделирования, как $\mathbf{Z_{ant}} = (\mathbf{Y_{ant}})^{-1}$. На рис. 2.2 показан модуль рассчитанной матрицы сопротивлений для решетки *APERTIF* на частоте 1 ГГц.

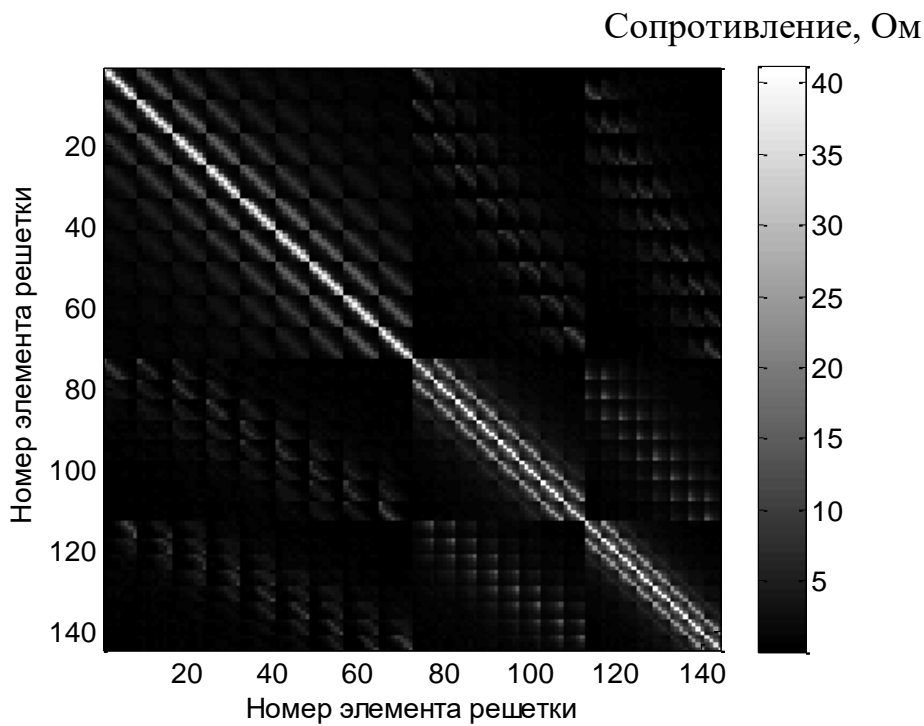

Рис. 2.2 — Рассчитанная матрица сопротивлений между входами элементов антенной решетки *APERTIF* на частоте 1 ГГц.

Все расчеты производились в программе *CAESAR*, в которой реализован метод ХБФ, и которая также имеет микроволновый симулятор.

На рис. 2.3 показан вид рассчитанного распределения тока в антенной решетке 9×8×2, составленной из элементов Вивальди, для случая одновременно возбужденных элементов решетки напряжением с равной амплитудой.

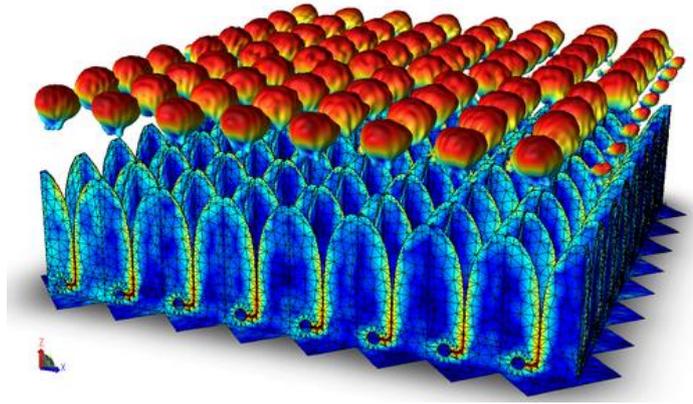

Рис.2.3 — Рассчитанные в программе *CAESAR* распределения тока и *I*-ДН
отдельных элементов решетки

На рис. 2.3 над каждым антенным элементом показана соответствующая
ему *I*-ДН для основной поляризационной составляющей. Синим цветом
показана меньшая плотность тока, красным — большая.

**Взаимосвязь *V*-ДН и *I*-ДН.**

Используемая в работе программа моделирования антенных решеток
*CAESAR* производит расчет токового распределения и характеристик
направленности при возбуждении элементов решетки *источниками
напряжения* с заданными напряжениями.

Для того чтобы рассчитать ДН каждого антенного элемента решетки в
присутствии остальных элементов, выполнялось *последовательное*
возбуждение каждого элемента решетки напряжением 1 В, в то время порты
остальных элементов были короткозамкнуты (т.е. возбуждались источником
напряжения с напряжением 0 В) (рис. 2.4). Для каждого возбуждения
рассчитывалось распределение токов и ДН *всей* решетки, а так как при этом
одновременно возбужден был лишь один антенный элемент, то ДН решетки
фактически является ДН этого элемента в присутствии остальных. В данной
работе ДН, рассчитанные при возбуждении элементов решетки источниками
напряжения называются *V*-ДН, и их количество равно количеству антенных
элементов в решетке.

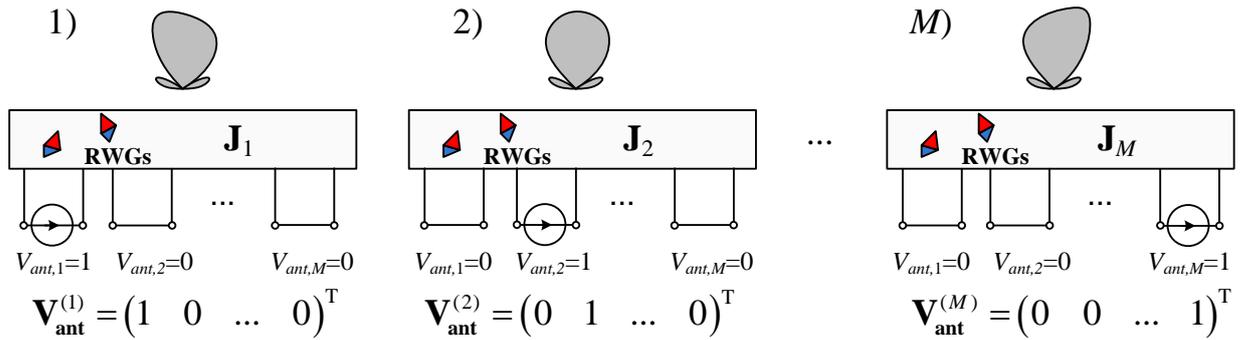

Рис. 2.4 — К определению $V$-ДН элементов решетки.

Однако для дальнейшего анализа антенной системы требуются $I$-ДН, то есть ДН, полученные при последовательном возбуждении элементов решетки источниками тока с амплитудой 1 А когда порты остальных элементов разомкнуты.

Покажем, как перейти от $V$-ДН к $I$-ДН элементов решетки.

Итак, дана антенная решетка, элементы которой возбуждаются как показано на рис. 2.5 (показан один случай, когда возбужден первый элемент). Также даны $M$ ($M$ — количество элементов в решетке) токовых распределений и соответствующие им ДН, рассчитанные в программе *CAESAR* (см. рис. 2.4).

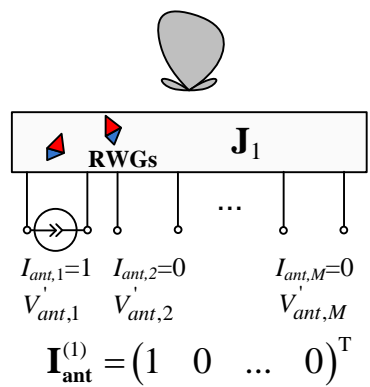

Рис. 2.5. Возбуждение первого элемента решетки источником тока при расчете $I$-ДН элементов решетки

Требуется найти $I$-ДН возбужденного элемента решетки.

При возбуждении одного элемента источником тока, как показано на рис. 2.5, вектор тока (ток всех через порты всех элементов решетки) будет равен $\mathbf{I}_{ant} = (1\ 0\ 0\ \ldots\ 0)^T$. Тогда, если представить решетку многополюсником, напряжение на всех портах элементов решетки будет равно

$$\mathbf{V}_{ant}^{'} = \mathbf{Z}\mathbf{I}_{ant}, \tag{A.1}$$

где $\mathbf{Z}$ — матрица сопротивлений антенной решетки, полученная в результате электромагнитного моделирования в программе *CAESAR*.

Зная напряжения на портах элементов решетки, соответствующие возбуждению источником тока ($\mathbf{V}_{ant}^{'}$), и зная все $V$-ДН, соответствующие последовательному возбуждению антенных элементов источником напряжения с амплитудой 1 В (рис. 2.4), можно рассчитать искомую $I$-ДН возбужденного элемента (рис. 2.5), как как суперпозицию $V$-ДН всех элементов решетки с коэффициентами $\mathbf{V}_{ant}^{'}$:

$$g_1^{(I)}(\theta, \varphi) = \sum_{i=1}^{M} g_i^{(V)}(\theta, \varphi) \cdot V_{ant,i}^{'}, \tag{A.2}$$

где $g_i^{(V)}$ — $V$-ДН $i$-го элемента решетки; $g_1^{(I)}$ — $I$-ДН 1-го элемента решетки.

Аналогичным образом, используя формулы (A.1) и (A.2), рассчитываются остальные ($M$–1) $I$-ДН.

## 2.2. Модель микрополоскового фидера для антенных решеток из щелевых антенн типа Вивальди

### 2.2.1 Микроволновая модель микрополоскового фидера

После того, как был вычислен (в программе *CAESAR*) импеданс между точками *A* и *B* каждого элемента решетки (см. рис. 2.1), к элементам подключался микрополосковый фидер, модель которого схематически изображена на рис. 2.6.

На рис. 2.6 использованы следующие обозначения: $L_a$, $L_b$, $L_1$, $L_{stub}$ —длины участков микрополосковой линии,   $Z_o$ — волновое сопротивление участков $L_a$, $L_b$ и $L_1$,   $Z_i$ — волновое сопротивление участков расширяющейся части микрополосковой линии, $Z_{slot}$ — сопротивление щели; $\varepsilon_{re}$ — эффективная диэлектрическая проницаемость диэлектрика микрополосковой линии; $p$ — коэффициент трансформации.

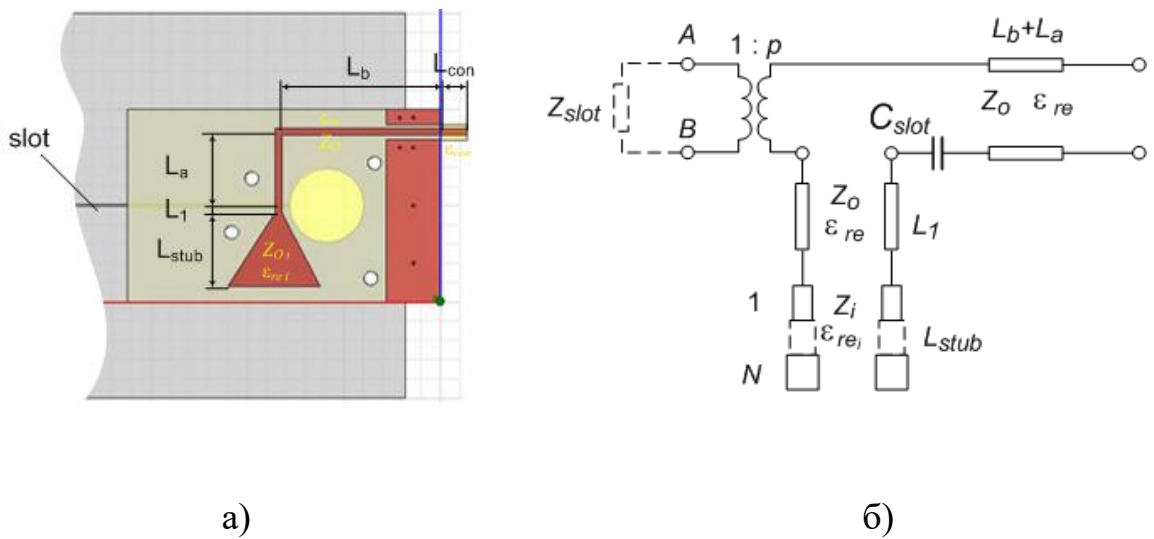

а)                                                б)

Рис. 2.6 Микрополосковый фидер для антенного элемента Вивальди:

а) схематическое изображение,  б) эквивалентная схема

На рис. 2.6, б показана эквивалентная схема МПФ, предложенная Кнорром [76], модифицированная для учета емкости щели. В данной модели переход микрополосок-щель моделируется трансформатором с коэффициентом трансформации $p = V_{sec}/V_{prim}$, где $V_{sec}$ и $V_{prim}$ — напряжения на вторичной (со стороны щели) и первичной (со стороны микрополосковой линии) обмотках соответственно. Последовательно с микрополосковой линией добавлена емкость, так как заземленный проводник этой линии прерывается щелью.

При отношении ширины микрополоска $W$ к высоте $h$ диэлектрика $W/h > 2$ параметры $Z_o$ и $\varepsilon_{re}$ можно рассчитать по следующим выражениям [77]:

$$Z_o = \frac{377}{\sqrt{\varepsilon_r}\left[\dfrac{W}{h} + 0{,}883 + \dfrac{\varepsilon_r + 1}{\pi\varepsilon_r}\left(\ln\left(\dfrac{W}{2h} + 0.94\right) + 1{,}451\right) + 0{,}165\dfrac{\varepsilon_r - 1}{\varepsilon_r^2}\right]};\ (2.2)$$

$$\varepsilon_{re} = \frac{\varepsilon_r + 1}{2} + \frac{\varepsilon_r - 1}{2}\frac{1}{\sqrt{1 + 12\dfrac{h}{W}}},\ (2.3)$$

где $\varepsilon_r$ — диэлектрическая проницаемость диэлектрика микрополосковой линии.

Следует отметить, что параметры $Z_o$ и $\varepsilon_{re}$ зависят от частоты. Но в случае, когда максимальная рабочая частота много меньше $f_p\,(\text{ГГц}) = 0{,}398\dfrac{Z_0}{h\,(\text{мм})}$, этой зависимостью можно пренебречь [77].

Потери в микрополосковой линии, как в металле, так и в диэлектрике, малы, и ими также можно пренебречь. Так, например, результаты моделирования показывают, что в рассматриваемой модели антенной решетки для частот $(1\ldots1{,}7)$ ГГц пренебрежение потерями приводит к ошибке при моделировании $S$-параметров решетки менее 0,2 дБ при уровне примерно –10 дБ.

Для увеличения полосы рабочих частот применен расширяющийся конец микрополосковой линии, или «заглушка» (рис. 2.7), которая может быть промоделирована короткими отрезками микрополосковой линии с увеличивающейся шириной [78]. И хотя этот метод не учитывает излучение линий и паразитные эффекты на стыках, он дает достаточно точный результат в указанном частотном диапазоне (см. рис. 2.8).

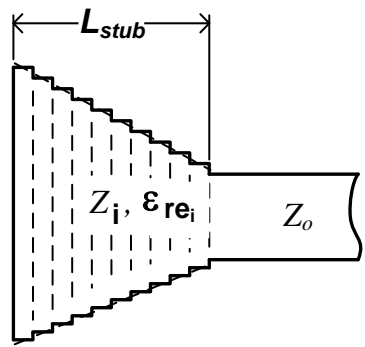

Рис. 2.7 Модель линии с расширяющимся концом

Для проверки адекватности такой модели рассчитано полное сопротивление треугольной заглушки в программе *HFSS*. На рис. 2.8а,б показано сравнение результатов моделирования треугольной заглушки в программе *HFSS* с результатами, полученными путем расчетов в программе *CAESAR* для двух видов модели заглушки: с одной и пятью секциями. Из рисунка видно, что пять секций уже достаточно, чтобы обеспечить необходимую точность моделирования. Модуль коэффициента отражения заглушки на рисунке не показан, так как $|S_{11}| > - 0,05$ дБ во всем диапазоне частот.

Была проверена адекватность модели всего МПФ. Для этого промоделирован и измерен коэффициент отражения для случая короткозамкнутого трансформатора (что соответствует отсутствию щели). Результаты показаны на рис. 2.8в,г. По рисунку видно, что и модуль, и аргумент $S_{11}$ модели хорошо согласуются с измерениями. Модель также учитывает потери в металле и диэлектрике МПФ, что также хорошо демонстрирует не нулевая величина модуля коэффициента отражения $S_{11}$ (рис. 2.8в). Эти потери будут вносить вклад в общую шумовую температуру системы, что приведет к некоторому уменьшению чувствительности.

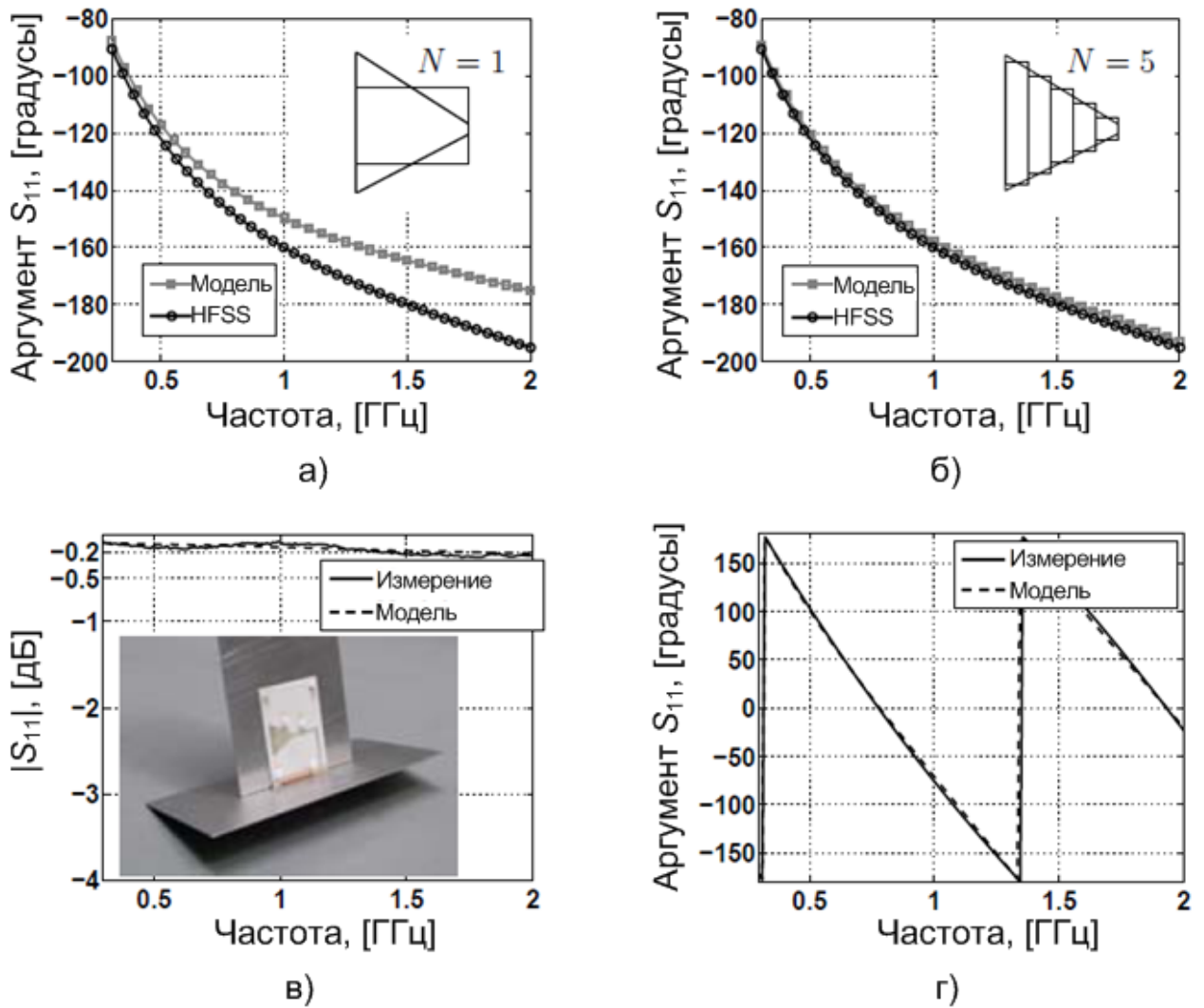

Рис. 2.8 — Результаты моделированния в *CAESAR* и *HFSS* треугольной заглушки:

а) аргумент $S_{11}$ для одной секции, б) аргумент $S_{11}$ для пяти секций, в) измеренный и моделированный модуль коэффициента отражения МПФ для случая короткозамкнутого трансформатора г) аргумент коэффициента отражения МПФ для случая короткозамкнутого трансформатора

## 2.2.2 Оптимизация модели микрополоскового фидера

Рассмотренная модель наряду со своей простотой обладает существенным недостатком: в ней отсутствует аналитическое выражение для расчета коэффициента трансформации $p$. Однако можно выполнить подбор $p$ так, чтобы результат моделирования наиболее адекватно отражал результат измерений. Это можно сделать, изготовив один антенный элемент или небольшую решетку и, используя их, выполнить оптимизацию $p$ так, чтобы ошибка между моделированными и измеренными $S$-параметрами была минимальна.

Кроме неизвестного значения $p$ в процедуру оптимизации целесообразно ввести еще одну величину — ширину участка щели антенны Вивальди под диэлектриком $W_{sl}$. Это необходимо для учета при ЭМ моделировании наличия диэлектрической подложки МПФ у металлической структуры антенны. Как будет показано далее, наличие диэлектрика уменьшает эквивалентную ширину щели ЭМ модели.

Алгоритм поиска оптимальных значений $p$ и $W_{sl}$ можно представить следующим образом:

1) производится ЭМ моделирование одного изготовленного антенного элемента Вивальди для диапазона значений $W_{sl}$ во всем рабочем диапазоне частот $f$, в результате которого получаем зависимость полного сопротивления виртуального порта антенного элемента Вивальди от частоты и ширины щели: $Z_{slot}\left(f, W_{sl}\right)$;

2) на основании найденной зависимости $Z_{slot}\left(f, W_{sl}\right)$, с помощью микроволнового моделирования (см. рис. 2.6б) рассчитывается матрица рассеяния $\mathbf{S}$ портов решетки с учетом МПФ для заданного диапазона изменения $p$, в результате чего получаем матрицу коэффициента отражения от порта элемента Вивальди с МПФ $S_{11}^{Model}\left(f, p, W_{sl}\right)$, зависящую от частоты, коэффициента трансформации и ширины щели;

3) выбирается целевая функция как функция ошибки между рассчитанным и измеренным коэффициентом отражения:

$$E\left(p,W_{sl}\right) = \frac{\left\|S_{11}^{Model}\left(f,p,W_{sl}\right) - S_{11}^{Meas}\left(f,p,W_{sl}\right)\right\|_2}{\left\|S_{11}^{Meas}\left(f,p,W_{sl}\right)\right\|_2} \cdot 100\% \,, \qquad (2.4)$$

где $S_{11}^{Model}$ и $S_{11}^{Meas}$ — соответственно промоделированный и измеренный комплексный коэффициент отражения, а оператор $\|A\|_2$ обозначает эвклидову норму матрицы ($\|A\|_2 = \sqrt{\sum_i |a_i|^2}$ );

4) находится минимум функции $E\left(p,W_{sl}\right)$; соответствующие минимуму величины $p$ и $W_{sl}$ являются оптимальными, и могут использоваться для расчета многоэлементной решетки; поиск минимума производится прямым методом (метод оптимизации нулевого порядка).

На рис. 2.9 показаны результаты оптимизации $p$ и $W_{sl}$ для одного элемента Вивальди.

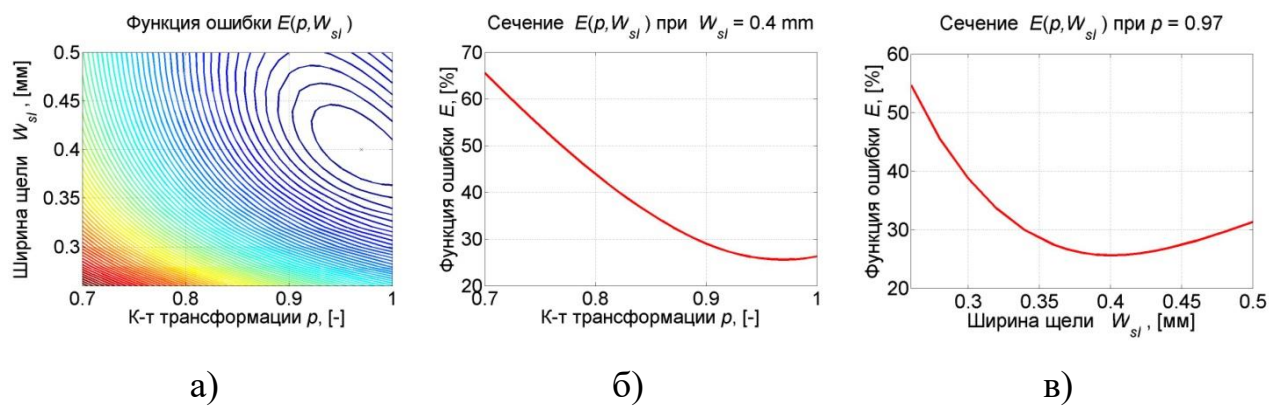

а)                           б)                           в)

Рис. 2.9 — Рассчитанная целевая функция для оптимизации МПФ: а) как функция двух неизвестных; б, в) ее сечения;

По рисункам видно, что оптимальными для данной модели элемента Вивальди с МПФ являются величины $p = 0{,}97$ и $W_{sl} = 0{,}4$ мм. Показательным является результат, демонстрирующий влияние диэлектрика на эквивалентную ширину щели, которая в данной модели равна 0,4 мм, при физической ширине 0,5 мм.

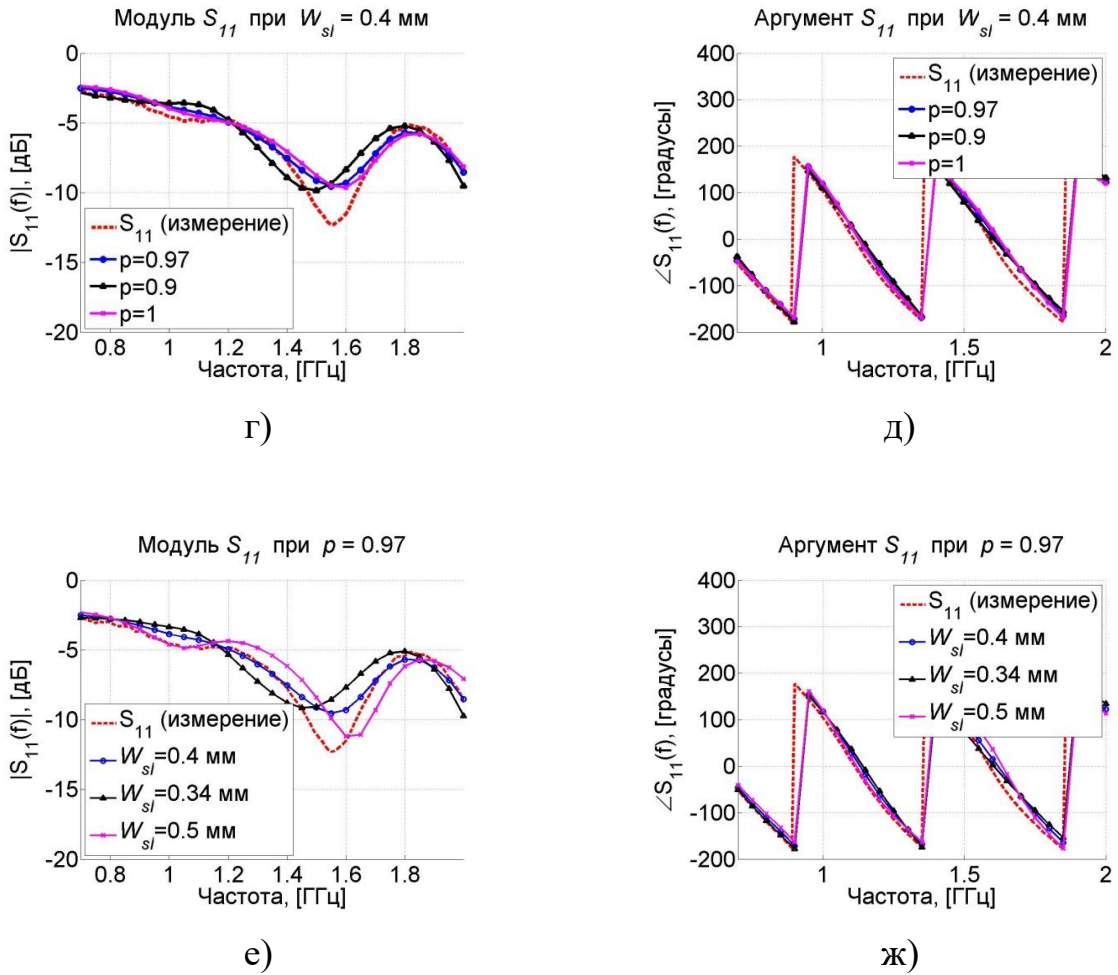

Рис. 2.10 — Результаты оптимизации МПФ: г), е) измеренные и моделированные для разных значений $p$ и $W_{sl}$ модули коэффициента отражения от порта элемента Вивальди с МПФ; д), ж) измеренные и моделированные для разных значений $p$ и $W_{sl}$ аргументы коэффициента отражения от порта элемента Вивальди с МПФ

## 2.3 Разработка алгоритма моделирования системы, состоящей из рефлектора, АФР, МПФ, МШУ и ФДН

В подразделах 2.1 и 2.2 был рассмотрена электромагнитная модель плотной конечной антенной решетки из элементов Вивальди, а также произведена экспериментальная проверка модели МПФ.

В данном подразделе все результаты предыдущих исследований объединяются в единый алгоритм расчета основных параметров системы.

### 2.3.1 Исходные данные

Исходными данными для расчетов являются:

1) первичные $I$-ДН отдельных элементов решетки $g_m(\theta,\varphi)$ с учетом их окружения (соседних элементов решетки);

2) матрица рассеяния $\mathbf{S_{arr}}$ и матрица сопротивлений $\mathbf{Z_{ant}}$ антенной решетки без МПФ, то есть чисто металлической структуры с виртуальными портами;

3) распределение шумовой температуры окружения $T_{env}(\theta,\varphi)$;

4) температура окружающей среды $T_{amb}$;

5) параметры МПФ;

6) параметры малошумящего усилителя (МШУ);

7) шумовая температура цепей приемника после МШУ $T_{sec}$ (secondary stage);

8) форма поля обзора телескопа и конфигурация лучей.

*Матрица сопротивлений* $\mathbf{Z_{ant}}$ антенной решетки без МПФ, а также ДН *встроенных элементов* $g_m(\theta,\varphi)$ моделировались методом характеристических базисных функций, как описано в подразделе 2.1.

*Матрица рассеяния* решетки $\mathbf{S_{arr}}$ рассчитывалась по известной формуле [79], в предположении, что все порты решетки нагружены на характеристическое сопротивление $Z_0 = 50$ Ом:

$$\mathbf{S_{arr}} = \left(\mathbf{Z_{ant}} + Z_0\mathbf{I}\right)^{-1}\left(\mathbf{Z_{ant}} - Z_0\mathbf{I}\right),$$ (2.5)

где $\mathbf{I}$ — единичная матрица.

В состав *шумовой температуры окружения* $T_{env}(\theta,\varphi)$ входят следующие компоненты (рис. 2.11):

а) при $0 < \theta < \theta_{sub}$ ($\theta_{sub}$ — половина угла раскрыва зеркала) температура $T_{env}$ равна температуре фонового излучения неба $T_{sky} = 5,3$ К, так как считается, что зеркало идеально непрозрачно на рабочих частотах;

б) при $\theta_{sub} < \theta < 90°$, температура $T_{env}$ равна температуре земли $T_{gnd} = 290$ К;

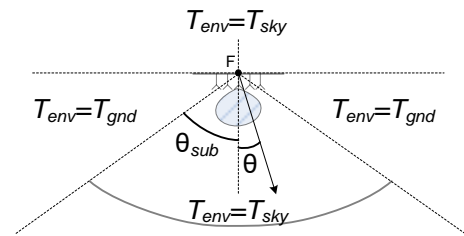

в) при $90° < \theta < 180°$ температура $T_{env}$ равна температуре фонового излучения неба $T_{sky} = 5,3$ К.

Рис. 2.11 Распределение шумовой температуры окружения

Следует отметить, что положение зеркала было выбрано именно как показано на рис. 2.11 (направленно в зенит), так как это наихудший случай с точки зрения приема сильных шумов земли.

*Температура окружающей среды* $T_{amb}$ используется для расчета составляющей шумовой температуры системы за счет потерь энергии в металле решетки и в МПФ, а также для корректировки измеренных величин $T_{sec}$ и $T_{LNA}$ (см. пункт 2.3.4).

*Параметры МПФ* были описаны в подразделе 2.2.

*Параметры МШУ*. Коэффициент шума $F$ МШУ можно записать при помощи следующего выражения [79]:

$$F = F_{min} + \frac{4R_n\left|\Gamma_s - \Gamma_{opt}\right|^2}{Z_0\left(1 - \left|\Gamma_s\right|^2\right)\left|1 + \Gamma_{opt}\right|^2}, \qquad (2.6)$$

где $F_{min}$ — минимальный коэффициент шума для данного МШУ; $R_n$ — эквивалентное шумовое сопротивление входного транзистора; $Z_0$ — характеристическое сопротивление (50 Ом); $\Gamma_s$ — коэффициент отражения сигнала от входа МШУ; $\Gamma_{opt}$ — оптимальный коэффициент отражения сигнала от входа МШУ, при котором коэффициент шума $F$ будет минимален и равен $F_{min}$.

Таким образом, МШУ характеризуется тремя шумовыми параметрами: $F_{min}$, $R_n$ и $\Gamma_{opt}$. В данной работе эти параметры получены путем измерений для конкретного типа МШУ, используемого в *APERTIF*.

*Шумовая температура цепей приемника после МШУ $T_{sec}$* представляет собой шумы следующих за МШУ каскадов усиления, а также шумы преобразователей частоты и аналого-цифровых преобразователей, приведенные ко входу МШУ. Она также была измерена, и составляет приблизительно 5 К.

*Форма поля обзора телескопа и конфигурация лучей.*

Так как рассматриваемая антенная система является многолучевой, лучи которой формируются одновременно, то необходимо задать форму поля обзора, конфигурацию лучей (то есть, определить как именно лучи будут заполнять поле обзора), а также количество лучей.

Для формирования поля обзора и конфигурации лучей исходными данными являются требуемая площадь поля обзора $A_{FOV}$ и максимально допустимое количество лучей $N_{max}$. Ограничение на количество лучей обуславливается стоимостью и вычислительной мощностью коррелятора интерферометра и компьютеров постобработки. Для проекта *APERTIF* требуемая площадь поля обзора задана $A_{FOV} = 8$ кв. градусов, а $N_{max}$ ограничено 40 одновременными лучами.

Следует отметить, что здесь под полем обзора понимается площадь неба, наблюдаемая при фиксированном положении зеркал интерферометра. Построение же карты всего неба происходит путем поворота зеркал и складывания «мозаики» из одиночных полей обзора.

Рассмотрены несколько способов размещения лучей внутри поля обзора (рис. 2.12). На рисунке показаны одиночное поле обзора (темная область) и соседние поля обзора при повороте зеркала телескопа. Окружностями показаны сечения лучей на уровне –3 дБ. Уровень –3 дБ задан в первом приближении и может изменяться при оптимизации непрерывности поля обзора (см. подраздел 2.3.4).

Самый простой способ формирования поля обзора, показанный на рис. 2.12, а, имеет небольшую площадь $A_{FOV}$, так как для обеспечения хорошего заполнения поля обзора лучами последние должны значительно перекрываться (на уровне более чем –3 дБ).

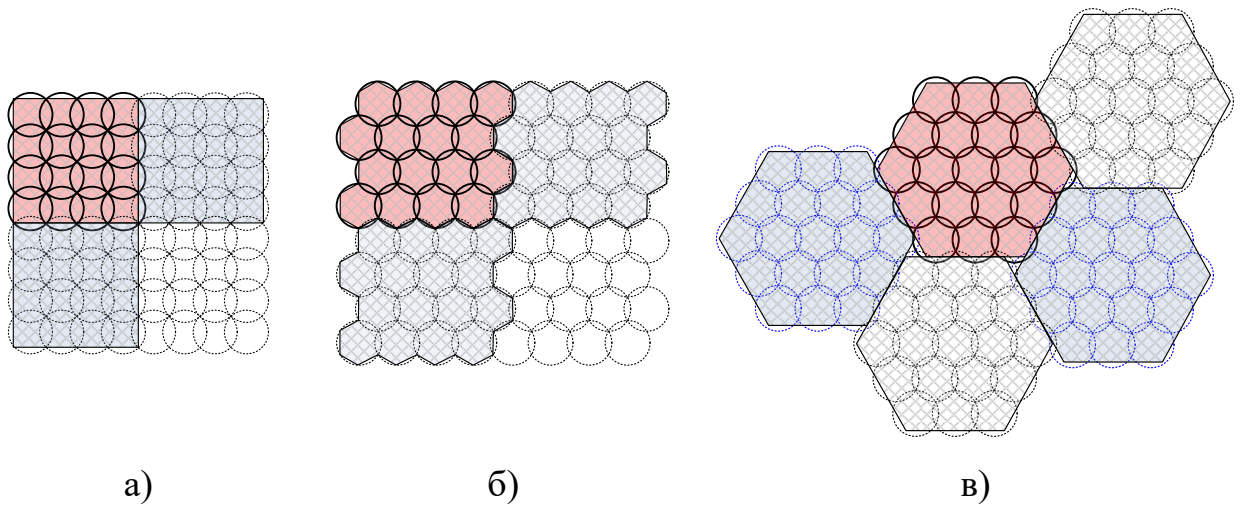

а)  б)  в)

Рис.2.12 — Рассмотренные конфигурации поля обзора телескопа

Несколько увеличить эффективность заполнения можно, использовав треугольную сетку расположения лучей, как показано на рис. 2.12,б. Но такой способ неприемлем, так как не позволяет формировать центральный луч, а угловые лучи значительно удалены от фокальной оси, что приведет к большему искажению их формы и уменьшению коэффициента направленного действия.

Чтобы избавиться от указанных недостатков, выбор был остановлен на гексогональной конфигурации поля обзора, которая показана на рис. 2.12, в. Такая конфигурация является более симметричной, а угловые лучи одинаково удалены от фокальной оси. В дальнейшем будет использоваться именно такое расположение лучей.

### 2.3.2 Первый этап: расчет системы без рефлектора

В общем виде весь процесс моделирования можно описать следующим образом.

Сначала рассматривается система без рефлектора. На рис.2.13,а показаны фотография одного антенного элемента решетки такой системы и его эквивалентная схема, а на рис.2.13,б — схема системы.

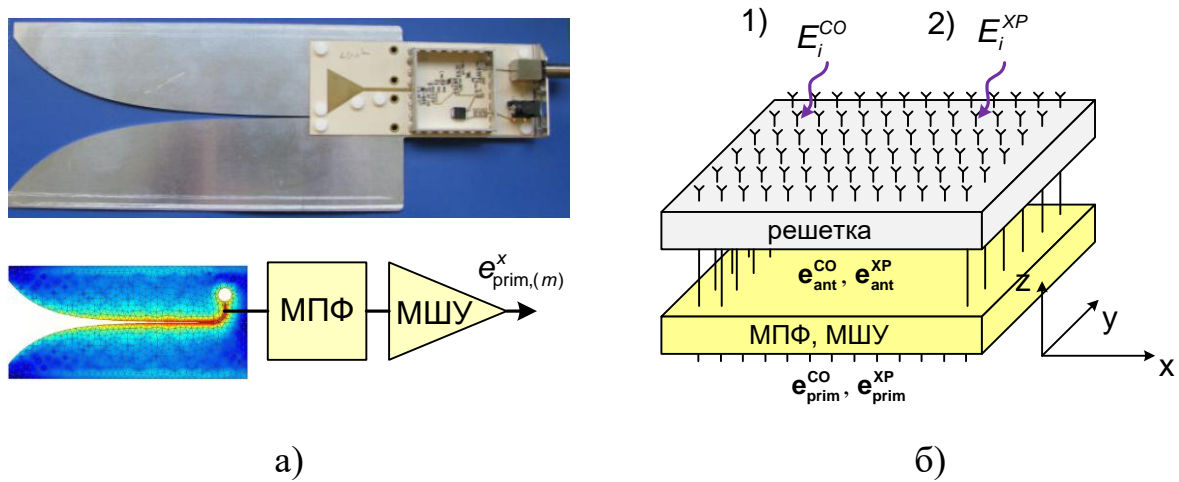

а)                                                        б)

Рис.2.13 — Система без рефлектора и ее части:

а) один антенный элемент системы, б) схема системы, включающая антенную решетку из отдельных антенных элементов

На рисунке использованы следующие обозначения: $E_i^{CO}$ и $E_i^{XP}$ — соответственно ко- и кроссполяризованная падающая с заданного направления волна с заданной напряженностью поля; $\mathbf{e}_{ant}^{CO}$ и $\mathbf{e}_{ant}^{XP}$ — векторы выходного сигнала на портах металлической структуры решетки, соответствующие падающей волне; $\mathbf{e}_{prim}^{CO}$ и $\mathbf{e}_{prim}^{XP}$ — сигнальный вектор с учетом МПФ и МШУ. Так как в дальнейшем часто будут рассматриваться векторы $\mathbf{e}_{prim}^{CO}$ и $\mathbf{e}_{prim}^{XP}$ в паре, обозначим поляризацию падающего поля как $x$, то есть $x \in \{CO, XP\}$. Здесь и далее под сигнальным вектором $\mathbf{e}$ понимается набор значений выходного сигнала с каждого элемента решетки (канала приема), то есть $\mathbf{e} = [e_1, e_2, \ldots, e_M]^T$, где $M$ — количество антенных элементов в решетке, верхний индекс Т означает транспонирование.

Данная система облучается со всех направлений падающей волной с напряженностью электрического поля $E_i$ составляющей 1 В/м и имеющую кополяризацию, а затем — кроссполяризацию по отношению к оси X координатной системы. Таким образом, рассчитываются два выходных сигнальных вектора на выходах МШУ: $\mathbf{e}_{prim}^{CO}$ и $\mathbf{e}_{prim}^{XP}$, или $\mathbf{e}_{prim}^{x}$. Такова физическая модель, примененная на первом этапе моделирования.

Опишем теперь модель математическую.

Рассчитывается напряжение на выходах металлической структуры решетки (без МПФ) без нагрузки, вектор $\mathbf{V}_{oc}^{x}(\theta, \varphi)$, элементы которого рассчитываются по формуле [4]

$$V_{oc,(m)}^{x}(\theta, \varphi) = \frac{1}{j\omega\mu}\left(E_i g_m^x(\theta, \varphi)\right), \qquad (2.7)$$

где $m = 1 \ldots M$, $M$ — количество элементов в решетке.

Сигнальный вектор на выходе металлической структуры $\mathbf{e}_{ant}^{x}$ рассчитывается следующим образом [4]:

$$\mathbf{e}_{ant}^{x} = \mathbf{L}\mathbf{V}_{oc}^{x},\qquad(2.8)$$

где $\mathbf{L} = \sqrt{Z_0}\mathbf{I} \cdot (\mathbf{Z}_{ant} + Z_0\mathbf{I})^{-1}$ — матрица преобразования напряжения на выходе открытых (без нагрузки) элементов решетки в выходной сигнальный вектор, который представляет собой выходную волну.

Далее, чтобы получить сигнальный вектор на выходе решетки с МПФ и МШУ, решетка представляется как $M$-полюсник с известной матрицей рассеяния (см. формулу (2.5)) и известной выходной сигнальной волной (формула (2.8)). К портам этого $M$-полюсника подключаются МПФ, эквивалентная схема которого показана на рис. 2.6,б, к выходу которого, в свою очередь, подключается МШУ. В результате получается микроволновая модель, показанная на рис. 2.14.

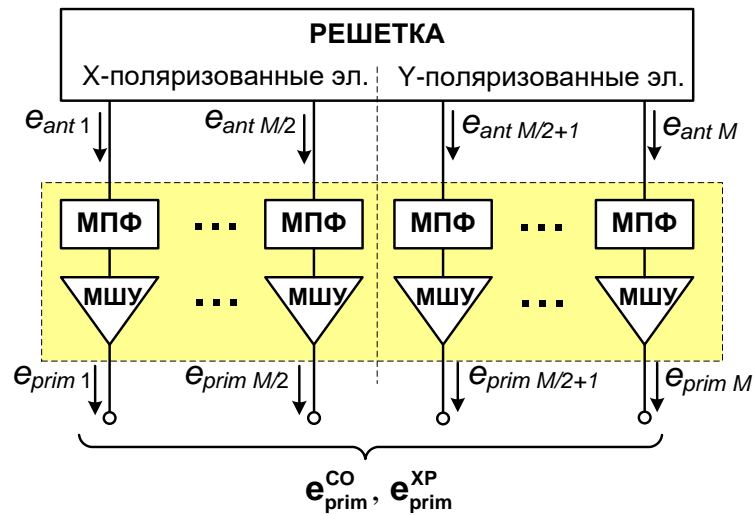

Рис. 2.14 — Микроволновая модель системы без рефлектора

Для численных расчетов приведенной микроволновой системы использовался микроволновый симулятор, входящий в комплект программы

CAESAR. Численные результаты моделирования прототипа фокальной решетки APERTIF будут приведены в разделе 3.

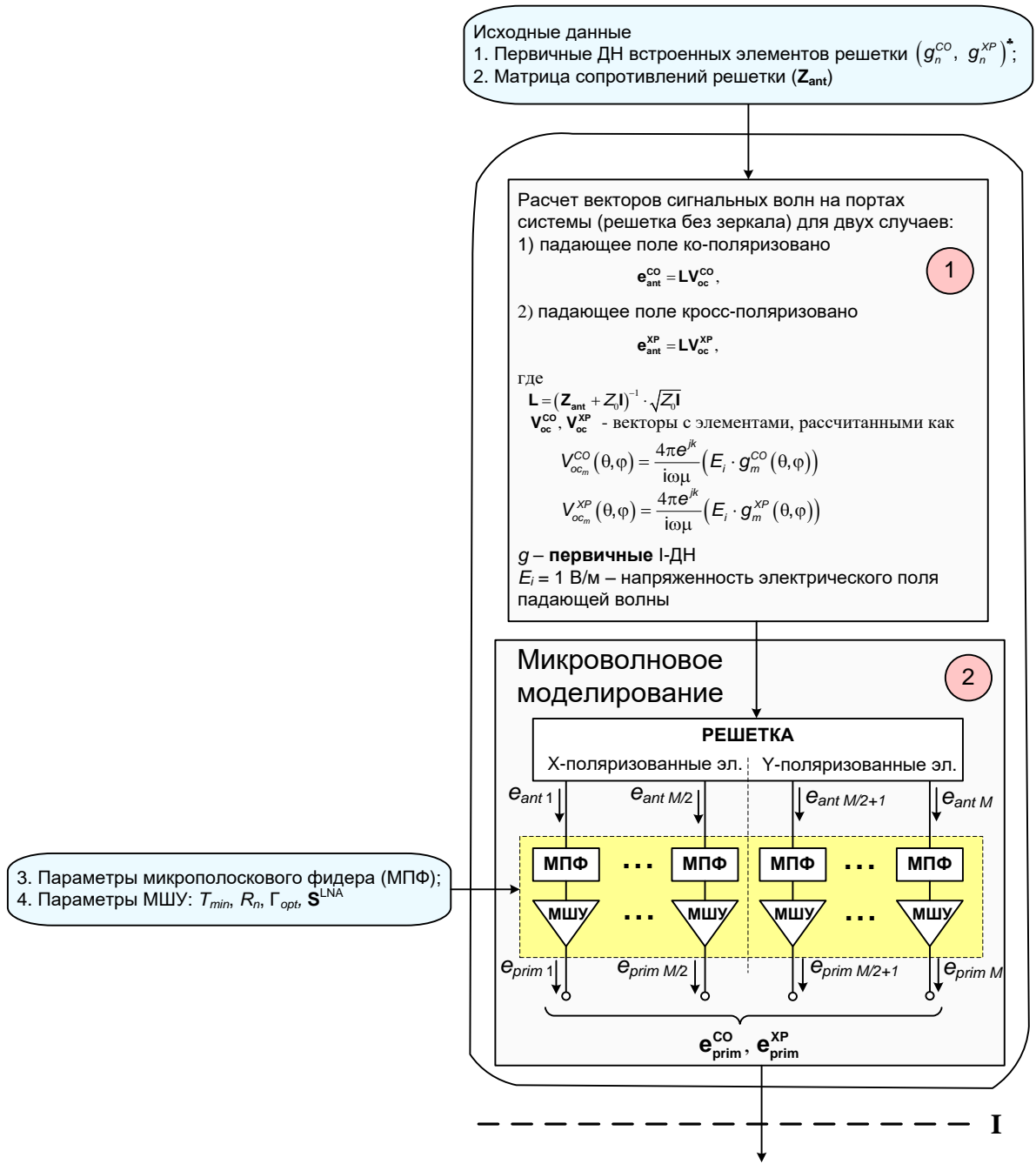

Рис. 2.15 — Алгоритм первого этапа моделирования: расчет выходного сигнала при приеме падающей плоской волны со всех направлений для антенной решетки без зеркала

Алгоритм первого этапа моделирования показан на рис. 2.15. На выходе второго блока (сечение I) имеется выходной сигнал от каждого элемента антенной решетки *без зеркала*, полученный от источника, формирующего падающее поле напряженностью $E_i = 1$ В/м, и последовательно расположенного на точках сферы вокруг решетки. Другими словами, мы имеем ненормированные ДН отдельных элементов решетки, с учетом взаимного влияния элементов друг на друга и с учетом эффектов, вносимых МПФ и МШУ. К этим эффектам относится, например, неидеальное согласование, частотная дисперсия сопротивлений элемента Вивальди, МПФ и МШУ, и т.д.

### 2.3.3 Второй этап: расчет системы с рефлектором

Второй этап моделирования — это определение выходного сигнального вектора, а также шумовой корреляционной матрицы (см. блок 5 на рис. 2.17), для случая системы с рефлектором (рис. 2.16).

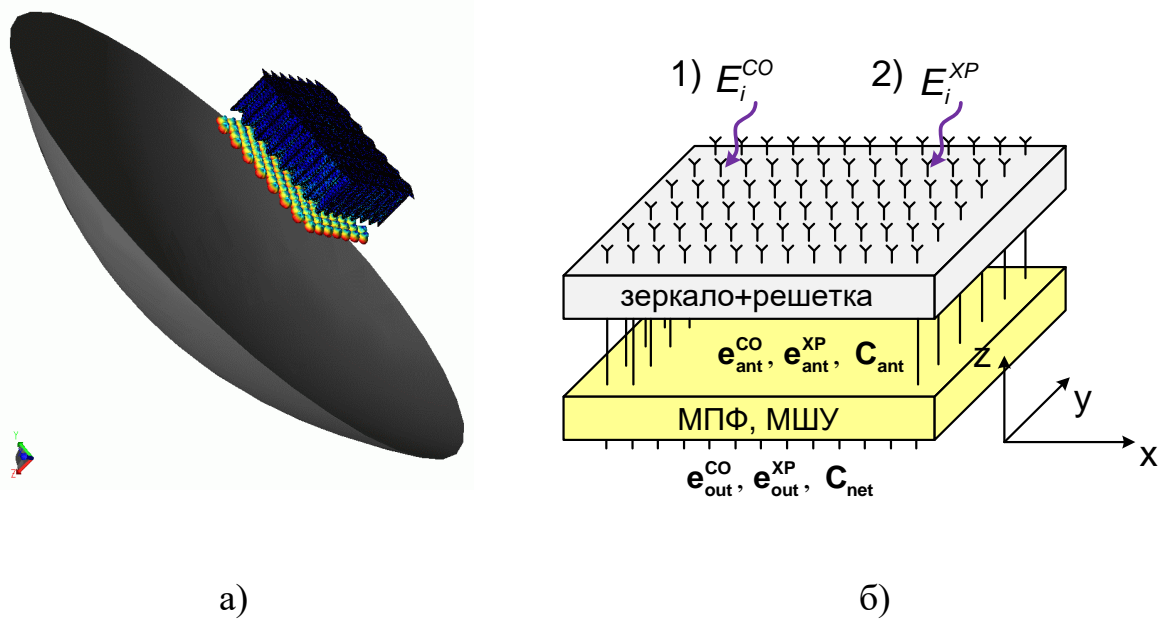

а)            б)

Рис. 2.16 — Модель системы с рефлектором:

а) зеркало с антенной решеткой (для наглядности диаметр зеркала уменьшен), б) схема системы с рефлектором

Алгоритм второго этапа показан на рис. 2.17.

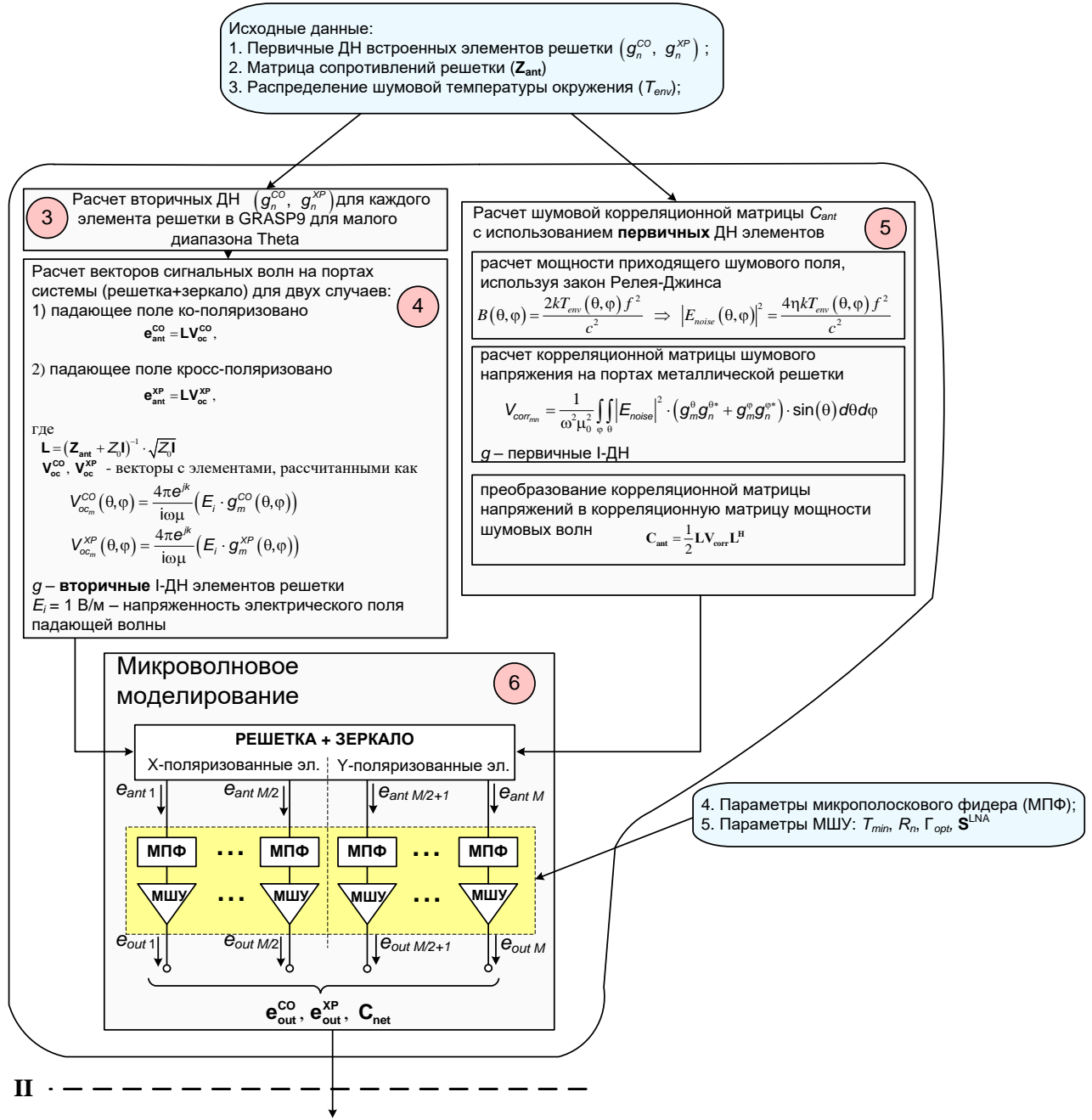

Рис. 2.17 — Алгоритм второго этапа моделирования: расчет выходного сигнала при приеме падающей плоской волны со всех направлений, а также расчет шумовой корреляционной матрицы для антенной решетки с зеркалом

Отличие алгоритма второго этапа моделирования от алгоритма для системы без зеркала (первый этап) заключается в наличии блоков 3 и 5.

В *блоке 3* рассчитываются вторичные ДН (то есть ДН после отражения от зеркала) через первичные ДН элементов решетки при помощи коммерческой программы *GRASP*9, в которой используется метод физической оптики (*PO*) и геометрическая теория дифракции (*GTD*) [80]. Для создания интерфейса между программами *CAESAR* и *GRASP*9 автором был написан специальный код в программе *MATLAB*.

Примеры рассчитанных для частоты 1,42 ГГц вторичных ДН элементов решетки показаны на рис. 2.18. На рисунке показаны нормированные зависимости КНД зеркала от направления, когда зеркало облучается одним антенным элементом решетки (в присутствии остальных антенных элементов). Единицы измерения — дБ.

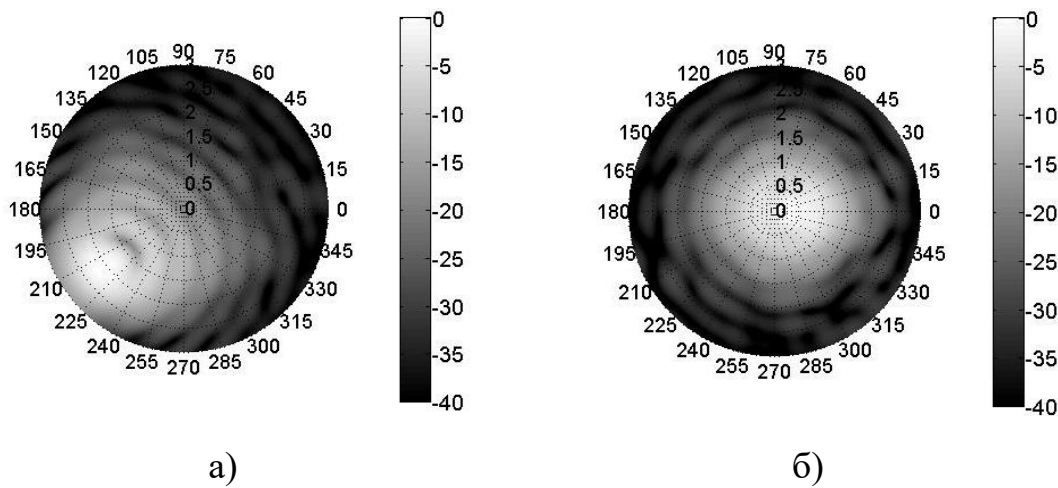

а)                              б)

Рис.2.18 Вторичные ДН для элементов решетки:
а) углового антенного элемента,  б) антенного элемента, близкого к центру решетки

Расчет выходного сигнального вектора (*блоки 4* и *6*) проводится аналогично первому этапу, только вместо первичных ДН элементов использовались вторичные.

В отличие от первого этапа моделирования, на котором считалось, что шумы в системе отсутствуют, и, соответственно, анализ шумов не проводился, на втором этапе рассматривается уже вся система в целом, и такой анализ необходимо провести. Для этого был введен *блок 5*. Рассмотрим этот блок подробнее.

Распределение шумовой температуры окружения антенной системы в зависимости от направления является заданной функцией: $T_{env}(\theta, \varphi)$. Как эта функция задавалась, было подробно описано в подразделе 2.3.1. Эти шумы принимаются каждым элементом решетки.

Очевидно, что шумы, принятые каждым элементом решетки, будут коррелированы, и корреляция будет тем сильнее, чем больше ДН элементов перекрываются по направлению. А раз шумы являются коррелированными, то уже нельзя просто складывать их мощности, и для расчета шумовых характеристик системы требуется производить расчет коэффициентов корреляции между шумами на выходе каждого антенного элемента решетки.

Коэффициенты корреляции, а также мощность шума на выходах каждого антенного элемента решетки *без МПФ и МШУ* удобно представить в виде шумовой корреляционной матрицы $\mathbf{C_{ant}}$ [4, с.179]:

$$\mathbf{C_{ant}} = \frac{1}{2} \mathbf{L} \mathbf{V_{corr}} \mathbf{L}^H , \qquad (2.9)$$

где $\mathbf{V_{corr}}$ — корреляционная матрица шумового напряжения на портах металлической структуры решетки; $\mathbf{L}$ — матрица преобразования напряжения в волны [4]: $\mathbf{L} = \left( \mathbf{Z_{ant}} + \mathbf{Z_0 I} \right)^{-1} \sqrt{\mathbf{Z_0 I}}$.

Элементы матрицы $\mathbf{V_{corr}}$ являются коэффициентами корреляции между напряжением на портах соответствующей пары элементов решетки при приеме только шумов окружения:

$$\mathbf{V_{corr}} = \overline{\mathbf{V}_{oc}\mathbf{V}_{oc}^{H}} = \begin{pmatrix} \overline{\left|V_{oc,1}\right|^2} & \overline{V_{oc,1}V_{oc,2}^*} & \cdots & \overline{V_{oc,1}V_{oc,M}^*} \\ \overline{V_{oc,2}V_{oc,1}^*} & \overline{\left|V_{oc,2}\right|^2} & \cdots & \overline{V_{oc,2}V_{oc,M}^*} \\ \vdots & \vdots & \ddots & \vdots \\ \overline{V_{oc,M}V_{oc,1}^*} & \overline{V_{oc,M}V_{oc,2}^*} & \cdots & \overline{\left|V_{oc,M}\right|^2} \end{pmatrix}, \qquad (2.10)$$

где оператор $\dots^{H}$ — это эрмитов оператор; $\dots^{*}$ — оператор комплексного сопряжения; $\overline{\dots}$ — оператор усреднения по времени.

Напряжение на открытом (без нагрузки) порту элемента решетки $V_{oc}$ при приеме сигнала от точечного источника с определенного направления рассчитывается по выражению (2.7). Но, так как шумы принимаются сразу со всех направлений и имеют различную интенсивность, то для расчета наведенного ими напряжения на выходе $m$-го антенного элемента решетки требуется вычислить интеграл

$$V_{oc,(m)} = \frac{1}{j\omega\mu}\int_{\Omega} E_{noise}(\Omega)\, g_m(\Omega)\mathrm{d}\Omega, \qquad (2.11)$$

где $E_{noise}(\Omega)$ — напряженность поля, создаваемого шумящим окружением.

С учетом (2.11) элементы матрицы $\mathbf{V_{corr}}$ будут представлять собой следующие интегралы:

$$\overline{V_{oc,(m)}V_{oc,(n)}^*} = \left(\frac{1}{j\omega\mu}\right)\left(\frac{1}{-j\omega\mu}\right)\int_{\Omega}\overline{E_{noise}(\Omega)E_{noise}^*(\Omega)}\cdot g_m(\Omega)g_n^*(\Omega)\mathrm{d}\Omega =$$

$$= \frac{1}{\omega^2\mu^2}\int_{\Omega}\overline{\left|E_{noise}(\Omega)\right|^2}\cdot g_m(\Omega)g_n^*(\Omega)\mathrm{d}\Omega. \qquad (2.12)$$

Так как $g_m$ — это поле, созданное $m$-ым антенным элементом решетки при возбуждении последнего источником тока, то $g_m$ можно разложить на две компоненты (две, так как рассматривается дальняя зона). Разложим $g_m$ на компоненты в сферической системе координат, то есть запишем выражение для $g_m$ в следующем виде: $g_m = g_m^\theta + g_m^\varphi$. Подставим это выражение для $g_m$ в выражение (2.12), и после упрощения получим

$$V_{corr_{mn}} = \overline{V_{oc,(m)} V_{oc,(n)}^*} =$$

$$= \frac{1}{\omega^2 \mu^2} \iint_{\varphi\ \theta} \overline{\left| E_{noise}(\theta, \varphi) \right|^2} \cdot \left[ g_m^\theta(\theta, \varphi) g_n^{\theta *}(\theta, \varphi) + g_m^\varphi(\theta, \varphi) g_n^{\varphi *}(\theta, \varphi) \right] \sin(\theta) \mathrm{d}\theta \mathrm{d}\varphi . \quad (2.13)$$

Чтобы рассчитать напряженность электрического поля, создаваемого шумящим окружением $E_{noise}(\Omega)$, воспользуемся законом Релея-Джинса, который устанавливает соответствие между спектральной плотностью потока энергии $B$ и температурой абсолютно черного тела, создающего этот поток:

$$B(\Omega) = 2k_B T_{env}(\Omega) \frac{f^2}{c^2} , \quad (2.14)$$

где $k_B = 1{,}38 \cdot 10^{-23}$ Дж/К — постоянная Больцмана, $f$ — частота, на которой ведется анализ; $c$ — скорость света.

Величина $B$ выражается в $\left[ \dfrac{\text{Вт}}{\text{м}^2 \cdot \text{ср} \cdot \text{Гц}} \right]$ [81]. В то же время величина $\dfrac{\overline{\left| E_{noise} \right|^2}}{2\eta}$ — это мощность падающего шумового поля, и выражается в $\left[ \dfrac{\text{Вт}}{\text{м}^2 \cdot \text{ср}} \right]$. Таким образом, $\dfrac{\overline{\left| E_{noise} \right|^2}}{2\eta}$ — это мощность, содержащаяся в полосе частот d$f$. Тогда можно записать следующее уравнение:

$$B\,\mathrm{d}f = \frac{\overline{\left|E_{noise}\right|^2}}{2\eta}, \tag{2.15}$$

где $\eta = 120\pi$ — сопротивление свободного пространства.

Подставив (2.14) в (2.15) и выразив $\overline{\left|E_{noise}\right|^2}$, получим

$$\left|E_{noise}(\theta,\varphi)\right|^2 = 2\eta\,B(\theta,\varphi)\,\mathrm{d}f = 4\eta k_B T_{env}(\theta,\varphi)\frac{f^2}{c^2}\,\mathrm{d}f. \tag{2.16}$$

Таким образом, используя формулы (2.16), (2.13), (2.10) и (2.9), можно рассчитать шумовую корреляционную матрица $\mathbf{C}_{ant}$ на выходе металлической структуры решетки за счет приема внешних шумов неба и земли. Полученная матрица используется в микроволновом симуляторе CAESAR для расчета шумовой корреляционной матрицы $\mathbf{C}_{net}$ на выходе уже всей системы, показанной на рис. 2.16, б.

Следует отметить, что при симулировании учитываются также шумы, генерируемые МПФ и МШУ, для чего в параметрах, описывающих МШУ и микрополосковые линии МПФ, задается температура окружающей среды $T_{amb}$.

Итак, итогом второго этапа моделирования (сечение II алгоритма) являются выходные сигнальные векторы $\mathbf{e}_{out}^{co}(\theta,\varphi)$ и $\mathbf{e}_{out}^{xp}(\theta,\varphi)$, а также шумовая корреляционная матрица $\mathbf{C}_{net}$ на выходе всей системы с рефлектором.

### 2.3.4 Третий этап: определение оптимальных весовых коэффициентов

В данной работе рассматривались и сравнивались три метода вычисления весовых коэффициентов:

    – метод согласования по полю [32, 82];

- метод максимизации чувствительности [72, 73];

- метод максимизации чувствительности с ограничениями по направлениям, известный также как *Linear Constrained Minimum Variance* (LCMV) [75].

Обзор этих методов был выполнен в подразделе 1.6. Рассмотрим их в приложении к гибридной антенной системе.

**Метод согласования по полю.**

Метод согласования по полю (*CFM — conjugate field matching*) — самый простой способ вычисления весовых коэффициентов, при котором антенная система принимает максимум падающей на нее энергии, то есть ведется согласованный прием.

Весовые коэффициенты $\mathbf{w_{opt}}$ в этом методе равны комплексно-сопряженной величине сигнала на выходе соответствующих элементов решетки при приеме сигнала от источника в заданном направлении, то есть

$$\mathbf{w_{opt}^{x}} = \left[ \mathbf{e_{out}^{x}} \left( \theta_0, \varphi_0 \right) \right]^*, \qquad (2.17)$$

где $\mathbf{w_{opt}^{x}} = \left[ w_1, w_2, \ldots, w_M \right]^T$ — оптимальные по заданному критерию весовые коэффициенты для приема падающей волны с поляризацией $x$; $\left( \theta_0, \varphi_0 \right)$ — направление требуемого максимума ДН системы.

Как отмечалось во введении, для радиоастрономии очень важным является такой параметр системы, как чувствительность $S$, которая определяется как отношение эффективной площади антенны $A_{eff}$ к шумовой температуре системы $T_{sys}$:

$$S = \frac{A_{eff}}{T_{sys}}. \qquad (2.18)$$

Очевидно, что чувствительность тем выше, чем больше КНД антенны $D_0$ в направлении источника. Это следует из формулы, связывающей КНД и эффективную площадь апертуры антенны [83, с.416]:

$$D_0 = \frac{4\pi}{\lambda^2} A_{eff},\qquad (2.19)$$

где $\lambda$ — длина волны; $A_{eff} = A_{phys}\eta_{ap}$ — эффективная площадь зеркальной антенны; $A_{phys}$ — физическая площадь апертуры зеркала; $\eta_{ap}$ — коэффициент использования поверхности (КИП), или апертурная эффективность (*aperture efficiency*).

Также известно, что при согласованном приеме принимается максимальная мощность излучения источника, и $D_0$ при этом максимален. Но в выражении для чувствительности (2.18) присутствует также шумовая температура системы $T_{sys}$, которую также необходимо учитывать при расчете весовых коэффициентов. Как было отмечено, шумы на выходах элементов решетки коррелированны, и, следовательно, их можно частично компенсировать путем оптимального выбора весовых коэффициентов.

Как будет показано в разделе 4, применение весовых коэффициентов, рассчитанных *CFM* методом, действительно не дает максимальной чувствительности.

**Метод максимизации чувствительности.**

Применительно к модели нашей системы, формула (1.12) примет вид:

$$\mathbf{w}_{opt}^x = \mathbf{C}_{net}^{-1}\mathbf{e}_{out}^x\left(\theta_0, \varphi_0\right).\qquad (2.20)$$

Расчеты, проведенные по данной методике, представлены в разделах 3 и 4.

**Метод максимизации чувствительности с ограничениями по направлениям.**

В случае использования многолучевой антенной системы перекрытие наблюдаемой области лучами должно осуществляться так, чтобы чувствительность в точке пересечения соседних лучей уменьшалась не более чем на (10…15) % (рис. 2.19). Для некоторых радиоастрономических наблюдений распределение чувствительности для всего поля обзора должно быть как можно ближе к равномерному.

Рассмотренный ранее метод максимизации чувствительности оптимизирует чувствительность *отдельных* лучей, и поэтому не позволяет регулировать ее неравномерность в области многих лучей.

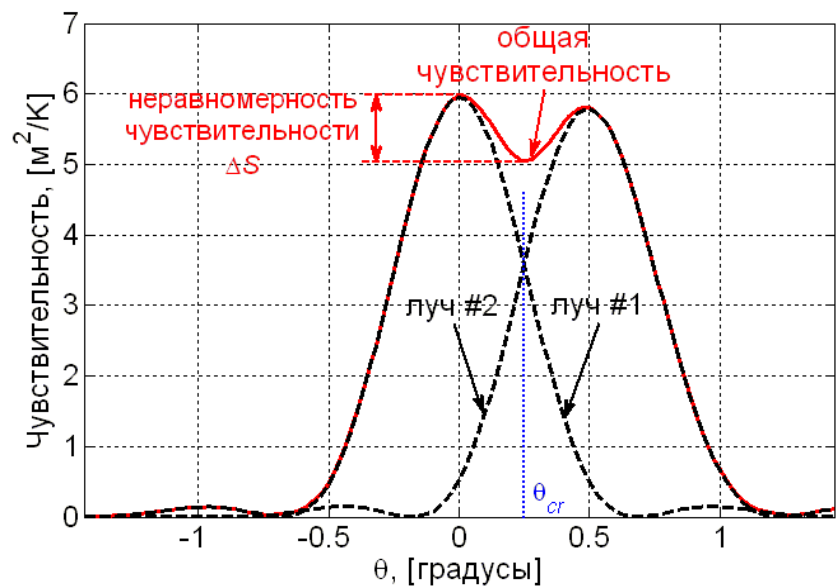

Рис. 2.19 — К объяснению проблемы неравномерности чувствительности внутри поля обзора телескопа

На рис. 2.19 пунктиром показана чувствительность одновременно формируемых двух лучей телескопа, оптимальных по критерию максимальной чувствительности в направлениях $\theta = 0°$ и $\theta = 0{,}5°$. Сплошной линией показана общая чувствительность всего интерферометра $S_{tot}$, которая рассчитывается как среднеквадратическое значение чувствительности всех лучей для направления θ:

$$S_{tot}(\theta) = \sqrt{\sum_{n=1}^{N} S_n^2(\theta)}, \qquad (2.21)$$

где $N$ — количество формируемых лучей.

Такой метод расчета общей чувствительности следует из следующих соображений. Чувствительность интерферометра вычисляется следующим образом [28]:

$$S_{tot} = \frac{A_{eff}}{T_{sys}} \cdot \frac{\sqrt{\Delta f \cdot t \cdot n}}{K_S}, \qquad (2.22)$$

где $\Delta f$ — полоса частот, в которой происходит наблюдение; $t$ — время интегрирования в корреляторе; $n$ — количество усредненных записей; $K_S$ — коэффициент чувствительности, который зависит от типа приемной системы и коррелятора.

Из (2.22) следует, что чувствительность увеличивается как корень квадратный от времени интегрирования. Но так как в направлении $\theta_{cr}$ наблюдение велось одновременно двумя лучами с одинаковой чувствительностью в течение одинакового времени, то в этом направлении общая чувствительность интерферометра будет в $\sqrt{2}$ раз выше.

Анализируя рис. 2.19, можно предложить два способа обеспечения чувствительности, близкой к постоянной в области поля обзора.

**Первый способ** основан на уменьшении расстояния между лучами. В этом случае чувствительность остается высокой, но при сближении лучей уменьшается общее поле обзора, а значит, для обеспечения заданного размера поля обзора необходимо увеличивать количество лучей, что ведет к усложнению и повышению стоимости коррелятора и повышению требований к вычислительной мощности компьютеров для последующей обработки полученных данных.

При проектировании многолучевой системы исходными данными (среди прочих) являются площадь поля обзора $A_{FOV}$ и максимальное количество лучей $N_{max}$, которыми можно заполнить это поле обзора. Первый способ устанавливает зависимость площади поля обзора от числа лучей, в то время как эти две величины должны быть независимыми.

Поэтому был рассмотрен **второй способ**, в котором расчет весовых коэффициентов $\mathbf{w_{opt}}$ производится методом *LCMV*, описанному в подразделе 1.6. Используя обозначения, примененные в модели, формула (1.13) примет вид:

$$\left(\mathbf{w}_{\mathbf{opt}}^{x}\right)^{H} = \mathbf{g}^{H}\left[\mathbf{G}^{H}\mathbf{C}_{\mathbf{net}}^{-1}\mathbf{G}\right]^{-1}\mathbf{G}^{H}\mathbf{C}_{\mathbf{net}}^{-1}, \qquad (2.23)$$

где $\mathbf{G}$ — матрица размером $M{\times}N_{dir}$, содержащая сигнальные вектора с $N_{dir}$ направлений ограничений ($M$ — количество элементов в решетке); $\mathbf{g}$ — вектор размерностью $N_{dir}{\times}1$, элементы которого устанавливают уровень ДН луча в каждом из $N_{dir}$ направлений.

Приведем пример. Предположим, необходимо рассчитать весовые коэффициенты для формирования луча № 1 (см. рис. 2.19) так, чтобы отношение чувствительности в направлении $\theta_{cr}$ к чувствительности в главном направлении луча ($\theta = 0°$) было выше, чем для случая максимизации чувствительности (без ограничений по направлениям). Для этого выбираем три направления ограничений ($N_{dir} = 3$): $\theta = \theta_0 = 0°$, $\theta = \theta_{cr}$ и $\theta = -\theta_{cr}$. При этом матрица $\mathbf{G}$ примет вид: $\mathbf{G} = \left[\mathbf{e}_{\mathbf{out}}^{x}\left(\theta_0\right), \ \mathbf{e}_{\mathbf{out}}^{x}\left(\theta_{cr}\right), \ \mathbf{e}_{\mathbf{out}}^{x}\left(-\theta_{cr}\right)\right]$, то есть содержит сигнальные векторы при приеме сигнала с выбранных трех направлений ограничения. Вектор $\mathbf{g}$ будет иметь три элемента, соответствующих трем выбранным направлениям. В направлении $\theta = \theta_0$ мы не будем вводить ограничения, поэтому первый элемент вектора **g** будет равен $g_1=1$. В направлениях $\theta = \theta_{cr}$ и $\theta = -\theta_{cr}$ мы хотим получить чувствительность (одинаковую вследствие симметрии осевого луча), которая должна быть меньше чувствительности в основном направлении луча $\theta_0$. Следовательно, элементы $g_2$ и $g_3$ вектора **g** будут одинаковы и равны некоторому значению $g_{const} < g_1$. Таким образом, вектор **g** будет равен $\mathbf{g} = \begin{bmatrix} 1, & g_{const}, & g_{const} \end{bmatrix}^T$. При этом величина $g_{const}$ должна определяться в процессе оптимизации неравномерности чувствительности.

Результаты расчета весовых коэффициентов по этому методу, оптимизация $g_{const}$ для получения максимальной гладкости чувствительности, а также анализ системы с применением рассчитанных весовых коэффициентов приведены в разделе 4.

### 2.3.4 Четвертый этап: расчет основных параметров системы

**Расчет вторичных ДН лучей.**

Зная весовые коэффициенты $\mathbf{w}_{\mathbf{opt}}^x$, оптимальные для приема $x$-поляризационной компоненты падающего поля, а также *вторичные* ДН (выходной сигнал) отдельных элементов решетки $\mathbf{e}_{\mathbf{out}}^y(\theta, \varphi)$ при приеме $y$-поляризационной компоненты падающего поля, можно вычислить общую *вторичную* ДН для $n$-го луча:

$$e_{out,tot,n}^{x,y}(\theta, \varphi) = \begin{bmatrix} \mathbf{e}_{\mathbf{out}}^y(\theta, \varphi) \end{bmatrix}^T \mathbf{w}_{\mathbf{opt},n}^x, \tag{2.24}$$

где $x \in \{CO, XP\}$ — индекс, указывающий, для приема какой поляризации поля были оптимизированы весовые коэффициенты; $y \in \{CO, XP\}$ — индекс, указывающий, для падающего поля какой поляризации ведется анализ; $e_{out,tot,n}^{x,y}(\theta, \varphi)$ — сигнал на выходе формирователя при приеме $y$- поляризационной компоненты падающего поля, и использованием весовых коэффициентов, оптимальных для приема $x$-поляризационной компоненты падающего поля.

Следует заметить, что для анализа системы рассматривается только линейная поляризация падающего поля, так как система содержит два формирователя лучей, которые должны выделять две линейные поляризационные составляющие. Рассчитанные поляриметрические характеристики системы приведены в разделе 4.

В описанном ранее процессе моделирования выходной сигнал от антенных элементов решетки зависит от напряженности падающего поля, величина которого была выбрана произвольно. При реальных измерениях напряженность поля также меняется в широком диапазоне. Чтобы исключить зависимость от интенсивности источника, необходимо нормировать общую вторичную ДН каждого луча к изотропному излучателю таким образом, чтобы общая мощность выходного сигнала со всех направлений была равна $4\pi$:

$$E_{out,tot,n}^{x,y}\left(\theta,\varphi\right) = \frac{e_{out,tot,n}^{x,y}\left(\theta,\varphi\right)}{\sqrt{GNF_n^x}}, \qquad (2.25)$$

где $\qquad GNF_n^x = \dfrac{1}{4\pi}\displaystyle\int\limits_0^{2\pi}\int\limits_0^{\pi}\left[\left|e_{out,tot,n}^{x,CO}\left(\theta,\varphi\right)\right|^2 + \left|e_{out,tot,n}^{x,XP}\left(\theta,\varphi\right)\right|^2\right]\sin\theta\;\mathrm{d}\theta\mathrm{d}\varphi$ —

нормализующий коэффициент, рассчитанный как общая мощность выходного сигнала $n$-го луча, деленная на общую мощность изотропного излучателя.

При этом, если взять квадрат модуля нормированной таким образом ДН, получим зависимость КНД от направления:

$$D_n^{x,y}\left(\theta,\varphi\right) = \left|E_{out,tot,n}^{x,y}\left(\theta,\varphi\right)\right|^2. \qquad (2.26)$$

Так как поле обзора небольшое ($A_{FOV} = 8$ кв. градусов), то максимальный угол сканирования при гексогональном расположении лучей не превышает 1,5°. Поэтому, для ускорения расчетов и экономии оперативной памяти компьютера, расчеты вторичных ДН проводились в диапазоне углов $0 \leq \theta \leq 3°$ и $0 \leq \varphi < 360°$. В этом случае нормировать вторичные ДН лучей описанным способом не получится (так как необходимо знать мощность выходного сигнала по всем направлениям). Однако, с учетом того, что мощность, содержащаяся в первичной и вторичной ДН одинакова (при условии, что вторичная ДН учитывает эффект переливания энергии облучателя за края зеркала, что учтено при расчете вторичных ДН элементов решетки в программе GRASP9), нормировка вторичных ДН лучей может быть произведена по выражению (2.25), в котором нормализующие коэффициенты $GNF_n^x$ рассчитываются через первичные ДН лучей:

$$GNF_n^x = \frac{1}{4\pi} \int\limits_0^{2\pi} \int\limits_0^{\pi} \left[ \left| e_{prim,tot,n}^{x,CO} (\theta,\varphi) \right|^2 + \left| e_{prim,tot,n}^{x,XP} (\theta,\varphi) \right|^2 \right] \cdot \sin\theta \, d\theta d\varphi, \quad (2.27)$$

где $e_{prim,tot,n}^{x,y}(\theta,\varphi) = \left[ \mathbf{e}_{prim}^y (\theta,\varphi) \right]^T \mathbf{w}_{opt,n}^x$ — сигнал на выходе формирователя, аналогично $e_{out,tot,n}^{x,y}(\theta,\varphi)$, но при приеме сигнала без зеркала.

Таким образом, результатом расчетов на данном шаге моделирования являются нормированные к изотропному излучателю ДН лучей до отражения от зеркала $E_{prim,tot,n}^{x,y}(\theta,\varphi)$ и после отражения — $E_{out,tot,n}^{x,y}(\theta,\varphi)$, а также зависимость КНД лучей от направления $D_n^{x,y}(\theta,\varphi)$.

**Расчет эффективной площади зеркальной системы и коэффициента использования поверхности.**

Зная КНД, из выражения (2.19) можно получить эффективную площадь зеркальной системы $A_{eff}$. В нашем анализе будем рассчитывать $A_{eff}$ для каждого набора весовых коэффициентов (*n*-го луча) и приема сигнала со всех направлений:

$$A_{eff,n}^{x,y}\left(\theta,\phi\right)=\frac{\lambda^2}{4\pi}D_n^{x,y}\left(\theta,\phi\right). \tag{2.28}$$

КИП вычисляется как отношение эффективной площади апертуры зеркальной системы $A_{eff}$ к физической площади зеркала $A_{ph}$:

$$\eta_{ap,n}^{x,y}\left(\theta,\phi\right)=\frac{A_{eff,n}^{x,y}\left(\theta,\phi\right)}{A_{ph}}. \tag{2.29}$$

**Расчет КПД антенной решетки.**

Рассмотрим режим передачи.

Так как решетка состоит из антенных элементов, выполненных из металла с конечной проводимостью, то, часть энергии, подведенной к портам решетки, будет рассеиваться в виде тепла. КПД металлической структуры $\eta_{rad,met}$ можно записать следующим образом:

$$\eta_{rad,met}=1-\frac{P_{ant}^{\text{dissipated}}}{P_{ant}^{accepted}}, \tag{2.30}$$

где $P_{ant}^{\text{dissipated}}$ — рассеянная в металле решетки мощность; $P_{ant}^{accepted}$ — принятая металлической структурой решетки мощность.

Кроме потерь в металле решетки существуют также потери в МПФ элементов решетки. КПД МПФ $\eta_{rad,MSF}$ можно записать как отношение

$$\eta_{rad,MSF}=\frac{P_{feed}^{out}}{P_{feed}^{accepted}}, \tag{2.31}$$

где $P_{feed}^{out}$ — мощность, выходящая из МПФ; $P_{feed}^{accepted}$ — мощность, принятая МПФ.

Общий КПД всей антенной решетки равен произведению КПД металлической структуры на КПД МПФ:

$$\eta_{rad} = \eta_{rad,met}\eta_{rad,MSF} . \tag{2.32}$$

Модель для расчета КПД антенной решетки показана на рис. 2.20.

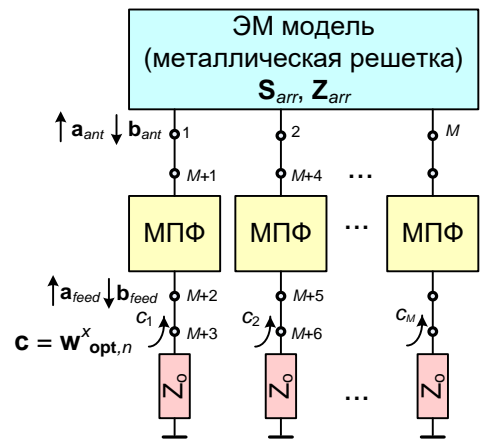

Рис. 2.20 Модель для расчета КПД антенной решетки

Представленная на рис. 2.20 микроволновая модель аналогична модели, показанной на рис. 2.14, только вместо МШУ ко входам МПФ подключены нагрузки с характеристическим сопротивлением $Z_0$, и модель рассматривается в режиме передачи.

На рис. 2.20 $\mathbf{a}_{ant}$ и $\mathbf{b}_{ant}$ — соответственно падающая на порты волна и отраженная волна от портов металлической структуры решетки. Аналогично, $\mathbf{a}_{feed}$ и $\mathbf{b}_{feed}$ — падающая и отраженная волна на входе МПФ.

Решетка возбуждается подачей волны **c** на выход нагрузок. Причем амплитуда этой волны для каждого элемента решетки пропорциональна соответствующим весовым коэффициентам, то есть выполняется формирование требуемого *n*-го луча: $\mathbf{c} = \mathbf{w}_{\mathbf{opt},n}^{x}$ .

Соотношение, связывающее волны **a**, **b** и **c** на *M*-полюснике, известно [79]:

$$\mathbf{b} = \mathbf{Sa} + \mathbf{c} . \tag{2.33}$$

Представленная модель рассчитывается в микроволновом симуляторе CAESAR, в результате чего находятся численные значения векторов $\mathbf{a}_{ant}$, $\mathbf{b}_{ant}$, $\mathbf{a}_{feed}$ и $\mathbf{b}_{feed}$.

Учитывая, что выходящая из МПФ мощность равна принятой металлической структурой мощности (см. рис. 2.20), $P_{feed}^{out} = P_{ant}^{accepted}$, выражение (2.31) можно привести к виду

$$\eta_{rad,MSF} = \frac{P_{feed}^{out}}{P_{feed}^{accepted}} = \frac{P_{ant}^{accepted}}{P_{feed}^{accepted}} = \frac{\mathbf{a}_{ant}^{H}\mathbf{a}_{ant} - \mathbf{b}_{ant}^{H}\mathbf{b}_{ant}}{\mathbf{a}_{feed}^{H}\mathbf{a}_{feed} - \mathbf{b}_{feed}^{H}\mathbf{b}_{feed}} . \tag{2.34}$$

Данное выражение является конечным для расчета КПД МПФ.

КПД металлической структуры можно рассчитать методами, описанными, например, в [84]. Однако точность этих методов составляет несколько процентов, что является недостаточным для анализа антенн с малыми потерями. Как будет показано в подразделе 3.2.2, КПД металлической структуры решетки превышает 99%. Недостаточная точность существующих методов также отмечена в [85]. Автор статьи провел расчеты КПД металлической решетки в программе *Ansoft HFSS*, и оказалось, что КПД меняется вплоть до 105%, что, естественно, говорит о недостаточной точности используемых методов. Поэтому авторами статьи [85] был предложен принципиально иной метод расчета КПД металлических антенн с малыми потерями, который и был использован для расчета металлической структуры нашей решетки.

Суть метода заключается в следующем. По известным распределению токов по поверхности металла и поверхностному сопротивлению последнего рассчитывается рассеянная мощность $P_{ant}^{\text{dissipated}}$. КПД определяется через отношение рассеянной мощности к мощности, принятой при возбуждении структуры. Покажем, как это делается применительно к исследуемой системе.

Согласно граничным условиям для бесконечно тонкого металлического листа тангенциальная составляющая электрического поля в любой точке на его поверхности равна $E_{tan} = Z_S J_S$, где $Z_S$ — поверхностное сопротивление металла, $J_S$ — плотность поверхностного тока. Тогда рассеянную мощность во всем листе можно рассчитать как интеграл по всей поверхности:

$$P_{ant}^{\text{dissipated}} = \frac{1}{2}\text{Re}\int_S J_S^* \cdot E_{tan}\, dS = \frac{1}{2}\text{Re}\int_S J_S^* \cdot Z_S J_S\, dS. \qquad (2.35)$$

Сложность прямого использования этого выражения заключается в том, что нам не известно распределение токов в решетке, а известны коэффициенты к $RWG$ функциям, аппроксимирующим эти токи. Поэтому, как показано в [85], выражение (2.35) можно переписать в следующем виде:

$$P_{ant}^{\text{dissipated}} = \frac{1}{2}\text{Re}\left[\mathbf{J}^H \mathbf{Z}^{IBC}\mathbf{J}\right], \qquad (2.36)$$

где $\mathbf{J}$ — вектор $N_{RWG} \times 1$, элементы которого представляют собой коэффициенты RWG функций при заданном возбуждении элементов решетки; $N_{RWG}$ — количество $RWG$ функций, на которые разбита структура всей решетки.

Элементы матрицы $\mathbf{Z}^{IBC}$ представляют собой скалярные произведения *m*-й и *n*-й *RWG* функций *f*, одна из которых умножена на $Z_S$:

$$Z_{m,n}^{IBC} = \int\limits_{S_m \cap S_n} f_m^* \cdot Z_S f_n \, \mathrm{d}S = \left\langle f_m, Z_S f_n \right\rangle . \qquad (2.37)$$

Так как материал решетки однородный, то $Z_S$ является константой, и ее можно вынести за знак интеграла. Таким образом, $\mathbf{Z}^{IBC}$ представляет собой произведения скалярной величины $Z_S$ на матрицу Грама $\mathbf{G}$ всех *RWG* функций:

$$\mathbf{Z}^{IBC} = Z_S \mathbf{G} . \qquad (2.38)$$

Так как геометрия решетки известна, и известно расположение в пространстве и геометрия для всех RWG базисных функций, то расчет матрицы Грама не составляет труда. То, что она является очень большой, размером $N_{RWG} \times N_{RWG}$ (в рассчитанной модели $N_{RWG} = 110905$), компенсируется тем, что она является разреженной, и пакет MATLAB в состоянии ее обработать на ПК среднего класса.

Таким образом, чтобы рассчитать $P_{ant}^{dissipated}$, достаточно найти поверхностное сопротивление металла решетки $Z_S$ и коэффициенты к RWG функциям $\mathbf{J}$ при заданном возбуждении элементов решетки.

Чтобы найти $\mathbf{J}$, учтем, что расчет ДН элементов производится по известному токовому распределению. Так, в результате электромагнитного моделирования в CAESAR, имеется *M* наборов коэффициентов к RWG функциям для случаев последовательного возбуждения каждого элемента решетки источником напряжения с амплитудой $V = 1$ В (рис. 2.21). Обозначим эти наборы коэффициентов $\mathbf{J}_{vec,1}, \ldots, \mathbf{J}_{vec,M}$ и объединим их в единую матрицу $\mathbf{J}_{vec}$ размером $N_{RWG} \times M$, как показано на рисунке. Другими словами, матрица $\mathbf{J}_{vec}$ состоит из $M$ векторов $\mathbf{J}_{vec,1}$, ..., $\mathbf{J}_{vec,M}$.

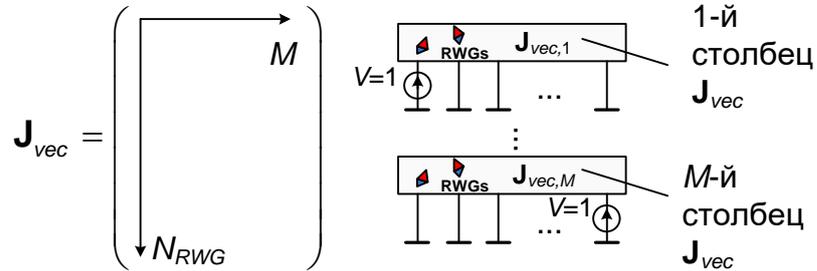

Рис. 2.21 Представление распределения токов в ЭМ модели

Чтобы получить распределение токов в металле структуры, соответствующее возбуждению решетки с помощью МПФ, необходимо возбудить одновременно все элементы решетки напряжениями, которые соответствует данному возбуждению, т.е. с учетом весовых коэффициентов (рис. 2.22).

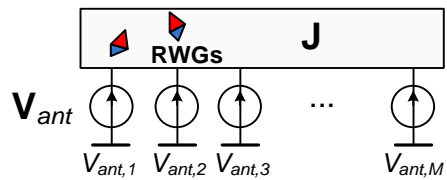

Рис. 2.22 К определению распределения токов в решетке при заданном возбуждении всех ее элементов с помощью МПФ

Напряжения на портах металлической структуры как функции падающей на $M$-полюсник (решетку) волны $\mathbf{a}_{ant}$, (была численно рассчитана ранее, при расчете КПД МПФ), и матрицы рассеяния этого $M$-полюсника имеют вид [79]:

$$\mathbf{V}_{ant} = \sqrt{Z_0} \left( \mathbf{I} + \mathbf{S}_{ant} \right) \mathbf{a}_{ant}. \tag{2.39}$$

Распределение токов в решетке $\mathbf{J}$, соответствующее возбуждению портов источниками напряжения $V_{ant,1}$, ..., $V_{ant,M}$, рассчитывается как суперпозиция распределений $\mathbf{J}_{vec,1}$, ..., $\mathbf{J}_{vec,M}$ с коэффициентами $V_{ant,1}$ ... $V_{ant,M}$. В матричной форме выражение для $\mathbf{J}$ записывается следующим образом:

$$\mathbf{J} = \mathbf{J}_{vec}\,\mathbf{V}_{ant}\,. \tag{2.40}$$

Поверхностное сопротивление $Z_S$ рассчитывается по формуле [86]:

$$Z_S = \frac{1-\mathrm{j}}{2\sigma\delta\tan\!\left(\left(1-\mathrm{j}\right)\dfrac{d}{2\delta}\right)}\,, \tag{2.41}$$

где $\delta = \sqrt{\dfrac{2}{\omega\mu_0\sigma}}$ — глубина скин-слоя; $\sigma$ — проводимость материала решетки; $d$ — толщина листа элемента Вивальди.

Рассеянная в металле мощность $P_{ant}^{dissipated}$ вычисляется в соответствии с (2.36) по рассчитанным $\mathbf{J}$ и $\mathbf{Z}^{IBC}$.

Выражение для принятой решеткой мощности $P_{ant}^{accepted}$ уже использовалось при вычислении КПД МПФ (см. (2.34)). Таким образом, КПД металлической структуры $\eta_{rad,met}$ и общий КПД антенной решетки определяются в соответствии с выражениями (2.30) и (2.32).

**Расчет шумовой температуры системы.**

Для расчета чувствительности по формуле (2.18) кроме эффективной площади зеркала необходимо также знать шумовую температуру системы, которая зависит от коэффициентов возбуждения, то есть должна рассчитываться для каждого луча отдельно.

В состав шумовой температуры системы входит большое количество компонент:

$$T_{sys,n}^{x} = T_{sp,n}^{x} + T_{sky} + k_T \frac{T_{LNA,n}^{x} + T_{sec} + \left(1 - \eta_{rad,n}^{x}\right) T_{amb}}{\eta_{rad,n}^{x}}, \qquad (2.42)$$

где $T_{sys,n}^{x}$ — шумовая температура системы при формировании $n$-го луча; $T_{sp,n}^{x}$ — шумовая температура за счет приема шумов земли из-за эффекта переливания энергии облучателя за края зеркала; $k_T$ — коэффициент температурной коррекции (см. далее); $T_{LNA,n}^{x}$ — вклад в шумовую температуру системы за счет внутренних шумов МШУ, а также за счет рассогласования МШУ с выходами антенных элементов; $T_{sec}$ — приведенная ко входу шумовая температура всех следующих за МШУ цепей (*secondary stage temperature*); $T_{amb}$ — температура окружающей среды; $\eta_{rad,n}^{x}$ — КПД антенной решетки (radiation efficiency).

Остановимся на каждой компоненте более подробно.

*Шумовая температура за счет переливания энергии облучателя за края зеркала $T_{sp}$.*

Данную шумовую компоненту можно вычислить, используя общие первичные ДН (первичные ДН лучей) $E_{prim,tot,n}^{x,y}(\theta, \varphi)$. Для этого температура земли $T_{gnd}$ умножается на отношение мощности ДН, содержащейся в диапазоне углов $\theta_{sub} \le \theta \le \pi/2$ (см. рис. 2.11), к общей мощности ДН:

$$T_{sp,n}^{x} = \frac{T_{gnd}}{4\pi} \int_{0}^{2\pi} \int_{\theta_{sub}}^{\pi/2} \left[ \left| E_{prim,tot,n}^{x,CO}(\theta, \varphi) \right|^2 + \left| E_{prim,tot,n}^{x,XP}(\theta, \varphi) \right|^2 \right] \cdot \sin\theta \, d\theta d\varphi. \quad (2.43)$$

*Коэффициент температурной коррекции* $k_T$ необходим для коррекции шумовых температур, измеренных или рассчитанных при нормальных условиях ($T_{amb} = 290$ К), и рассчитывается как отношение $k_T = T_{amb} / 290$.

*Приведенная ко входу шумовая температура всех следующих за МШУ цепей* $T_{sec}$ была измерена в институте ASTRON на одной из антенн Вестерборгского радиотелескопа: $T_{sec} \approx 5$ К.

*Эквивалентная шумовая температура за счет всех МШУ* $T_{LNA,n}^x$ рассчитывается по методике, описанной в [8], которая заключается в приведении системы «решетка+МШУ+формирователь» к эквивалентной одноэлементной антенне с МШУ и нагрузкой.

Приведем выражения для расчета $T_{LNA,n}^x$.

Температура $T_{LNA,n}^x$ при формировании *n*-го луча рассчитывается по следующей формуле:

$$T_{LNA,n}^x = \frac{1}{\eta_{mis}^x \left( \mathbf{w}_{opt,n}^x \right)^H \mathbf{w}_{opt,n}^x} \sum_{m=1}^{M} T_{LNA,n,m}^x \left| w_{opt,n,m}^x \right|^2 \left( 1 - \left| \Gamma_{act,n,m}^x \right|^2 \right), \qquad (2.44)$$

где $\eta_{mis}^x$ — коэффициент рассогласования МШУ с элементами решетки; $T_{LNA,n,m}^x$ — шумовая температура *m*-го МШУ; $w_{opt,n,m}^x$ — *m*-ый элемент вектора оптимальных коэффициентов возбуждения для *n*-го луча; $\Gamma_{act,n,m}^x$ — активный коэффициент отражения от входа *m*-го элемента решетки в режиме передачи.

Коэффициент рассогласования МШУ с элементами решетки $\eta_{mis}^x$ рассчитывается через матрицу рассеяния решетки $\mathbf{S_{arr}}$ и весовые коэффициенты:

$$\eta_{mis}^{x} = 1 - \frac{P_{refl}}{P_{inc}} = 1 - \frac{\left(\mathbf{w}_{opt,n}^{x}\right)^{H}\left(\mathbf{I} - \mathbf{S}_{arr}^{H}\mathbf{S}_{arr}\right)\mathbf{w}_{opt,n}^{x}}{\left(\mathbf{w}_{opt,n}^{x}\right)^{H}\mathbf{w}_{opt,n}^{x}}, \qquad (2.45)$$

где $P_{refl}$ — отраженная от портов антенных элементов решетки мощность в режиме передачи; $P_{inc}$ — падающая мощность на элементы решетки.

Шумовая температура $m$-го МШУ $T_{LNA,n,m}^{x}$ рассчитывается по известной формуле для активного шумящего двухполюсника (2.6), которая в приложении к описываемой модели системы принимает вид [87, 88]:

$$T_{LNA,n,m}^{x} = T_{min} + \frac{4R_nT_0}{Z_0}\frac{\left|\Gamma_{act,n,m}^{x} - \Gamma_{opt}\right|^2}{\left|1 + \Gamma_{opt}\right|^2\left(1 - \left|\Gamma_{act,n,m}^{x}\right|^2\right)}, \qquad (2.46)$$

где $T_{min}$ — минимальная шумовая температура для данного типа МШУ; $T_0 = 290$ К.

Под активным коэффициентом отражения от входа $m$-го элемента решетки $\Gamma_{act,m}$ в режиме передачи понимается отношение отраженной волны $b_m$ к падающей волне $a_m$ на данном элементе с учетом соседних элементов. То есть волна $b_m$ представляет собой сумму собственной отраженной волны $m$-го элемента и волн, наведенных другими элементами решетки. Таким образом, для $n$-го луча нашей системы $\Gamma_{act,n,m}^{x}$ определяется как

$$\Gamma_{act,n,m}^{x} = \frac{1}{w_{opt,n,m}^{x}}\sum_{k=1}^{M}S_{arr,m,k}w_{opt,n,k}^{x}. \qquad (2.47)$$

Произведя расчеты по формулам (2.43) — (2.47), можно определить общую шумовую температуру системы $T_{sys,n}^{x}$ по формуле (2.42).

**Расчет эффективности облучения зеркала**

При облучении неидеального зеркала любым неидеальным облучателем эффективная площадь зеркала будет меньше физической площади апертуры за счет нескольких эффектов. КИП $\eta_{ap}$ в приведенном методе рассчитывается как произведение

$$\eta_{ap} = \eta_{tap}\eta_{ph}\eta_{sp}\eta_{block}\eta_{transp}, \qquad (2.48)$$

где $\eta_{tap}$ — коэффициент, учитывающий неравномерность амплитудного распределения поля в апертуре; $\eta_{ph}$ — коэффициент, учитывающий отклонение фазового распределения поля в апертуре от заданного (меняется при сканировании); $\eta_{sp}$ — коэффициент перехвата (используется обычно только для приемных антенн); $\eta_{block}$ — коэффициент, учитывающий блокировку падающего поля облучателем и опорами; $\eta_{transp}$ — коэффициент, учитывающий неидеальность формы зеркала, а также его прозрачность (так как большие зеркала часто изготавливаются из металлической сетки, и на высоких частотах могут стать частично прозрачными).

В данной работе $\eta_{transp}$ был исключен ввиду его близости к единице в рабочем диапазоне частот, а $\eta_{block}$ на данный момент находится на стадии исследования, и был принят равным 0,98.

Коэффициент $\eta_{sp}$ определяется как отношение мощности облучателя, перехваченной зеркалом, к общей мощности облучателя, которая при анализе была нормирована к мощности изотропного облучателя, создающего напряженность поля равную 1 В/м на сфере радиусом 1 м (общая мощность такого излучателя равна $4\pi$):

$$\eta_{sp,n}^{x} = \frac{1}{4\pi} \int\limits_{0}^{2\pi} \int\limits_{0}^{\theta_{sub}} \left[ \left| E_{prim,tot,n}^{x,CO}(\theta,\varphi) \right|^2 + \left| E_{prim,tot,n}^{x,XP}(\theta,\varphi) \right|^2 \right] \cdot \sin\theta \, d\theta d\varphi . \quad (2.49)$$

Выведем формулу для расчета коэффициента, учитывающего неравномерность амплитудного распределения в апертуре зеркала $\eta_{tap}$.

Из [89, стр.326] известно, что

$$\eta_{tap}\eta_{ph} = \frac{1}{A} \frac{\left| \int\limits_{A} E(x,y) \, dA \right|^2}{\int\limits_{A} \left| E(x,y) \right|^2 dA}, \quad (2.50)$$

где $A$ — площадь апертуры зеркала; $E(x,y)$ — распределение поля в апертуре.

Чтобы исключить из (2.50) влияние распределения фазы, возьмем $E$ в числителе по модулю. Тогда, аналогично показанному в [89], для случая круглых зеркал (2.50) преобразуется к виду

$$\eta_{tap} = \frac{1}{\pi R_0^2} \frac{\left( \int\limits_{0}^{2\pi} \int\limits_{0}^{R_0} \left| E(R,\varphi) \right| R \, dR d\varphi \right)^2}{\int\limits_{0}^{2\pi} \int\limits_{0}^{R_0} \left| E(R,\varphi) \right|^2 R \, dR d\varphi}, \quad (2.51)$$

где $R_0$ — радиус апертуры зеркала; $(R,\varphi)$ — координаты точки в апертуре в полярной системе координат.

Чтобы перейти от распределения поля в апертуре $E(R,\varphi)$ к ДН облучателя $E_f(\theta,\varphi)$, рассмотрим рис. 2.23.

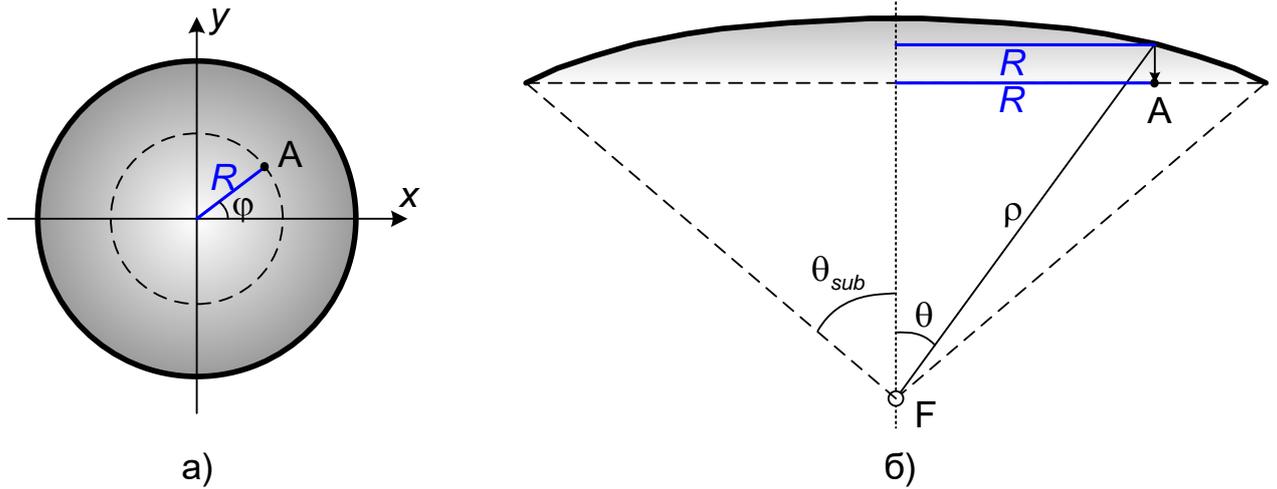

Рис. 2.23 — К определению эффективности амплитудного распределения

Координата $\varphi$ для точки А в апертуре зеркала и для ДН облучателя является одинаковой, поэтому остается найти только зависимость, связывающую координату $R$ точки в апертуре и угол $\theta$.

Профиль параболического зеркала в полярных координатах задается уравнением

$$\rho(\theta) = \frac{2F}{1+\cos(\theta)}, \qquad (2.52)$$

где $F$ — фокусное расстояние зеркала.

По рис. 2.23,б видно, что

$$R(\theta) = \rho\sin(\theta) = \frac{2F\sin(\theta)}{1+\cos(\theta)} = 2F\tan\left(\frac{\theta}{2}\right). \qquad (2.53)$$

Тогда дифференциал d$R$ в уравнении (2.51) будет равен

$$\mathrm{d}R = 2F\,\frac{1}{\cos^2\left(\dfrac{\theta}{2}\right)}\cdot\frac{1}{2}\,\mathrm{d}\theta = 2F\,\frac{1}{1+\cos(\theta)}\,\mathrm{d}\theta. \qquad (2.54)$$

Подставив (2.53) и (2.54) в (2.51), и выполнив необходимые математические преобразования, получим формулу для оценки эффективности амплитудного распределения:

$$\eta_{tap} = \frac{1}{\pi \tan^2 \dfrac{\theta_{sub}}{2}} \frac{\left( \displaystyle\int_0^{2\pi}\int_0^{\theta_{sub}} \left| E_f\left(\theta,\varphi\right) \right| \dfrac{\sin\theta}{\left(1+\cos\theta\right)^2} \, d\theta d\varphi \right)^2}{\displaystyle\int_0^{2\pi}\int_0^{\theta_{sub}} \left| E_f\left(\theta,\varphi\right) \right|^2 \dfrac{\sin\theta}{\left(1+\cos\theta\right)^2} \, d\theta d\varphi}. \qquad (2.55)$$

В выводе формулы считалось, что напряженность электрического поля в точке апертуры А равна напряженности поля, создаваемой облучателем в направлении $\left(\theta,\varphi\right)$, то есть $E\left(R,\varphi\right) = E_f\left(\theta,\varphi\right)$.

Следует отметить, что выражение (2.55) является оценочным, так как оно не учитывает интерференцию поля облучателя и отраженного поля, а также искажение поля в апертуре за счет эффектов, возникающих на краях зеркала.

Коэффициент $\eta_{ph}$, учитывающий отклонение фазового распределения поля в апертуре от заданного, ввиду сложности прямого расчета, рассчитывался из известного КИП $\eta_{ap}$ следующим образом:

$$\eta_{ph} = \frac{\eta_{ap}}{\eta_{tap}\eta_{sp}}. \qquad (2.56)$$

## 2.4. Выводы по разделу

2.4.1. Произведено ЭМ моделирование металлической структуры решетки методом ХБФ, в результате которого были рассчитаны $V$-ДН антенных элементов решетки $g_m(\theta,\varphi)$ с учетом влияния остальных элементов. Рассчитана матрица сопротивлений $\mathbf{Z_{ant}}$ антенной решетки без микрополоскового фидера (МПФ), то есть сопротивление между портами каждого из металлических элементов решетки. Написан соответствующий код в программе MATLAB.

2.4.2. Разработана модифицированная модель микрополоскового фидера, примененного для питания антенн типа Вивальди, а также выполнена оптимизация одного из его параметров (коэффициента трансформации) прямым методом. В качестве целевой функции была выбрана разность между измеренным и промоделированным коэффициентом отражения $S_{11}$. Найденная оптимальная величина коэффициента трансформации для рассмотренной системы равна $p=0{,}97$. Также в результате проведенной оптимизации была найдена эффективная ширина щели антенного элемента Вивальди, которая оказалась меньше физического размера за счет влияния диэлектрической подложки МПФ, и равна 0,4 мм, в то время как физическая ширина равна 0,5 мм. Полученное значение ширины щели использовалась для ЭМ моделирования решетки, а коэффициент трансформации использовался как постоянный параметр модели МПФ.

2.4.3. Разработан алгоритм моделирования антенной системы с фокальной решеткой, выбрана конфигурация лучей, разработана методика расчета оптимальных по нескольким критериям (максимальная принятая мощность, максимальная чувствительность, максимальная чувствительность с ограничениями по направлениям) весовых коэффициентов для антенной решетки и даны выражения для расчетов важнейших ее характеристик. В частности, для каждого из формируемых лучей (или, для каждого набора весовых коэффициентов) получены выражения для расчета:

— комплексных ДН лучей системы с учетом МПФ;

— выходного сигнала системы для весовых коэффициентов, рассчитанных для приема *co-* и *xp*-компонент падающего поля (это необходимо для анализа поляризационных свойств системы);

— зависимости чувствительности системы от направления;

— коэффициентов эффективности системы ($\eta_{ap}$, КПД, $\eta_{sp}$, $\eta_{tap}$, $\eta_{ph}$);

— шумовой температура системы $T_{sys}$, а также ее составляющих: $T_{sp}$, $T_{LNA}$, $T_{sec}$;

Численные результаты моделирования представлены в разделах 3 и 4.

# РАЗДЕЛ 3

## ЧИСЛЕННОЕ МОДЕЛИРОВАНИЕ И ЭКСПЕРИМЕНТАЛЬНАЯ ПРОВЕРКА ПАРАМЕТРОВ АНТЕННОЙ РЕШЕТКИ

В данном разделе проводится численное моделирование отдельных частей системы и приводятся промежуточные результаты, производится их экспериментальная проверка с целью подтверждения адекватности разработанной модели.

В ходе работы было рассмотрено несколько вариантов антенных решеток, построенных в институте *ASTRON*. Первый вариант — это решетка *VALARRAY* [99], которая является предшественником решетки для радиотелескопа нового поколения *SKADS/EMBRACE* [90, 91] (апертурная решетка). Второй и третий варианты — два прототипа решетки *APERTIF* (фокальная решетка). Все три варианта решеток состоят из антенных элементов Вивальди, различаются конструктивным исполнением и рабочим диапазоном частот.

Промежуточными результатами, приведенными в данном разделе, являются рассчитанные значения

- коэффициента трансформации перехода «микрополосок-щель» в модели МПФ;

- матрицы рассеяния на портах решетки, показывающая коэффициенты отражения от портов элементов, а также степень взаимного влияния элементов друг на друга;

- ДН отдельных элементов антенной решетки.

Также в данном разделе проведено сравнение двух конструкций антенного элемента Вивальди, разработаны меры по улучшению механической прочности антенной решетки;

### 3.1. Исследование влияния коэффициента трансформации в модели микрополоскового фидера на матрицу рассеяния решеток

Прежде чем переходить к моделированию большой решетки, целесообразно проверить адекватность модели только решетки с МПФ, а также выяснить, как влияет коэффициент трансформации перехода «микрополосок-щель» в модели МПФ на характеристики решетки в целом. В качестве контрольной характеристики была выбрана матрица рассеяния решетки **S**, так как ее элементы являются наиболее легко измеряемыми параметрами.

Были промоделированы, построены и измерены три конструкции антенных решеток, схематическое расположение элементов которых показано на рис. 3.1.

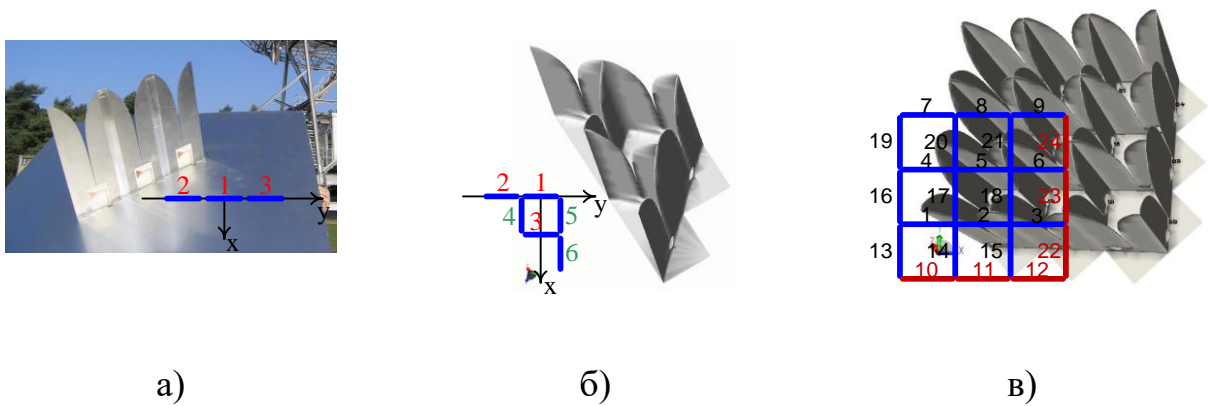

а)                              б)                              в)

Рис. 3.1 Схематическое расположение антенных элементов Вивальди в исследуемых решетках:

а) 3-элементная решетка из элементов *VALARRAY*; б) 6-элементная решетка из элементов первого прототипа *APERTIF*; в) 24-элементная решетка из элементов второго прототипа *APERTIF*

Матрица рассеяния для всех структур рассчитывалась путем микроволнового моделирования для схемы рис. 2.6 и сравнивалась с измеренной для различных значений коэффициента трансформации *p*.

Для учета того факта, что положение плоскости, от которой анализатор цепей начинает отсчет фазы при измерении, неизвестно, в модель (см. рис. 2.6) была введена дополнительная неизвестная величина — длина разъема $L_{con}$, которая может быть представлена как длина продолжения микрополосковой линии.

По известной зависимости каждого элемента матрицы рассеяния $S_{mk}$ от частоты вычислялся коэффициент корреляции $R$ между его модулем для модели и эксперимента, а также интегральная относительная ошибка $Err_{PH}$ в их аргументе [92]:

$$R\left(p, L_{con}\right) = \frac{\text{cov}\left(\left|S_{mk}^{Model}\left(p, L_{con}, f\right)\right|, \left|S_{mk}^{Meas}\left(f\right)\right|\right)}{\sqrt{\text{cov}\left(\left|S_{mk}^{Model}\left(p, L_{con}, f\right)\right|, \left|S_{mk}^{Model}\left(p, L_{con}, f\right)\right|\right) \cdot \text{cov}\left(\left|S_{mk}^{Meas}\left(f\right)\right|, \left|S_{mk}^{Meas}\left(f\right)\right|\right)}}, \quad (3.1)$$

$$Err_{PH}\left(p, L_{con}\right) = \sum_{f=F\min}^{F\max}\left|\arg\left(S_{mk}^{Model}\left(p, L_{con}, f\right)\right) - \arg\left(S_{mk}^{Meas.}\left(f\right)\right)\right|, \quad (3.2)$$

где $\text{cov}\left(\left|S_{mk}^{Model}\right|, \left|S_{mk}^{Meas}\right|\right)$ — ковариация между модулями моделированных $S$-параметров $S_{mk}^{Model}\left(p, L_{con}, f\right)$ и измеренных функций $S_{mk}^{Meas}\left(f\right)$.

Для оптимизации коэффициента трансформации $p$ по модулю матрицы рассеяния использовался коэффициент корреляции, так как он сильно зависит от *формы* кривых, а $|S_{mk}|$, как правило, имеет ярко выраженный резонансный характер. Так как фаза меняется по закону, близкому к линейному, для оптимизации по $\arg(S_{mk})$ был использован "разностный" критерий (см. (3.2)).

На рис. 3.2 показаны результаты расчетов по приведенным выше формулам (3.1) и (3.2) для коэффициента отражения $S_{11}$ от первого элемента трехэлементной решетки, показанной на рис. 3.1а.

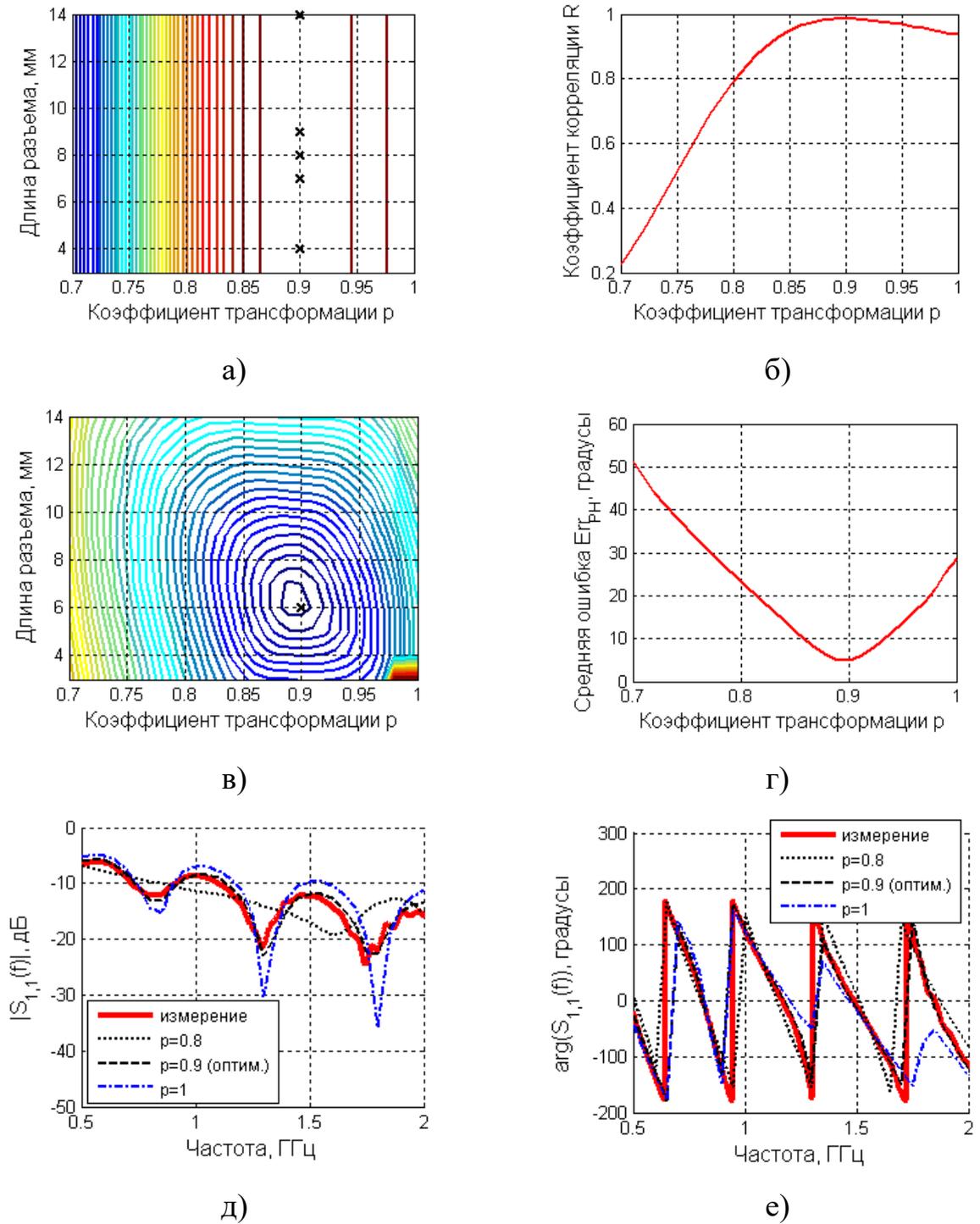

Рис. 3.2 Коэффициент корреляции между значениями, полученными в результате моделирования, и измеренными значениями $|S_{1,1}|$:

а) сечение коэффициента корреляции при $L_{con}$=6 мм ; б) ошибка по фазе $S_{1,1}$; в) сечение ошибки по фазе при $L_{con}$ = 6 мм; г), измеренные (пунктирные линии) и полученные в результате моделирования (сплошные линии) для разных значений коэффициента трансформации $p$; д) модуль $S_{1,1}$; е) аргумент $S_{1,1}$

Из рис.3.2а видно, что модуль коэффициента отражения не зависит от длины разъема, как и следовало ожидать, а зависимость есть только для аргумента коэффициента отражения. Максимум коэффициента корреляции наблюдается при коэффициенте трансформации $p = 0{,}9$. Из рис. 3.2в следует, что минимум ошибки в фазах моделированного и измеренного $S_{1,1}$ достигается также при $p = 0{,}9$, что подтверждает адекватность модели. Однако, использованная модель обеспечивает хорошее совпадение только до частот порядка (5…10) ГГц. На более высоких частотах начинает сказываться излучение микрополосковой структуры и дополнительные потери в проводнике и диэлектрике.

На рис. 3.2д, е показаны измеренные и полученные в результате моделирования для нескольких значений $p$ зависимости коэффициента отражения $S_{1,1}$ от частоты. Из рис. 3.2д, е видно, что даже при незначительном (около 10%) отклонении в модели коэффициента трансформации $p$ от его реальной (оптимальной) величины наблюдается значительное изменение коэффициента отражения: происходит не только увеличение ошибок в аргументе и изменение уровня модуля коэффициента отражения, но и смещение и сглаживание максимумов.

Для того, чтобы экспериментально проверить разработанную модель, было изготовлено несколько прототипов решеток с конфигурацией, показанной на рис. 3.1б, в.

На рис. 3.3 показаны результаты моделированных $S$-параметров при различных значениях коэффициента трансформации $p$ для структуры 6-элементной решетки *APERTIF* (см. рис. 3.1б) и сравнение их с измеренными.

Из рис. 3.3б и рис. 3.3в видно, что оптимальный коэффициент трансформации, сообветствующий минимуму ошибки для двух случаев составляет 0,84 и 0,86.

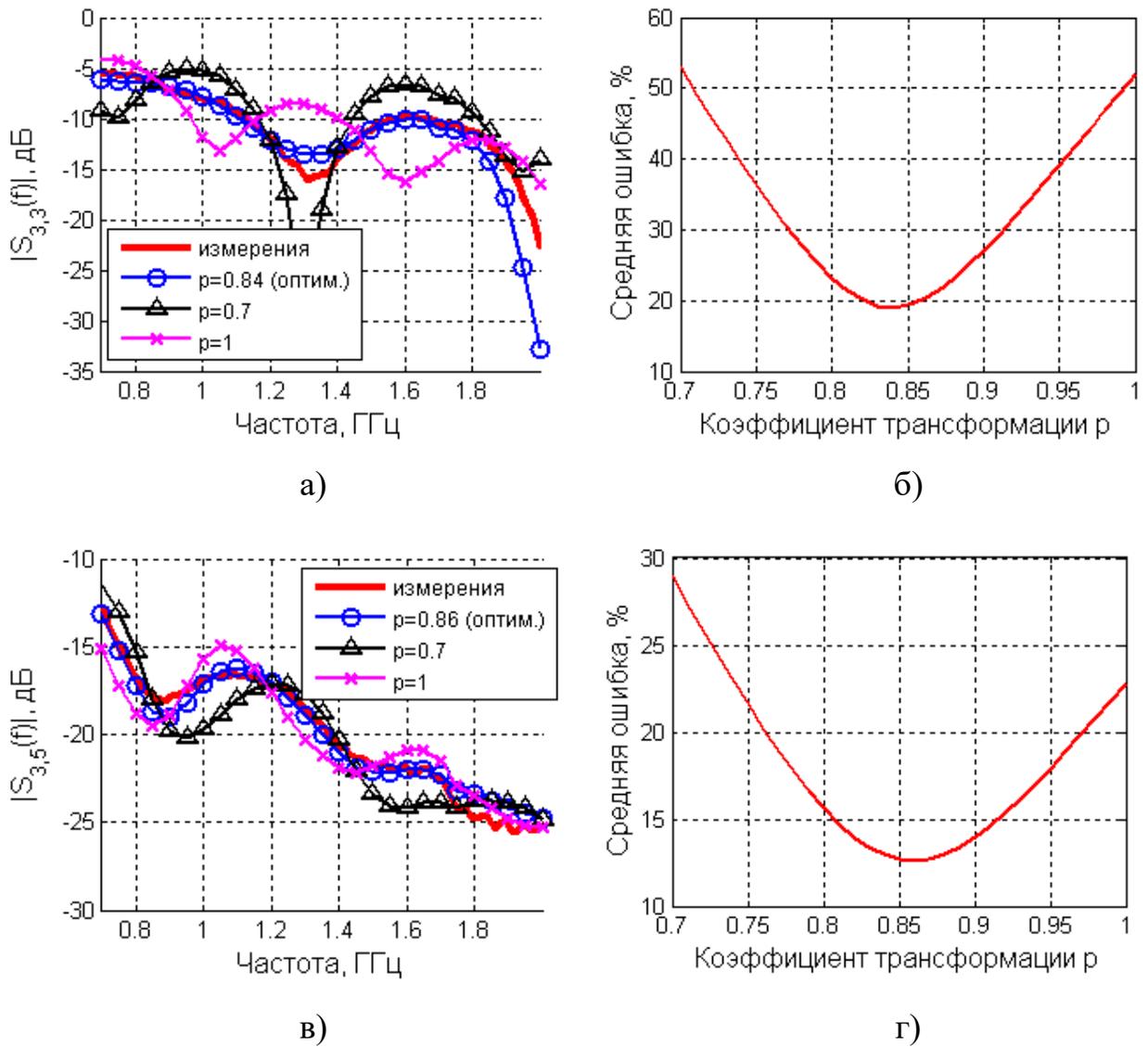

Рис. 3.3 Моделированные и измеренные *S*-параметры 6-элементной
антенной решетки:

а) коэффициент отражения $S_{3,3}$ от 3-го элемента при разных значениях *p*,

б) интегральная функция ошибки между измеренными и расчетными

значениями коэффициента отражения $S_{3,3}$ как зависимость от *p*, в), г) то же

для коэффициента связи $S_{3,5}$

В табл. 3.1 показаны оптимальные значения $p$, то есть значения $p$, для которых ошибка между измеренным и рассчитанным $S_{m,k}$ минимальна.

Таблица 3.1

Коэффициенты трансформации, оптимальные для $S_{m,k}$

| m \ k | 1 | 2 | 3 | 4 | 5 | 6 |
|-------|------|------|------|------|------|------|
| 1 | 0,83 | 0,85 | 0,87 | 0,85 | 0,85 | 0,84 |
| 2 | - | 0,83 | 0,85 | 0,83 | 0,85 | 0,84 |
| 3 | - | - | 0,84 | 0,87 | 0,86 | 0,83 |
| 4 | - | - | - | 0,85 | 0,87 | 0,83 |
| 5 | - | - | - | - | 0,85 | 0,87 |
| 6 | - | - | - | - | - | 0,86 |

Из таблицы 3.1 видно, что коэффициент трансформации принимает значения от 0,83 до 0,87. Как будет показано далее, непостоянство коэффициента трансформации обусловлено в основном механической нестабильностью между МПФ и элементом Вивальди для первого прототипа APERTIF. Для симуляции большой решетки было выбрано значение $p = 0,85$.

На рис. 3.4 показаны некоторые результаты нахождения $p$ для 24-элементной решетки второго прототипа APERTIF (см. рис. 3.1в).

Анализируя результаты, показанные на рис. 3.3 и рис. 3.4, можно сделать вывод, что значение коэффициента трансформации $p$ значительно влияет лишь на коэффициенты отражения $S_{5,5}$, в то время как коэффициенты передачи между элементами при изменении $p$ меняются незначительно. Поэтому значения $p$ вычислялись через коэффициенты отражения $S_{5,5}$.

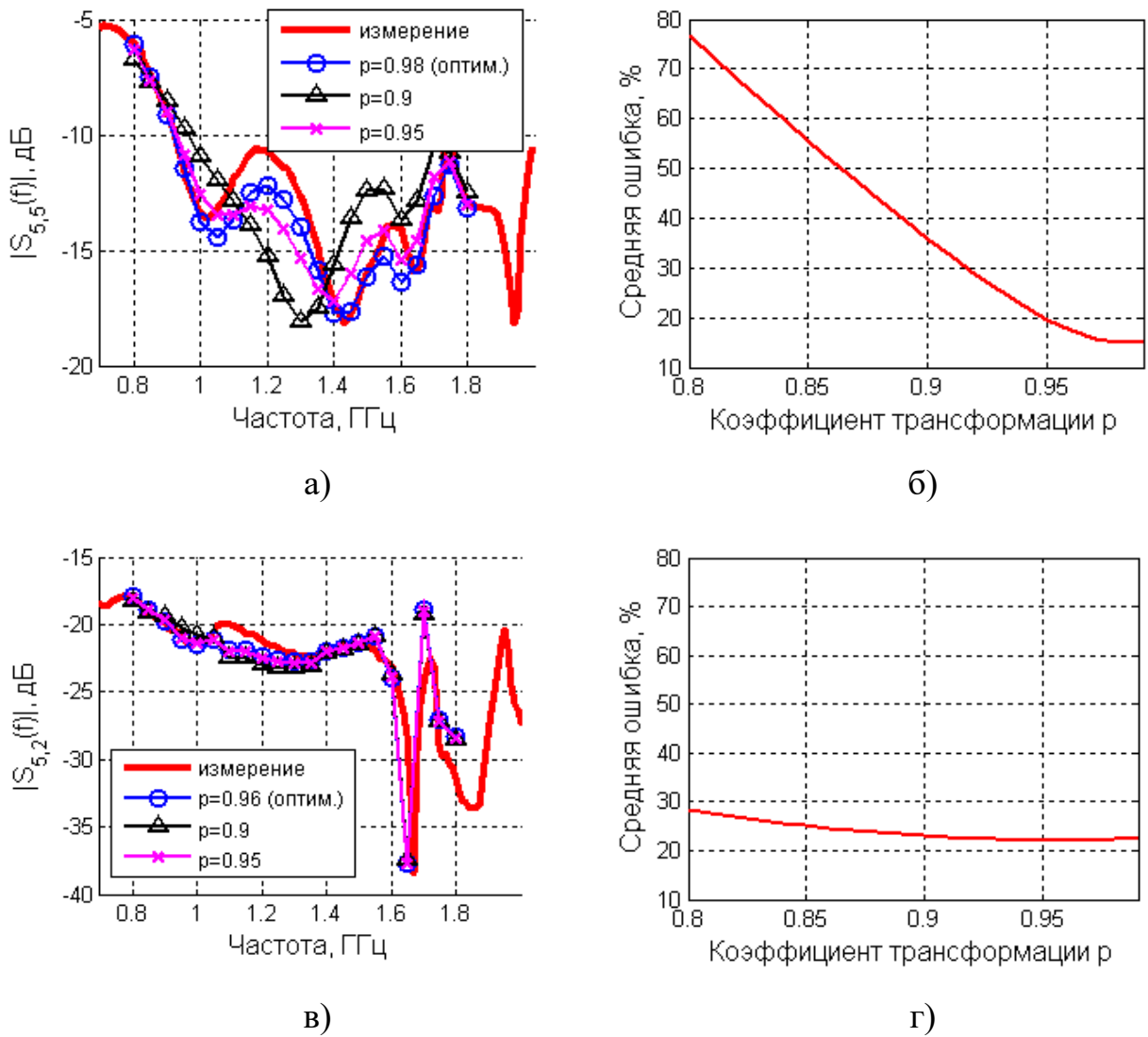

Рис. 3.4 — $S$-параметры 24-элементной решетки:

а) зависимость коэффициента отражения $S_{5,5}$ от частоты (при отражении от 5-го элемента) при разных значениях $p$; б) зависимость интегральной функции ошибки между измеренным и расчетным значениями коэффициента отражения $S_{5,5}$ от $p$; в) зависимость коэффициента связи $S_{5,2}$ от частоты (при отражении от 5-го элемента) при разных значениях $p$; г) зависимость интегральной функции ошибки между измеренным и расчетным значениями коэффициента связи $S_{5,2}$ от $p$

В таблице 3.2 показаны оптимальные значения $p$, вычисленные по коэффициентам отражения $S_{m,m}$.

<div align="right">Таблица 3.2</div>

<div align="center">Коэффициенты трансформации, оптимальные для $S_{m,m}$</div>

| $m$ | 1 | 2 | 3 | 4 | 5 | 6 | 7 | 8 | 9 | 10 | 11 | 12 |
|---|---|---|---|---|---|---|---|---|---|---|---|---|
| $p$ | 0,97 | 0,99 | 0,97 | 0,99 | 0,98 | 0,99 | 0,96 | 0,99 | 0,97 | 0,96 | 0,97 | 0,98 |
| $m$ | 13 | 14 | 15 | 16 | 17 | 18 | 19 | 20 | 21 | 22 | 23 | 24 |
| $p$ | 0,99 | 0,97 | 0,99 | 0,99 | 0,98 | 0,98 | 0,98 | 0,98 | 0,98 | 0,96 | 0,99 | 0,97 |

Из таблицы 3.2 видно, что $p$ для второго прототипа решетки меняется в значительно меньших пределах, чем для первого прототипа, то есть робастность второго прототипа выше.

Увеличение значения $p$ по сравнению с первым прототипом обусловлено различием у прототипов как самих элементов Вивальди, так и МПФ (см. подраздел 3.2).

**3.2. Улучшение параметров отдельного элемента решетки с микрополосковым фидером. Сравнение конструкций первого и второго прототипов APERTIF.**

Первая конструкция решетки оказалась неудачной по следующим причинам. Во-первых, измеренная шумовая температура системы с такой решеткой оказалась довольно высокой, примерно (100…130) К, хотя по требованиям к системе *APERTIF* (см. подраздел 1.5.1) она не должна превышать 55 К. Во-вторых, наблюдалась нестабильность измерений $S$-параметров из-за особенностей способа изготовления элементов решетки и крепления к ним МПФ. Поэтому была изготовлена вторая конструкция, в которой был сделан ряд изменений [9, 93], описанных ниже.

### 3.2.1. Первый прототип антенной решетки

Структура решетки первого типа показана на рис. 3.5.

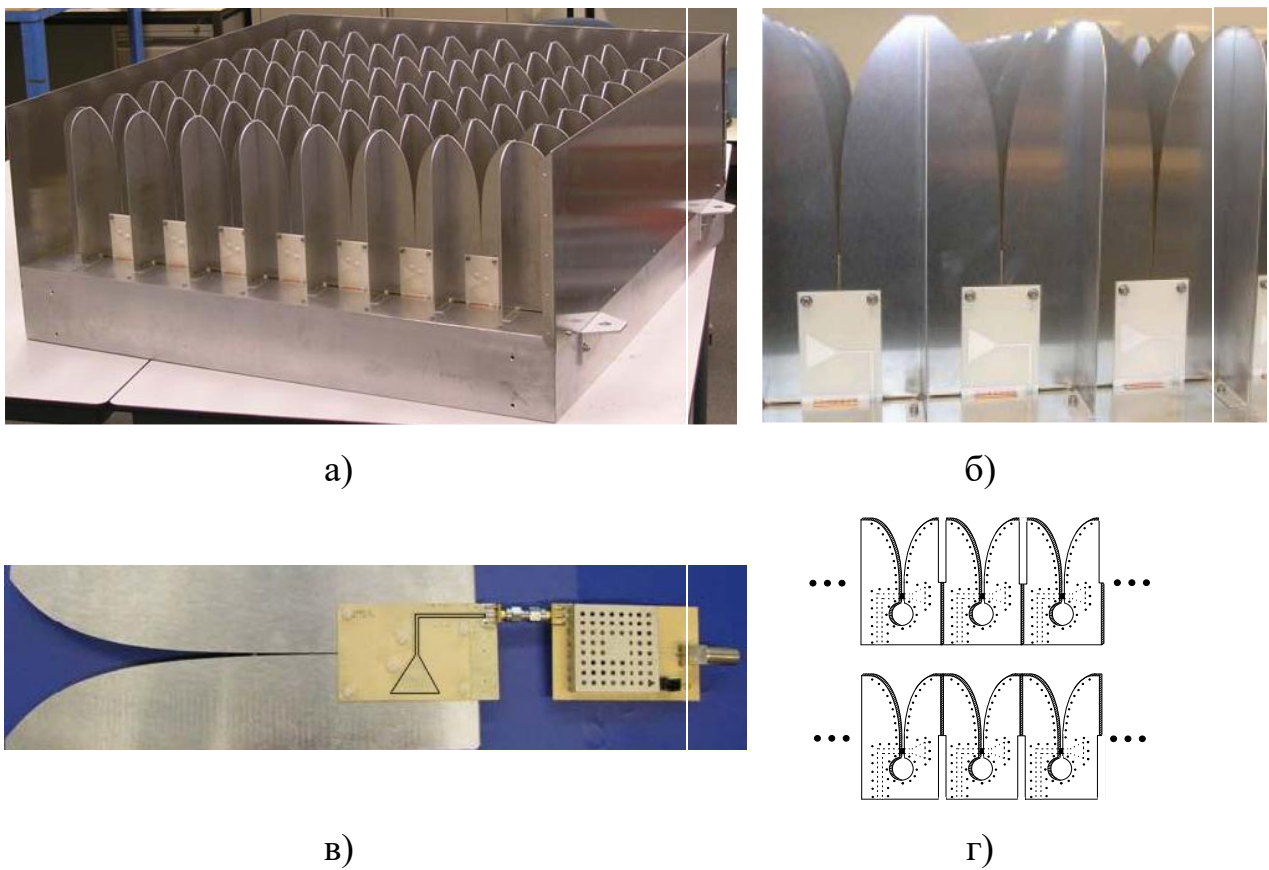

а)  б)

в)  г)

Рис. 3.5 — Первый прототип антенной решетки: а) общий вид собранной решетки; б) расположение МПФ и подключение к нему МШУ (в составе решетки); в) расположение МПФ и подключение к нему МШУ (отдельный элемент); г) пластины из элементов Вивальди для приема ортогональных компонент падающего поля (в решетке пластины перпендикулярны друг другу)

Решетка выполнена из алюминиевых пластин, в которых лазером вырезаны элементы Вивальди. Причем, каждая пластина состоит из 7 или 8 смежных элементов (см. рис. 3.5,г). В пластинах вырезаны щели между элементами Вивальди таким образом, чтобы, собирая пластины «крест-накрест», можно было собрать двухполяризационную решетку, показанную рис. 3.5,а. МПФ крепился к элементу Вивальди при помощи пластиковых заклепок. МШУ подключался к МПФ при помощи SMA разъема (см. рис. 3.5,в).

### 3.2.2 Улучшение шумовых характеристик

Как было отмечено в подразделе 3.2, измеренная шумовая температура системы с решеткой первого типа оказалась довольно высокой: (100…130) К. Для того, чтобы выяснить, что приводит к столь значительной шумовой температуре системы, были проведены измерения шумовых характеристик используемого МШУ, а также рассчитаны и измерены потери, возникающие в металле решетки и МПФ для диапазона частот (т.е., фактически, определялся КПД). Все измерения были выполнены в институте *ASTRON*.

Оказалось, что значительную часть шумов (около 50%) является следствием большого значения параметра $T_{min}$ МШУ (это минимальная шумовая температура, которую вносит МШУ после его полного *шумового* согласования). В институте *ASTRON* был разработан новый МШУ, у которого $T_{min}$ вдвое меньше, и составлял примерно 30 К при нормальных условиях.

Соединитель МПФ и МШУ (см. рис. 3.5в) увеличивает шумовую температуру системы на 3 К. Чтобы исключить эту составляющую, МШУ был выполнен на одной подложке с МПФ (рис. 3.8б), которая изготовлена из диэлектрика с низкими потерями.

Шумы, обусловленные потерями в решетке и МПФ, могут быть уменьшены использованием металла с более высокой проводимостью или охлаждением решетки криогенной установкой. Использование такого подхода привело бы к значительному удорожанию конструкции, поэтому он не был использован. Потери в МПФ были уменьшены за счет уменьшения длины микрополосковой линии. При этом был доработан антенный элемент Вивальди, в котором был перемещен круглый резонатор и закруглена щель, идущая к резонатору (рис. 3.8б). Также была перемещена «заглушка» МПФ. Это позволило значительно укоротить микрополосок МПФ, тем самым уменьшить потери в нем. На рис. 3.6а показаны измеренная шумовая температура, вносимая исходным вариантом МПФ и новым, модифицированным, МПФ. Видно, что на частоте 1,4 ГГц выигрыш составил примерно 2 К.

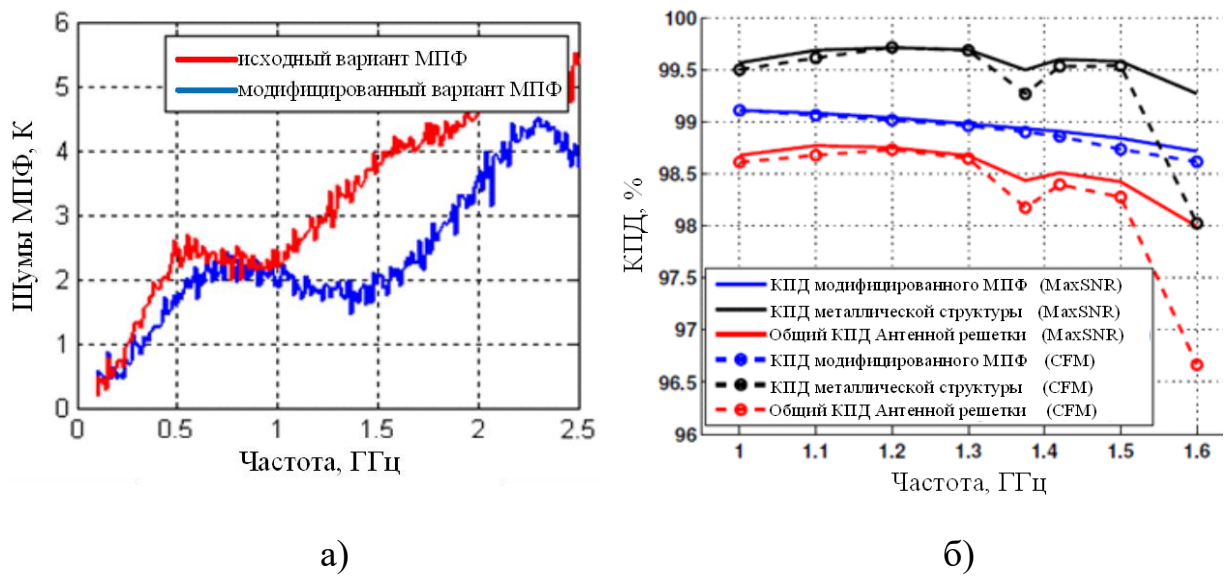

а)                                                        б)

Рис. 3.6 Потери в МПФ и металле решетки: а) измеренная шумовая температура, вносимая исходным вариантом МПФ и новым, модифицированным, МПФ; б) рассчитанное значение КПД за счет потерь в металле и модифицированном МПФ для осевого луча и двух способов формирования луча

На рис. 3.6,б показаны зависимости КПД от частоты, рассчитанные для случаев учета потерь только в металле, только в модифицированном МПФ, совметных потерь в металле и МПФ, для двух ситуаций формирования осевого луча: согласования по полю и обеспечения максимальной чувствительности.

За счет указанных мер удалось уменьшить шумовую температуру на (35…37) К.

### 3.2.3 Улучшение стабильности электрических характеристик антенной решетки за счет улучшения жесткости конструкции

Кроме высокой шумовой температуры первый прототип обладал еще одним недостатком — недостаточной жесткостью конструкции, которая выражалась в плохой повторяемости измерений $S$-матрицы решетки, а также значительным различием коэффициента отражения от элементов решетки с одинаковым окружением. На рис. 3.7,а показаны коэффициенты отражения, измеренные для краевых элементов решетки первого прототипа. Из рисунка видно, что повторяемость измерений не обеспечивается в достаточной степени.

Так как наблюдается значительная нестабильность измерений коэффициента отражения, то можно ожидать также изменение во времени ДН антенной решетки из-за изменения механических нагрузок на решетку, которые существуют в зеркальной антенной системе. Однако изменения ДН в процессе измерений (между калибровками инструмента) недопустимы, так как это исказит карту небесного объекта. Поэтому необходимо было улучшить стабильность измеряемых параметров решетки.

Экспериментальным путем было выяснено, что нестабильность была вызвана несколькими факторами. Во-первых, недостаточно плотное крепление МПФ к элементу Вивальди приводило к изменению коэффициента трансформации перехода «микрополосок-щель», что приводило к эффектам, описанным в подразделе 3.1. Во-вторых, из-за технологии изготовления и сборки решетки возникал частичный разрыв электрического контакта между смежными элементами решетки, что изменяло распределение токов, и, как следствие, приводило к изменению коэффициентов отражения и искажению откалиброванной ДН.

Чтобы исключить влияние описанных факторов, было сделано несколько модификаций во втором прототипе антенной решетки. Во-первых, для улучшения контакта МПФ и элементов Вивальди крепление МПФ было выполнено в нескольких местах вдоль щели. Во-вторых, каждый элемент Вивальди был выполнен в виде отдельной структуры, которая устанавливалась в специальные стойки (см. рис. 3.8в).

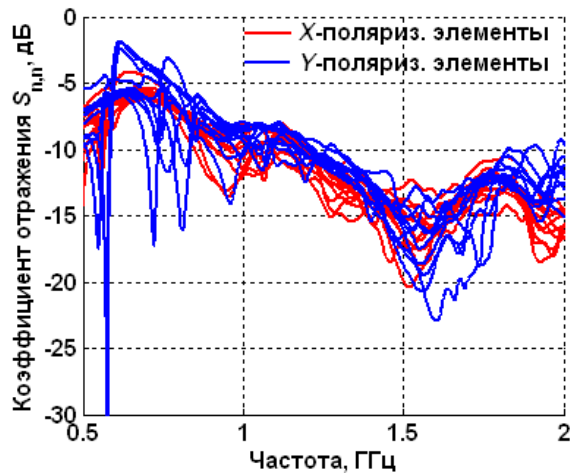

а)

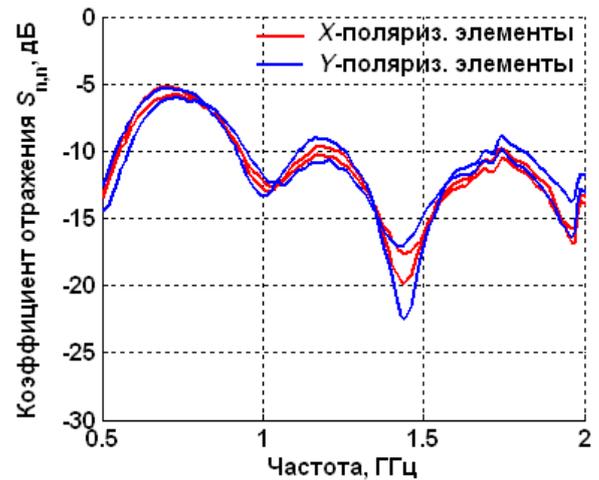

б)

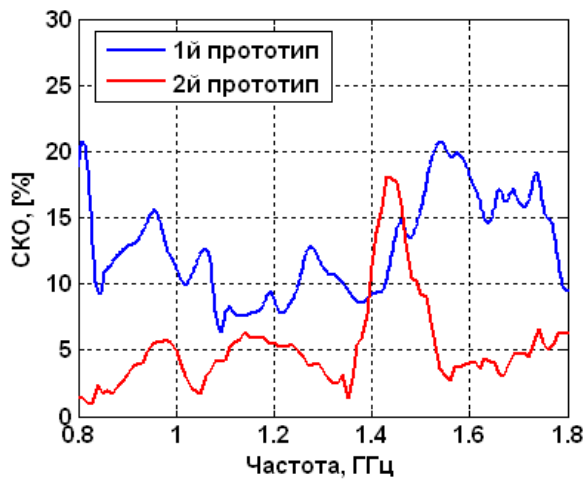

в)

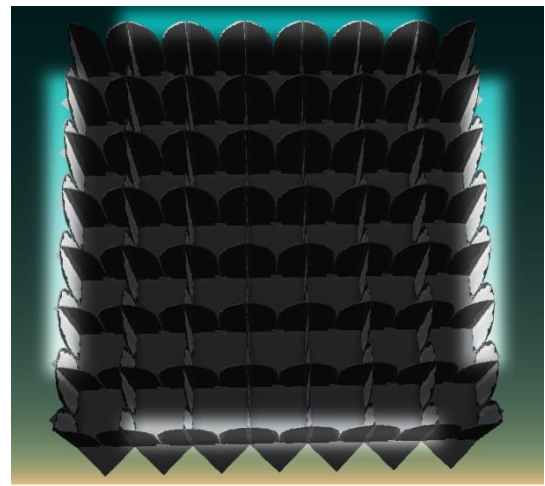

г)

Рис. 3.7 Нестабильность измерений коэффициента отражения элементов с похожим окружением: а) коэффициент отражения от краевых элементов для первого прототипа APERTIF; б) коэффициент отражения от краевых элементов для второго прототипа; в) среднеквадратическое отклонение коэффициента отражения краевых элементов для двух прототипов; г) элементы, коэффициенты отражения которых измерялись.

Данные меры заметно улучшили механическую прочность антенной решетки и стабильность измеряемых характеристик. На рис. 3.7б показаны зависимости коэффициента отражения от частоты, измеренные для краевых элементов решетки второго прототипа, а на рис. 3.7в показано среднеквадратическое отклонение (СКО) коэффициента отражения краевых элементов для двух прототипов решеток в процентах. Видно, что СКО коэффициента отражения для второго прототипа значительно меньше.

### 3.2.4. Второй прототип антенной решетки

Описанные выше изменения были реализованы во втором прототипе решетки, показанном на рис. 3.8.

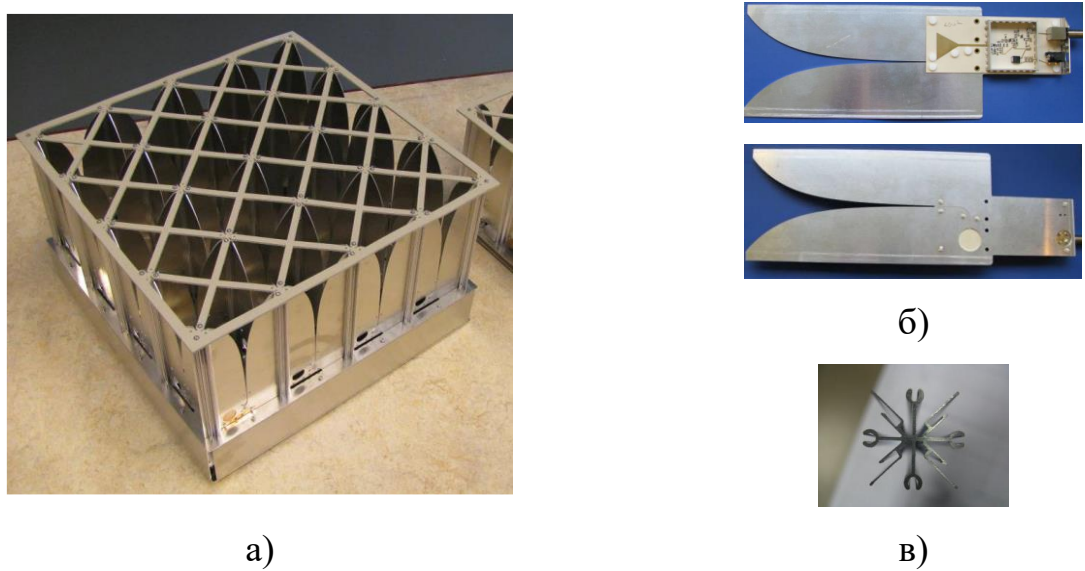

а)

б)

в)

Рис. 3.8 Второй прототип антенной решетки: а) уменьшенный вариант решетки; б) элемент решетки; в) профиль стоек, в которых крепятся элементы

## 3.3. Сравнение рассчитанной матрицы рассеяния решетки с измеренной

На рис. 3.9 показаны коэффициенты связи $S_{29,m}$ элемента № 29 решетки (центральный элемент) с остальными ко-поляризованными элементами [99].

Измерение                    Моделирование

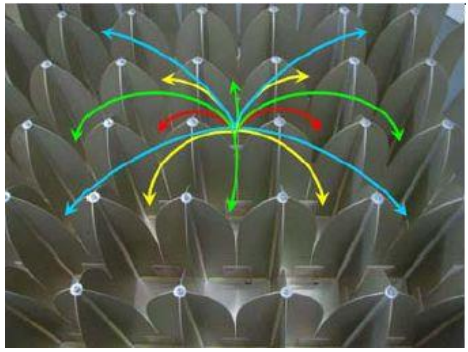          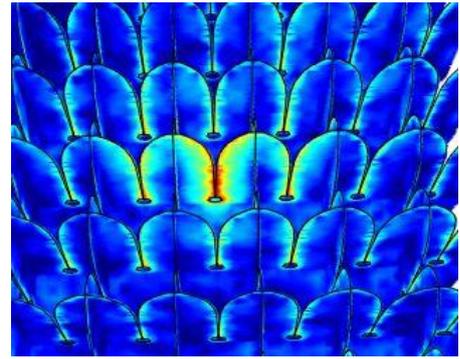

а)                           б)

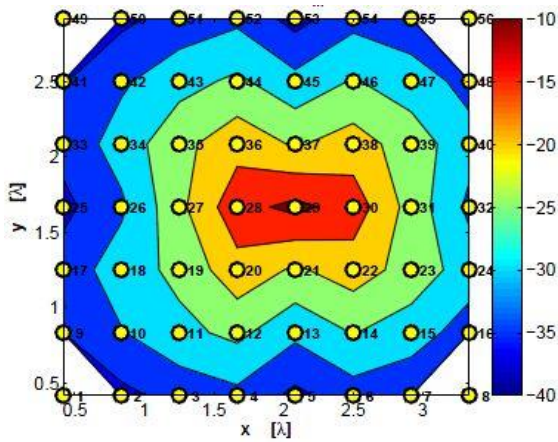          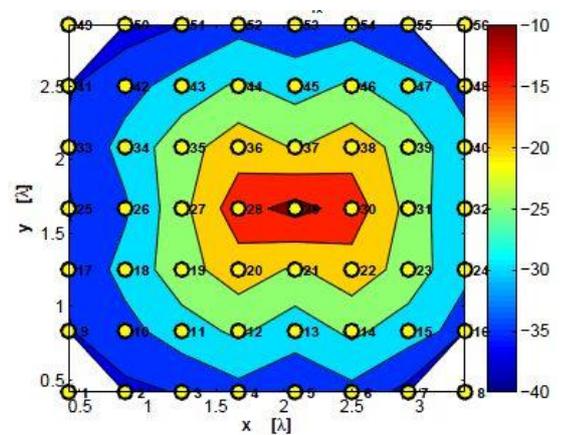

в)                           г)

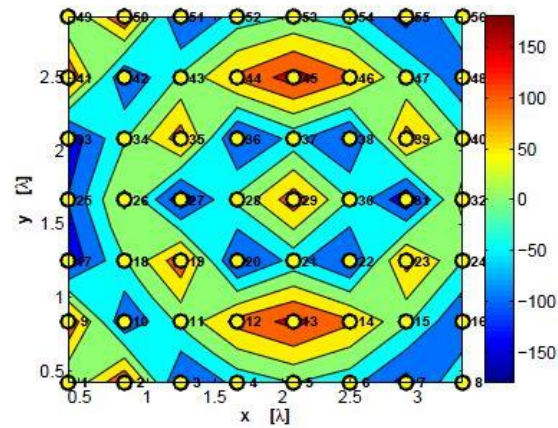          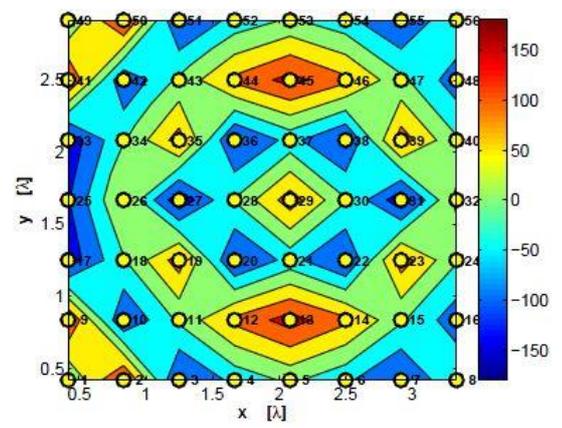

д)                           е)

Рис. 3.9 Коэффициенты связи $S_{29,m}$ элемента № 29 с остальными ко-поляризованными элементами решетки на частоте 1 ГГц: а) вид решетки сверху; б) распределение токов в решетке (логарифмический масштаб, 60 дБ динамический диапазон) при возбуждении элемента № 29; в) измеренные $|S_{29,m}|$, дБ (в); г) моделированные $|S_{29,m}|$, дБ; д) измеренные аргументы $S_{29,m}$, градусы; е) моделированные аргументы $S_{29,m}$, градусы

Каждая точка на контурном графике соответствует одному элементу решетки в его позиции в решетке, а цвет контуров в этой позиции соответствует модулю (рис. 3.9в, г) или аргументу (рис. 3.9д, е) коэффициента связи элемента № 29 с данным элементом.

Из рис. 3.9 видно хорошее совпадение результатов моделирования и измерений, как для модуля, так и для аргумента коэффициентов связи.

### 3.4. Расчет диаграмм направленности элементов решетки

Точное определение диаграмм направленности каждого элемента решетки в присутствии остальных элементов является важной задачей при моделировании, так как от них зависят оптимальные (независимо от вида критерия) весовые коэффициенты.

На рис. 3.10 показаны рассчитанные ДН элементов решетки при возбуждении каждого из ее элементов по отдельности.

Из рис. 3.10,б видно, что кросс-поляризационная компонента ДН краевых элементов значительно выше, чем у ДН центральных элементов. Из рис. 3.10,в,г видно, что ДН центральных и краевых элементов значительно отличаются как в диапазоне углов, перехватываемых зеркалом (границы этих углов показаны штриховыми линиями), так и за их пределами. Поэтому, для правильного расчета КИП, коэффициента перехвата и коэффициента поляризационной дискриминации нельзя считать, что ДН всех элементов одинаковы. Требуется рассматривать ДН каждого элемента по отдельности.

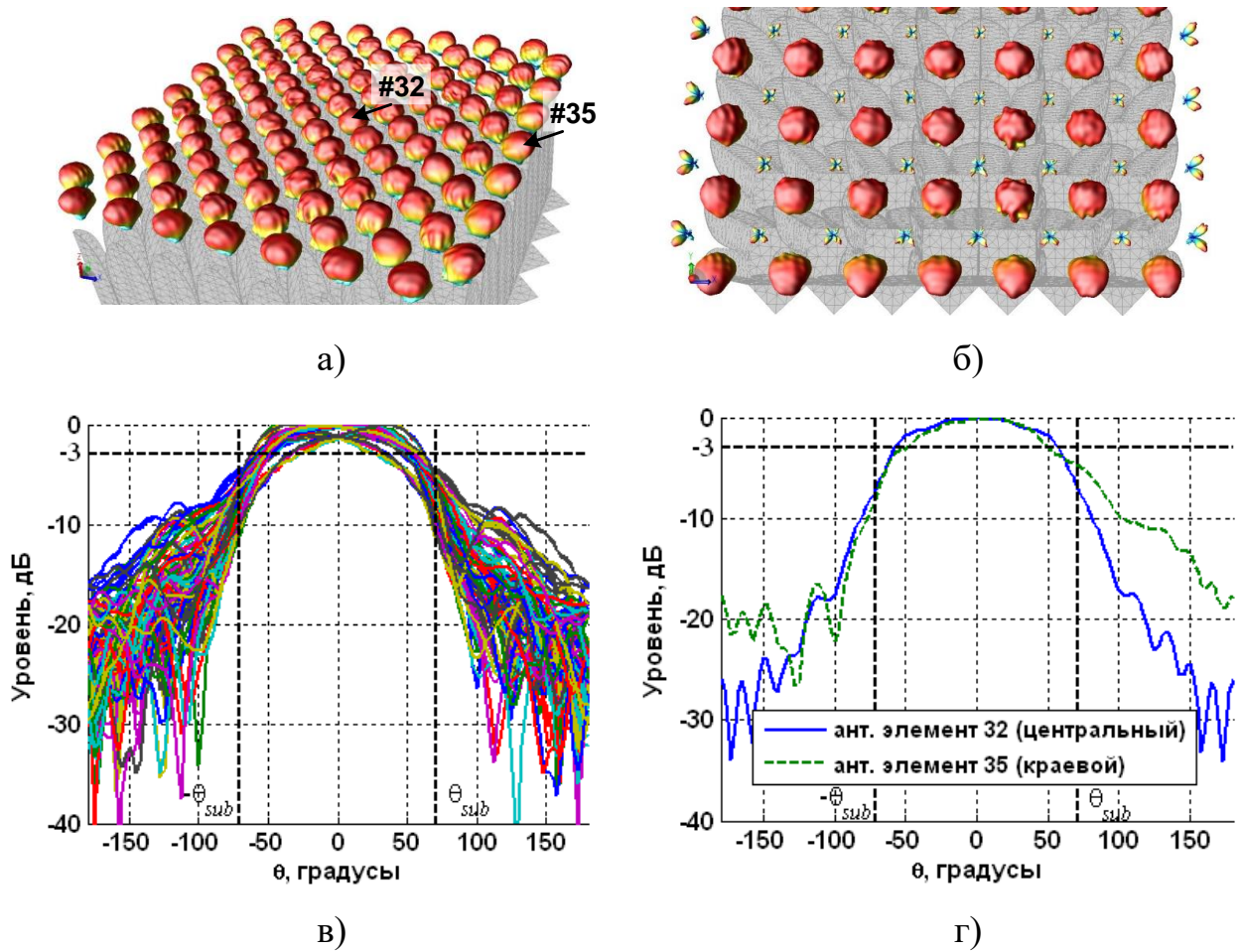

Рис. 3.10 — ДН элементов решетки на частоте 1,42 ГГц: а) объемные мощностные ДН всех элементов; б) объемные ДН элементов решетки для ко-поляризационной компоненты падающего поля; в) сечение всех мощностных ДН при φ = 0; г) сечение мощностных ДН центрального элемента № 32 и краевого элемента № 35

## 3.5. Экспериментальное определение весовых коэффициентов на телескопе. Проверка модели путем расчета фокусного расстояния зеркала

При моделировании оптимальные по критерию максимальной чувствительности весовые коэффициенты определяются через известные сигнальный вектор **e** при приеме сигнала с заданного направления луча и шумовую корреляционную матрицу **C**. Эти параметры могут быть определены также экспериментально [94].

При этом должны быть выполнены следующие этапы.

1. Телескоп направляется на яркую звезду и на выходе коррелятора системы получают полную корреляционную матрицу $\mathbf{C_{on}} = \mathbf{P} + \mathbf{C}$, которая состоит из корреляционной матрицы сигнала от звезды **P** и шумовой корреляционной матрицы **C**. Таким образом, в $\mathbf{C_{on}}$ входят как шумы (внутренние и внешние по отношению к антенной системе), так и сигнал от источника.

2. Предполагается, что за время измерений система не меняет своих параметров. Тогда, если направить телескоп на пустой участок неба, то на выходе коррелятора будет только шумовая корреляционная матрица **C**, обусловленная внутренними и внешними шумами.

3. Путем вычитания матрицы **C** из матрицы $\mathbf{C_{on}}$ получают сигнальную корреляционную матрицу **P**. Теперь, если найти собственный вектор этой матрицы, соответствующий максимальному собственному значению матрицы, то он и будет искомым сигнальным вектором **e**.

4. Теперь, зная **e** и **C**, можно рассчитать оптимальные весовые коэффициенты **w** по выражению (2.20).

Если требуется определить весовые коэффициенты для критерия максимальной чувствительности с ограничениями по направлениям (см. выражение (2.23)), то этапы 1 и 3 повторяют при смещении телескопа так, чтобы источник попал в каждое из $N_{dir}$ направлений ограничений. Но, так как лучи частично перекрываются, реальное число измерений будет значительно меньше, чем $N \cdot N_{dir}$ (где $N$ — число лучей), так как некоторые направления являются общими для нескольких лучей.

Этапы 1, 3 и 4 повторяются для каждого луча.

Описанным способом в институте *ASTRON* были измерены весовые коэффициенты **w**, которые будут использованы далее, и на которые будем ссылаться как на «измеренные весовые коэффициенты».

На рис. 3.11,а показано расположение координатных систем при моделировании. Координатная система зеркала находится в вершине параболоида, ось $z$ направлена к точке фокуса *F*. Начало координатной системы решетки находится в центре основания решетки и совпадает с точкой фокуса зеркала. Ось $z$ координатной системы решетки направлена к вершине параболоида. При таком взаимном расположении зеркала и решетки рассчитывались вторичные ДН, а также все характеристики антенной системы, которые приведены в разделе 4.

В антенной системе телескопа, ввиду конструктивных особенностей, решетка смещена на величину $\Delta z = 19$ см (рис. 3.11,б).

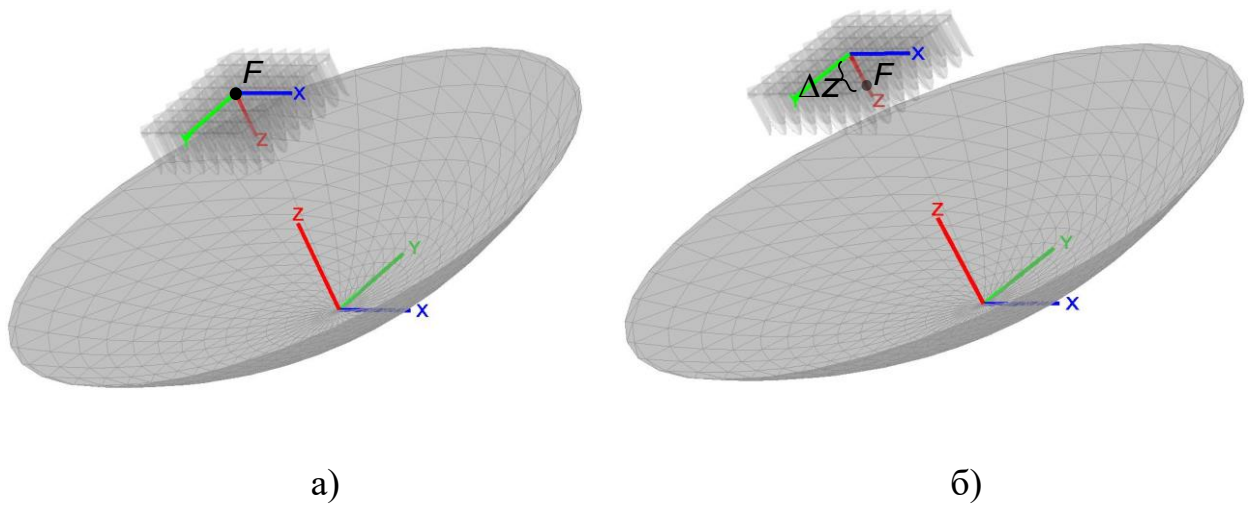

а)                                                б)

Рис. 3.11 Взаимная ориентация координатных систем зеркала и решетки:
а) при моделировании; б) при измерениях

Для того, чтобы проверить расчетную модель, измеренные весовые коэффициенты были применены к рассчитанным ДН элементов, как это сделано в выражении (2.24), и таким образом была рассчитана общая ДН осевого луча $E_f(\theta, \varphi)$.

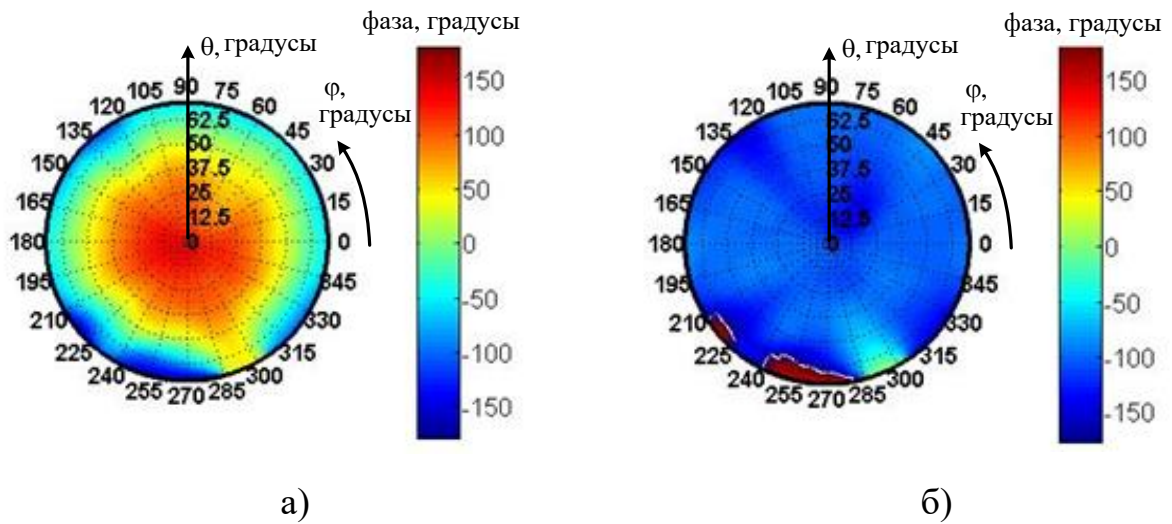

а)                                              б)

Рис. 3.12 Рассчитанное фазовое распределение в апертуре зеркала при применении к промоделированным ДН элементов измеренных весовых коэффициентов: а) $\Delta z = 0$ см; б) $\Delta z = 17{,}5$ см

Так как был выбран осевой луч, то фазовое распределение в апертуре зеркала должно быть равномерным. При применении к системе рис. 3.11а оптимальных для системы рис. 3.11б весовых коэффициентов можно ожидать неравномерного распределения фазы в апертуре. На рис. 3.12а видно, что фазовое распределение, рассчитанное указанным способом, меняется более чем на 270 градусов от центра зеркала к краям.

Для того, чтобы найти оптимальное значение $\Delta z$ для системы, показанной на рис. 3.11а, при котором фазовое распределение становится наиболее равномерным, была введена целевая функция, которую можно записать следующим образом:

$$\eta_{ph}\left(\Delta z\right) = \frac{\left|\int\limits_{0}^{2\pi}\int\limits_{0}^{\theta_{sub}} E_f\left(\theta,\varphi,\Delta z\right)\tan\left(\frac{\theta}{2}\right)d\theta d\varphi\right|^2}{\left(\int\limits_{0}^{2\pi}\int\limits_{0}^{\theta_{sub}} \left|E_f\left(\theta,\varphi,\Delta z\right)\tan\left(\frac{\theta}{2}\right)\right|d\theta d\varphi\right)^2}, \qquad (3.3)$$

где $\eta_{ph}(\Delta z)$ — фазовая эффективность как функция от смещения $\Delta z$; $E_f(\theta, \varphi, \Delta z)$ — комплексная ДН поля облучателя (решетки), которая также является функцией смещения $\Delta z$ (формула справедлива только для дальней зоны):

$$E_f(\theta, \varphi, \Delta z) = E_f(\theta, \varphi) \cdot e^{-jk \cdot \Delta z \cdot \cos(\theta)}. \tag{3.4}$$

Выражение (3.3) представляет собой формулу для расчета фазовой эффективности, которая равна единице при равномерном фазовом распределении, и меньше единицы при неравномерном. Таким образом, выражение применимо для расчета фазовой эффективности только для осевого луча, так как при формировании сканирующего луча фаза в апертуре заведомо неравномерна, и, в идеале, должна меняться линейно в направлении проекции вектора луча на апертуру зеркала. Поэтому для расчета $\Delta z$ был выбран осевой луч.

После максимизации целевой функции (3.3) относительно переменной $\Delta z$, было найдено, что фазовая эффективность максимальна при $\Delta z = 18{,}1$ см (напомним, что для системы рис. 3.11,б $\Delta z = 19$ см). Следовательно, погрешность $\Delta z$ имеет значение 1,5 см, что составляет 0,11% от фокусного расстояния зеркала и 4% от длины волны. Таким образом, точность определения $\Delta z$ достаточно высока, что подтверждает адекватность выбранной модели.

### 3.6 Выводы по разделу

3.6.1. В разделе приведены численные результаты моделирования антенной решетки вне зеркала и произведена их экспериментальная проверка с целью подтверждения адекватности разработанной модели.

3.6.2. Показано, что при незначительном (около 10%) отклонении коэффициента трансформации $p$ в модели от его реальной (оптимальной) величины наблюдается значительное изменение коэффициента отражения: происходит не только увеличение ошибки в его аргументе и изменение значения модуля, но и смещение и сглаживание максимумов частотной зависимости. Для рассмотренной антенной решетки ошибка между моделированным и измеренным значением модуля $S_{11}$ достигает 60% в диапазоне частот (0,7…1,7) ГГц при изменении $p$ на 10%, в то время как для оптимальной величины $p$ ошибка не превышает 20%. Поэтому точное определение $p$ является важной задачей при использовании описанной модели.

3.6.3. Коэффициент трансформации $p$ намного сильнее влияет на коэффициент отражения, чем на коэффициент передачи, поэтому для определения $p$ лучше использовать коэффициент отражения.

3.6.4. Было предложено улучшение конструкция антенной решетки, что привело к улучшению стабильности электрических характеристик решетки, а также уменьшению шумовой температуры. Среднеквадратическое отклонение коэффициента отражения для краевых элементов решетки было уменьшено с (7…20) % до (3…7) % в диапазоне частот (0,8…1,8) ГГц. Шумовая температура МПФ была уменьшена на 5 К за счет исключения соединителя МПФ и МШУ, а также за счет укорочения длины микрополоска МПФ.

3.6.5. Промоделирована и измерена полная матрица рассеяния решетки 7×8×2 элементов и показано хорошее соответствие между результатами моделирования и измерений, что также подтверждает адекватность модели. Ошибка между измеренными и промоделированными значениями модуля коэффициента передачи от центрального элемента решетки к остальным элементам не превышает 1 дБ. Ошибка в аргументе этих коэффициентов не превышает 15°.

3.6.6. Показано значительное различие между ДН центральных и краевых элементов решетки (разница достигает 5 дБ в диапазоне углов, перекрываемых зеркалом), поэтому нельзя применять методы бесконечных решеток для электромагнитного моделирования антенной решетки.

3.6.7. *Экспериментально* определены весовые коэффициенты, оптимальные по критерию максимальной чувствительности. С использованием этих значений весовых коэффициентов проверена модель путем расчета смещения решетки $\Delta z$ относительно точки фокуса зеркала. Хорошее совпадение полученного значения ($\Delta z = 18{,}1$ см) с реальной величиной ($\Delta z = 19$ см) также подтверждает адекватность модели.

## РАЗДЕЛ 4

## МОДЕЛИРОВАННИЕ И ОПТИМИЗАЦИЯ ОСНОВНЫХ ПАРАМЕТРОВ ПРОТОТИПА APERTIFF И ИХ ИЗМЕРЕНИЕ НА ТЕЛЕСКОПЕ

В данном разделе показаны результаты расчетов параметров антенной системы в целом, а также сравнение некоторых из них с измеренными на одной из антенн Вестерборгского радиоинтерферометра.

В *первом подразделе* показаны такие параметры антенной системы как:

— ДН всей системы;

— системная шумовая температура и ее составляющие;

— коэффициенты эффективности системы (КИП, коэффициент перехвата, коэффициенты эффективности амлитудного и фазового распределений в апертуре зеркала).

Все результаты представлены как зависимости от частоты и номера луча (угла сканирования).

Во *втором подразделе* будет произведен поиск оптимального параметра $g_{const}$ для критерия максимальной чувствительности с ограничениями (*LCMV*) (см. подраздел 2.3.4).

В *третьем подразделе* приведены результаты расчета чувствительности системы (как отдельных лучей, так и чувствительности внутри многолучевого поля обзора) и ее сравнение с измерениями. Также приведено сравнение чувствительности и ее неравномерности при использовании различных критериев оптимизации весовых коэффициентов.

В *четвертом подразделе* показаны свойства системы как поляриметра.

## 4.1. Результаты моделирования основных параметров системы

### 4.1.1. Расчет весовых коэффициентов

На рис. 4.1 показаны модули весовых коэффициентов для осевого луча (луч № 19), рассчитанные по выражениям (2.17), (2.20) и (2.23). На рисунках использованы сокращения: *CFM* — критерий согласования по полю, *MaxSNR* — критерий максимальной чувствительности без ограничений по направлениям, *LCMV* — критерий максимальной чувствительности с ограничениями по направлениям. Весовые коэффициенты показаны для антенных элементов решетки одной поляризации и выражены в децибелах. Из рисунка видно, что число сильно возбужденных (|w| > − 10 дБ) антенных элементов уменьшается при усложнении критерия оптимизации. Это должно приводить к уменьшению принятой мощности сигнала при усложнении критерия оптимизации, и, следовательно, уменьшению КИП и эффективной площади апертуры $A_{eff}$.

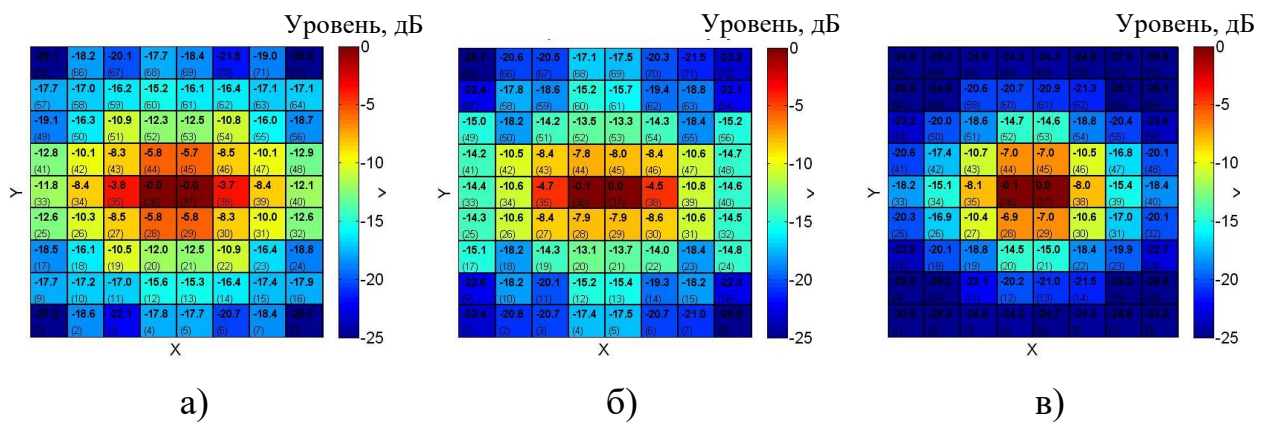

Рис. 4.1 Весовые коэффициенты для различных критериев оптимизации:
а) *CFM*; б) *MaxSNR*; в) *LCMV*

4.1.2. Расчет диаграмм направленности

В проекте *APERTIF* поле обзора содержит 37 одновременно формируемых лучей. Конфигурация лучей внутри поля обзора, а также их нумерация, были показаны на рис. 1.14.

Все трехмерные ДН показаны в *True-View* системе координат с нанесением угловых координат для удобства восприятия.

На рис. 4.3а...4.3в показаны ДН по мощности для решетки для трех критериев оптимизации весовых коэффициентов, на рис. 4.3г...4.3и — сечения ДН решетки на нескольких частотах, причем сплошной линией показана ко-поляризованная к элементам решетки компонента поля, а штриховой — кроссполяризованная. На рис. 4.4 показаны соответствующие вторичные ДН (по мощности). Все ДН приведены для осевого луча (луч № 19).

Как можно видеть из рис. 4.3а...4.3и, первичные ДН для критериев *CFM* и *MaxSNR* довольно широкие, но несколько изрезанные в области облучения зеркала (угол раскрыва зеркала показан малой черной окружностью в объемных ДН и вертикальной пунктирной линией в сечениях). Также первичные ДН значительно меняются при изменении частоты.

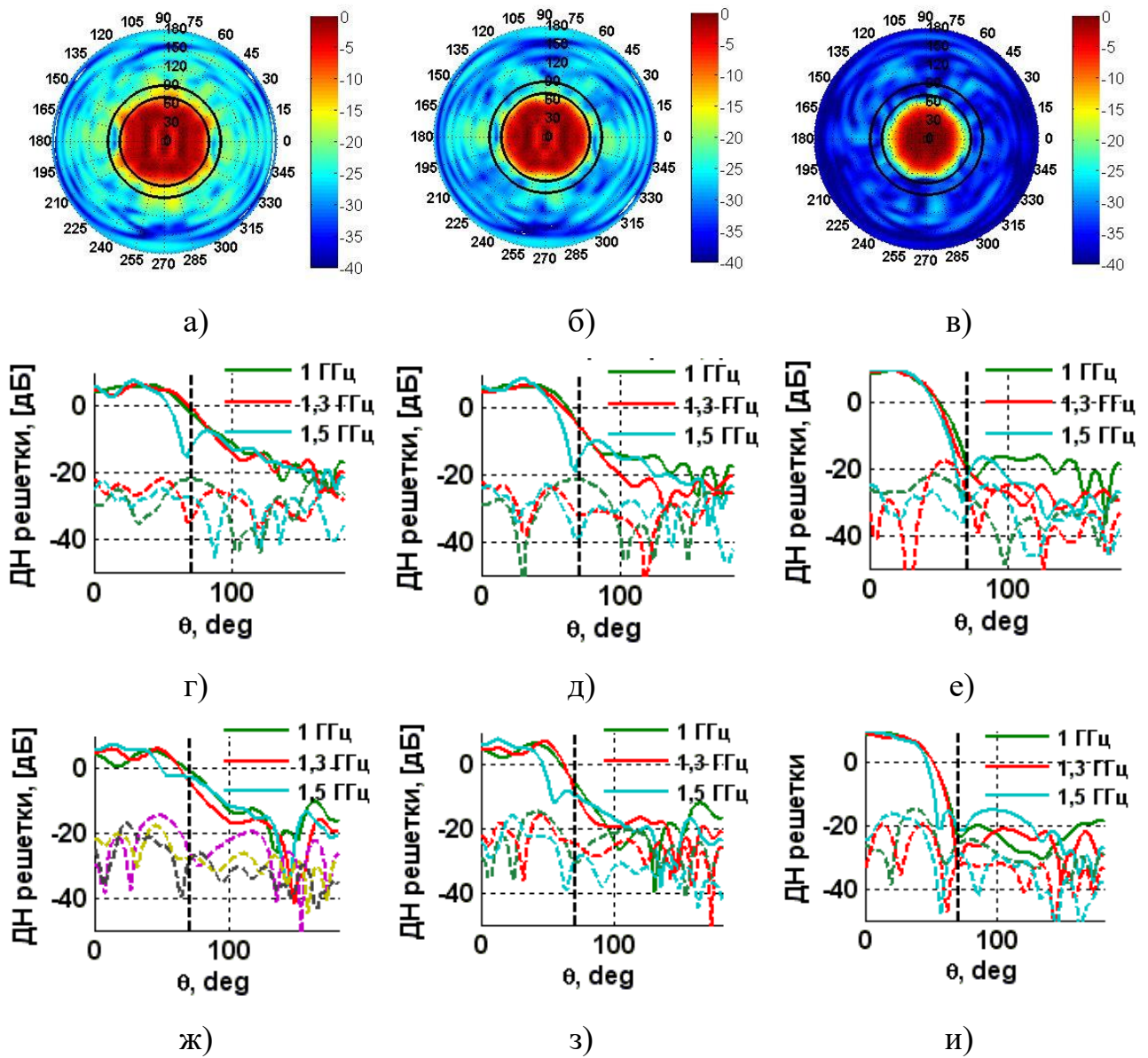

Рис. 4.3 — ДН антенной решетки по мощности: а) критерий *CFM*; б) критерий *MaxSNR*; в) критерий *LCMV*; г) сечение ко- и кроссполяризованных компонент ДН при φ = 0 градусов, критерий *CFM*; д) сечение ко- и кроссполяризованных компонент ДН при φ = 0 градусов, критерий *MaxSNR*; е) сечение ко- и кроссполяризованных компонент ДН при φ = 0 градусов, критерий *LCMV*; ж) сечение ко- и кроссполяризованных компонент ДН при φ = 90 градусов, критерий *CFM*; з) сечение ко- и кроссполяризованных компонент ДН при φ = 90 градусов, критерий *MaxSNR*; и) сечение ко- и кроссполяризованных компонент ДН при φ = 90 градусов, критерий *LCMV*

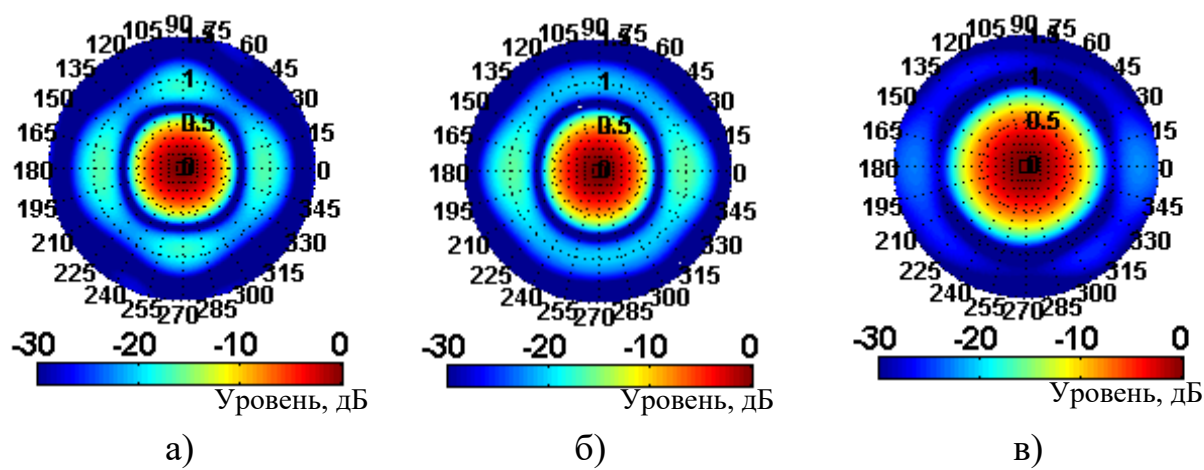

а)          б)          в)

Рис. 4.4 — Вторичные ДН: а) для критерия *CFM*; б) для критерия
*MaxSNR*; в) для критерия *LCMV*

В случае критерия максимальной чувствительности с ограничениями по направлениям (рис. 4.3в, рис. 4.3е, рис. 4.3и и рис. 4.4в) можно наблюдать меньшую изрезанность ДН и лучшую стабильность в диапазоне частот. Также из указанных рисунков видно, что луч уже, чем для первых двух случаев. Вторичная ДН более широкая, что является следствием более узкой первичной ДН. Из-за принудительного расширения вторичной ДН общий КНД системы уменьшился примерно на 2 дБ (рис. 4.5), но, как будет показано в подразделе 4.3, улучшилось перекрытие лучей (уменьшилась неравномерность чувствительности в поле обзора).

На рис. 4.5 показана зависимость КНД осевого луча от частоты для трех выбранных критериев оптимизации весовых коэффициентов.

Как и ожидалось, при критерии *CFM* КНД максимален, так как при этом ведется согласованный прием и принимается максимально возможная мощность.

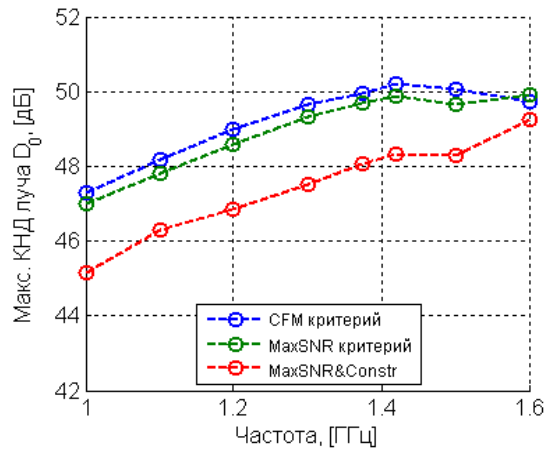

Рис. 4.5 Зависимость КНД в направлении луча от частоты для осевого луча и трех критериев оптимизации

При критерии *MaxSNR* КНД немного ниже, что объясняется тем, что метод оптимизации по этому критерию изменяет аплитудно-фазовое распределение поля в апертуре зеркала так, чтобы максимизировать отношение сигнал/шум на выходе приемника. Естественно, что при измененном аплитудно-фазовом распределении (по отношению к методу *CFM*) КНД будет меньше из-за уменьшенного КИП. Но, несмотря на это, чувствительность при этом методе оптимизации максимальна, что будет показано в подразделе 4.3.

Как уже отмечалось, КНД при критерии *LCMV* заметно меньше, чем для первых двух случаев, и это приводит к уменьшению чувствительности, но в то же время дает выигрыш в равномерности чувствительности в многолучевом поле обзора.

Из рис. 4.5, видно, что КНД растет с увеличение частоты, что является естественным, так как увеличивается электрический размер зеркала. КНД увеличивается примерно до частоты 1,4 ГГц, затем рост замедляется, и при дальнейшем увеличении частоты происходит незначительное уменьшение. Особенно это заметно для критерия *CFM*. Такое поведение объясняется тем, что расстояние между элементами в решетке равно 11 см, что соответствует половине длины волны на частоте 1,36 ГГц. То есть, при частоте более 1,36 ГГц расстояние между элементами становится больше половины длины волны, и не выполняется условие Найквиста, из-за чего резко уменьшается КИП, и, как следствие, КНД.

### 4.1.3. Расчет коэффициентов эффективности и шумовой температуры

На рис. 4.6 и 4.7 показаны зависимости рассчитанной шумовой температуры системы $T_{sys}$ и ее главных составляющих от направления сканирования и частоты соответственно.

На этих рисунках $T_{sp}$ обозначает вклад в шумовую температуру за счет приема шумов земли из-за эффекта «переливания энергии» за края зеркала, $T_{lna}$ — эквивалентная шумовая температура малошумящих усилителей. Кроме собственно шумов самих МШУ в $T_{lna}$ входит составляющая $T_{coup}$, которая представляет собой пересчитанные в температуру потери из-за рассогласования элементов решетки с МШУ. Расчет этих потерь производился методом эквивалентной одноэлементной антенны с МШУ [8], в котором используется *активные* коэффициенты отражения, то есть учитывается сильное взаимное влияние элементов решетки друг на друга.

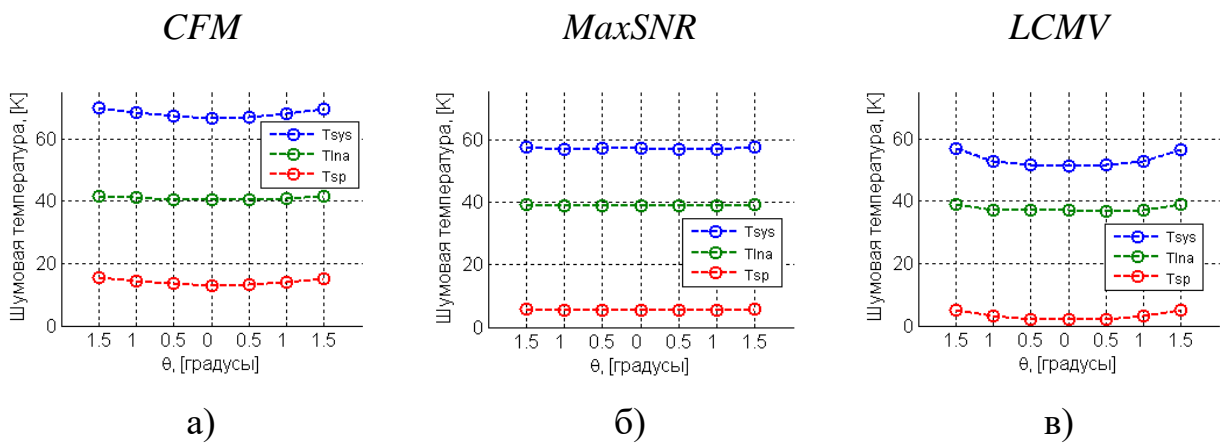

Рис. 4.6 Зависимости шумовой температуры системы и ее составляющих от направления сканирования на частоте 1,42 ГГц

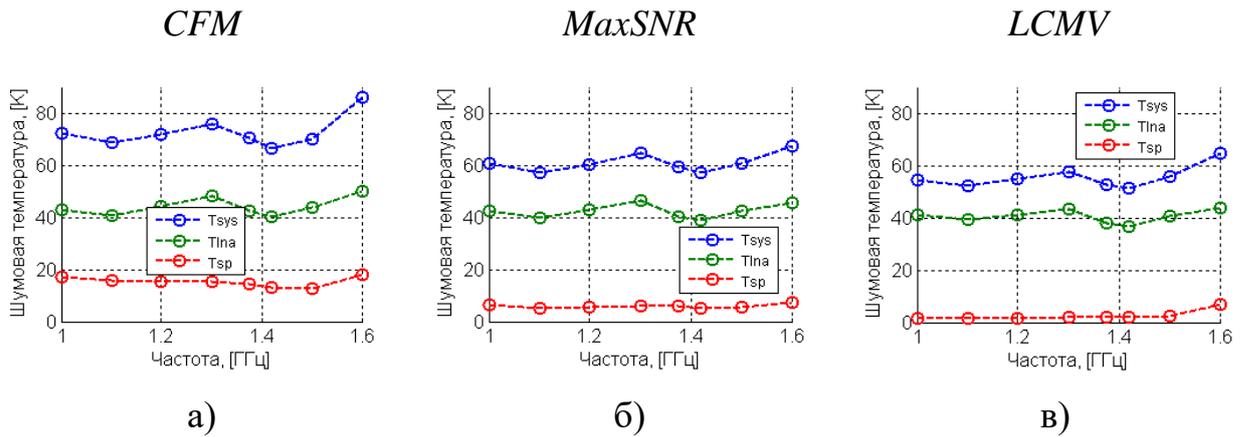

Рис. 4.7 Зависимости шумовой температуры системы и ее составляющих от частоты для осевого луча

Как и ожидалось, схема формирования лучей *CFM* (рис. 4.6а, рис. 4.7а) приводит к наибольшей шумовой температуре системы, так как в ней максимизируется принятая мощность сигнала от источника без учета внешних и внутренних шумов. Поэтому, несмотря на максимальный КНД для этого случая (см. рис. 4.5), можно ожидать уменьшения чувствительности по сравнению с другими схемами формирования. Схемы оптимизации *MaxSNR* и *LCMV* приводят к меньшим шумам, так как учитывают принимаемые и внутренние шумы.

Из рисунков также видно, что наибольший вклад в $T_{sys}$ вносят шумы МШУ, которые практически одинаковы для всех трех схем формирования лучей. Снижение $T_{sys}$ для схем *MaxSNR* и *LCMV* происходит, в основном, из-за уменьшения принимаемых шумов земли ($T_{sp}$).

На рис. 4.6 изображена зависимость $T_{sys}$ от угла сканирования, причем если для схемы *MaxSNR* $T_{sys}$ изменяется на (1…2) %, то для схемы *LCMV* это изменение достигает 15 %, за счет этого можно ожидать более быстрый спад чувствительности при приближении к краю поля обзора.

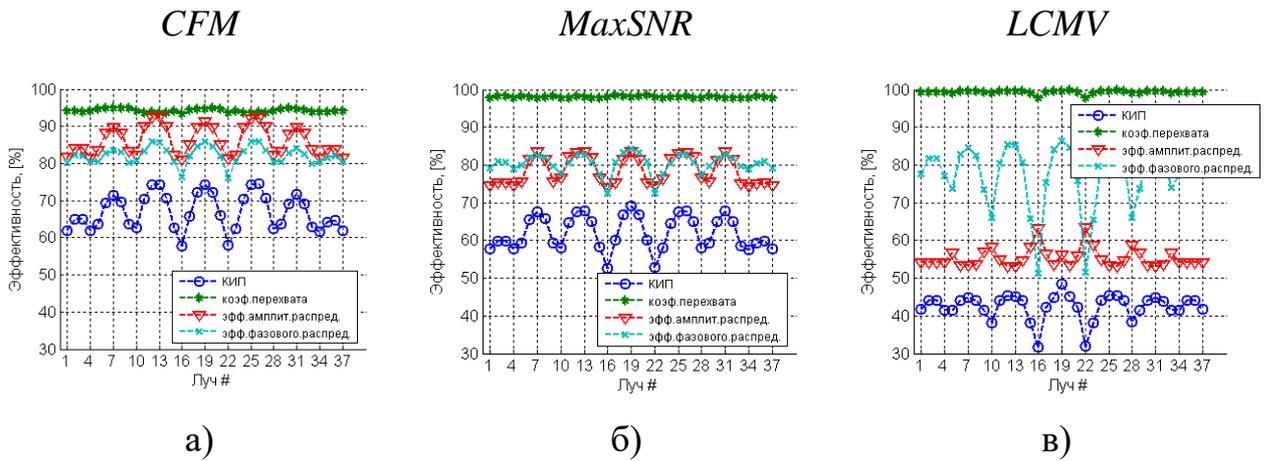

Рис. 4.8 Зависимость КИП и его составляющих от направления

сканирования (номера луча) на частоте 1,42 ГГц

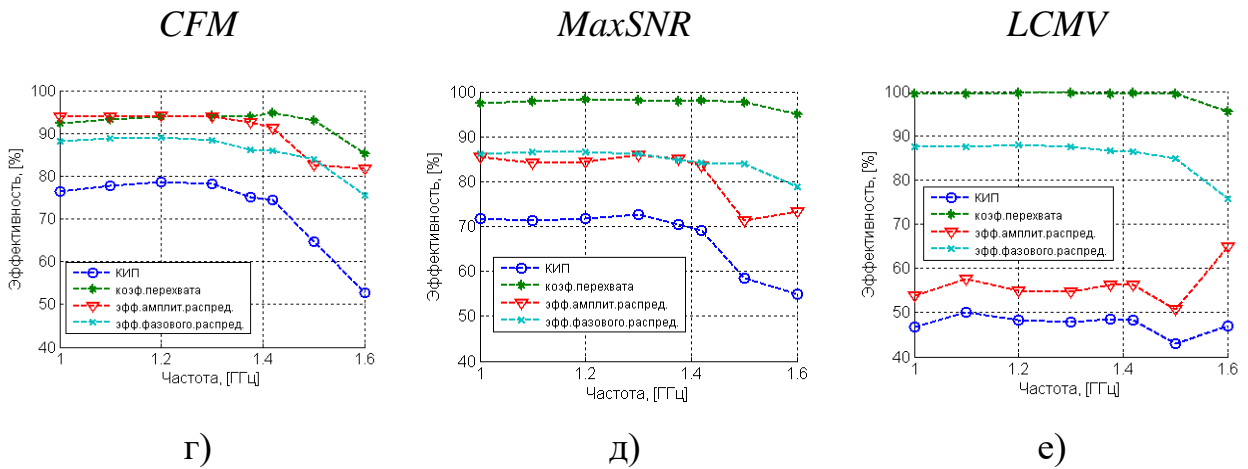

Рис. 4.9 Зависимости КИП и его составляющих для осевого луча от

частоты

На рис. 4.8 и 4.9 показаны зависимости рассчитанного КИП и его основных составляющих (см. выражение (2.48)) от направления сканирования и частоты соответственно. Направление сканирования представлено номером луча в многолучевом поле обзора (см. рис. 4.2).

Из рис. 4.8 и 4.9 видно, что КИП максимален для схемы *CFM* и минимален для схемы *LCMV*. Соответственно меняется и КНД лучей (см. рис. 4.5). Для коэффициента перехвата ситуация обратная, поэтому в случае схем *MaxSNR* и *LCMV* будет меньше принимаемых шумов земли, чем для схемы *CFM*.

Также можно наблюдать значительное уменьшение КИП для лучей, находящихся на краю поля обзора (лучи 1, 4, 10, 16, …). Это происходит из-за ограниченного размера решетки, и при сканировании на 1,5° (лучи 1, 4, 10, 16, …) часть принятой зеркалом энергии не перехватывается решеткой.

На рис. 4.9, можно наблюдать резкий спад КИП на частотах выше 1,4 ГГц. Как уже было отмечено в предыдущем подразделе, это происходит из-за невыполнения условия Найквиста на этих частотах.

### 4.2. Оптимизация весовых коэффициентов по критерию LCMV

Чтобы рассчитать весовые коэффициенты по методу *LCMV* (максимизация чувствительности с ограничениями по направлениям), необходимо задаться направлениями в пределах каждого луча, в которых будет производиться ограничение чувствительности, а также величиной этих ограничений, задаваемой в векторе-константе **g** (см. (2.23)).

На рис. 4.10 черными точками показаны выбранные направления ограничений на примере лучей 18…20. Это значит, что для каждого луча задано по 7 направлений: одно в максимуме луча и 6 — в точках пересечения с соседними лучами. Таким образом, $N_{dir} = 7$, и размерность матрицы **G** в соотношении (2.23) равна 144×7. Таким образом, **G** содержит комплексную величину сигнала, принимаемого с каждого из 7 направлений каждым из 144 элементов решетки. Нормированный вектор **g** имеет размерность 7×1 и содержит желаемую величину ограничений уровня в каждом из 7 направлений луча. Направления ограничений для остальных лучей задавались аналогичным образом.

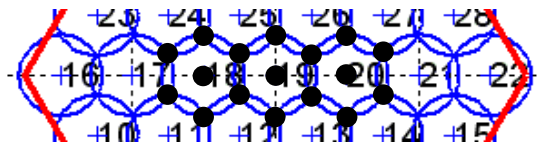

Рис. 4.10 Направления ограничений для лучей 18…20

Матрица **G** определяется в результате моделирования, а вектор **g** выбирается таким образом, чтобы достичь компромисса между чувствительностью и равномерностью во всем поле обзора.

Так как вектор **g** нормированный, его первый элемент, соответствующий направлению максимума луча, равен 1. Так как требуется, чтобы поле обзора было заполнено как можно более равномерно, примем, чтобы остальные 6 элементов вектора **g** одинаковы и равны некоторой оптимальной величине $g_{const}$. При этом для поиска оптимальной величины $g_{const}$ вводится следующий критерий: максимальная чувствительность в многолучевом поле обзора не должна уменьшиться более, чем на 10%.

На рис. 4.11 показана зависимость нормированной чувствительности (закрашенная область, правая ось ординат) и ее неравномерности в пределах поля обзора (кривая с круглыми маркерами, левая ось ординат) от параметра $g_{const}$. Чувствительность нормировалась к ее максимальному значению для схемы формирования лучей *MaxSNR*. Расчет производился по методу, описанному во 2 разделе диссертации.

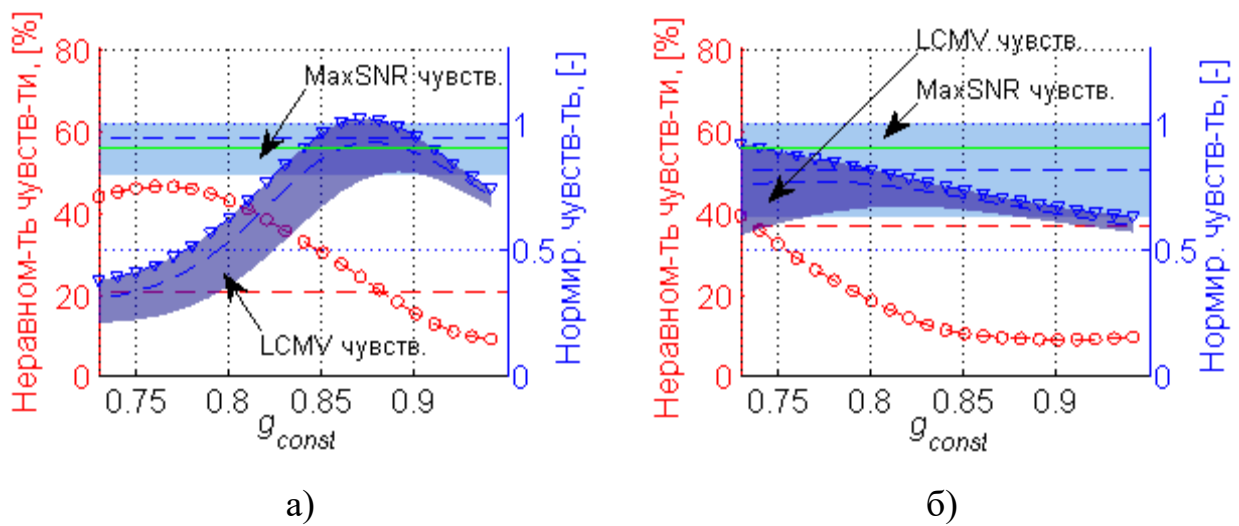

Рис. 4.11 Нормированная чувствительность и ее неравномерность в пределах поля обзора (целевая функция): а) для частоты 1 ГГц; б) для частоты 1,6 ГГц

Из графиков рис. 4.11 видно, что максимальная чувствительность для схемы *LCMV* равна 0,9 от максимальной чувствительности для схемы *MaxSNR* при $g_{const} = 0,91$ и $g_{const} = 0,75$ для частот 1 ГГц и 1,6 ГГц соответственно.

Таблица 4.1

Результаты оптимизации $g_{const}$

| $F$, ГГц | $g_{const}$ | $\dfrac{S_{LCMV}^{max}}{S_{MaxSNR}^{max}}$ | $\Delta S_{MaxSNR}$, % | $\Delta S_{LCMV}$, % | Выигрыш в неравн., % |
|---|---|---|---|---|---|
| 1,000 | 0,910 | 0,900 | 20,6 | 12,8 | 37,9 |
| 1,100 | 0,885 | 0,900 | 21,0 | 16,0 | 23,8 |
| 1,200 | 0,870 | 0,900 | 20,5 | 17,8 | 13,4 |
| 1,300 | 0,850 | 0,900 | 25,5 | 21,1 | 17,1 |
| 1,375 | 0,830 | 0,900 | 27,6 | 23,8 | 13,9 |
| 1,420 | 0,820 | 0,890 | 28,9 | 25,0 | 13,5 |
| 1,500 | 0,820 | 0,900 | 26,6 | 20,7 | 22,2 |
| 1,600 | 0,750 | 0,890 | 36,7 | 32,2 | 12,3 |

В таблице 4.1 приведены результаты оптимизации параметра $g_{const}$ для восьми частот рабочего диапазона. Кроме оптимального значения $g_{const}$ в таблице показаны отношения максимальных чувствительностей для схем формирования лучей *LCMV* и *MaxSNR*, их неравномерности $\Delta S$ и полученный выигрыш в неравномерности при использовании схемы *LCMV*.

## 4.3. Расчет чувствительности системы и неравномерности чувствительности в пределах поля обзора

На рис. 4.12а…в показана рассчитанная чувствительность приемной антенной системы во всем поле обзора (37 одновременно формируемых лучей) для трех критериев оптимизации весовых коэффициентов. Черной линией обозначена граница поля обзора. Сечения чувствительностей лучей плоскостью φ = 0° приведены на рис. 4.12г…е. Штриховой линией показана чувствительность отдельных лучей, сканирующих от –1,5° до 1,5°, сплошной линией показана общая чувствительность, полученная после обработки в корреляторе сигналов от всех лучей.

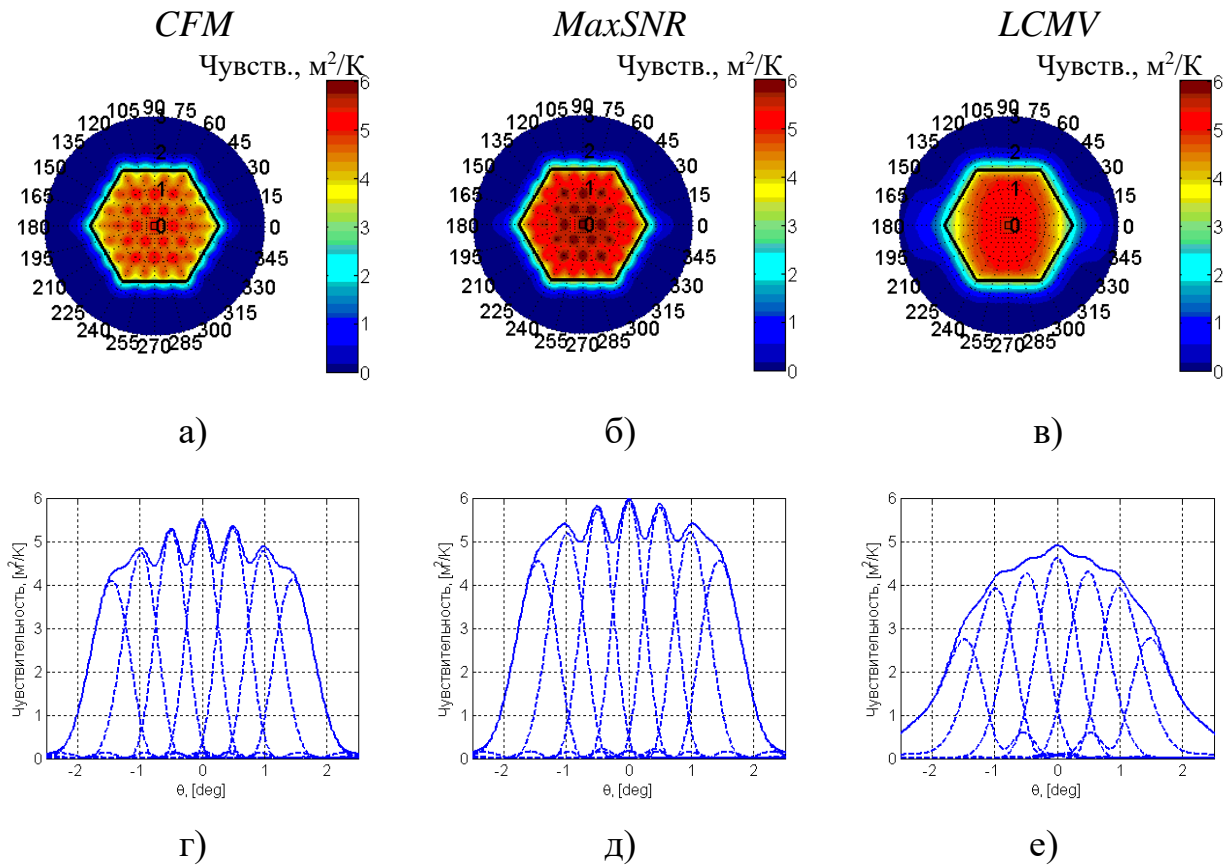

Рис. 4.12 Чувствительность в поле обзора антенной системы для трех схем формирования лучей

В случае, если исследуемые радиотелескопом небесные объекты превышают его поле обзора (часто встречающийся случай), производится сканирование путем механического поворота зеркала. При этом изображение складывается из смещенных полей обзора (см. рис. 2.11в). Чувствительность полученной объединенной области показана на рис. 4.13.

Все результаты, показанные на рис. 4.12 … 4.13, рассчитаны для частоты 1,42 ГГц.

Сравнивая распределение чувствительности на рис. 4.12 … 4.13 для трех схем формирования лучей, можно отметить, что схемы *CFM* и *MaxSNR* обладают существенной неравномерностью чувствительности. Схема *LCMV* дает более равномерное распределение чувствительности в поле обзора, но при этом уменьшается чувствительность на (10…15) %.

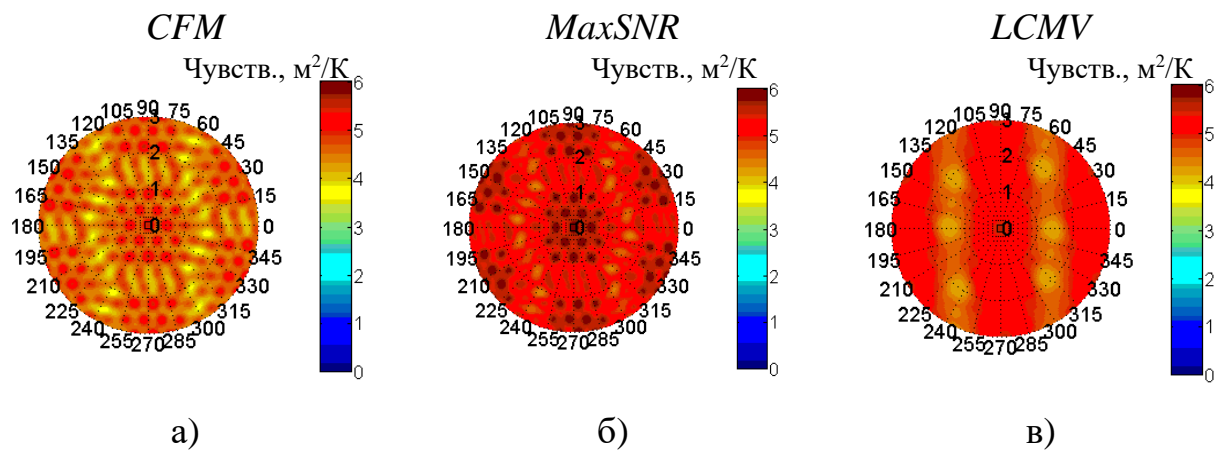

а)                   б)                   в)

Рис. 4.13 Чувствительность антенной системы для трех схем формирования лучей при нескольких направлениях механического сканирования

На рис. 4.14 показаны гистограммы распределения чувствительности внутри 37-лучевого поля обзора на трех частотах. Из этого рисунка видно, что для получения равномерности чувствительности целесообразно использовать схему *LCMV* (для этой схемы гистограммы более узкие и высокие).

Зависимость неравномерности чувствительности от частоты показана на рис. 4.15.

Из рис. 4.15 видно, что происходит увеличение неравномерности чувствительности при увеличении частоты, что можно было ожидать, так как с увеличением частоты лучи становятся уже. Так как направления максимумов лучей остаются неизменными, то увеличиваются провалы усиления, и, следовательно, провалы чувствительности между лучами, что приводит к увеличению неравномерности.

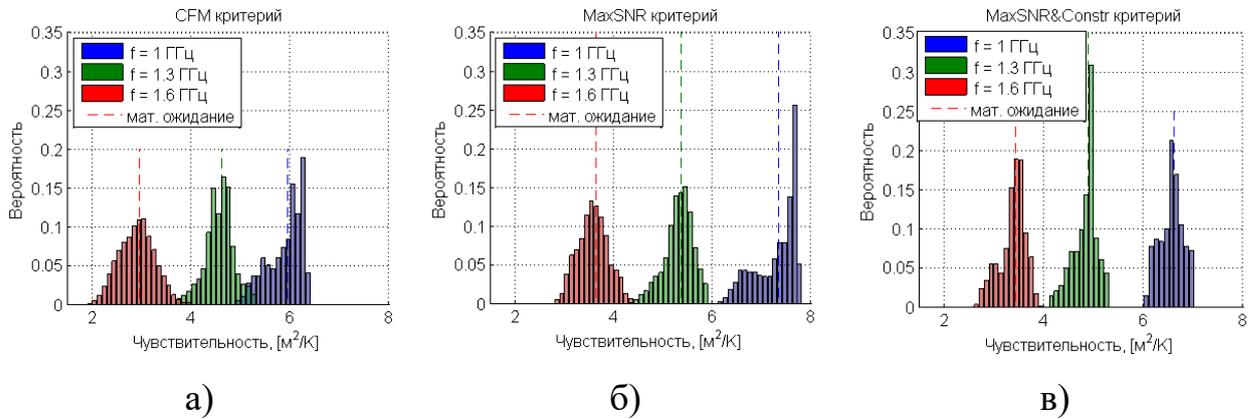

| а) | б) | в) |

Рис. 4.14 Гистограммы распределения чувствительности внутри поля обзора

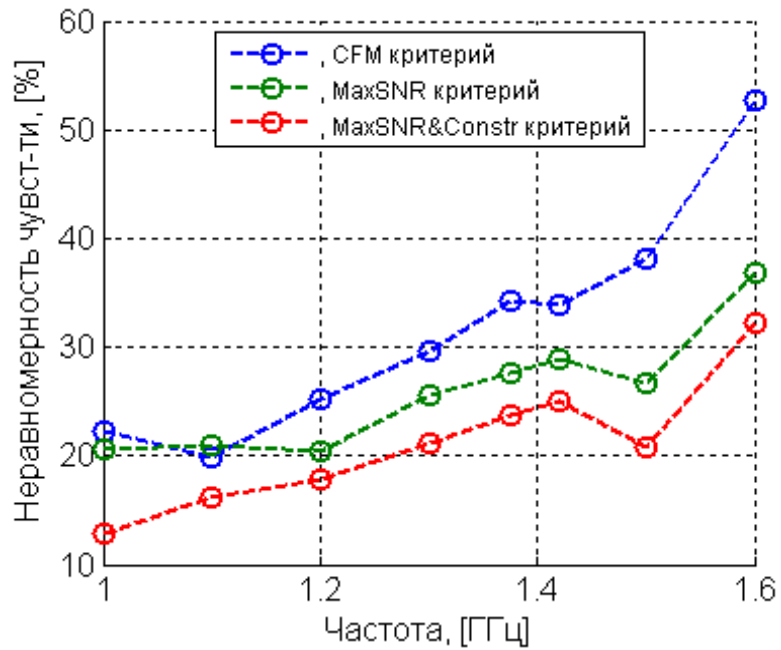

Рис. 4.15 Зависимость неравномерности чувствительности от частоты для трех схем формирования лучей

## 4.4. Сравнение рассчитанной чувствительности с измеренной

Для проверки приведенных результатов моделирования были выполнены несколько измерений на одной из антенн Вестерборгского радиотелескопа, в фокусе которой была установлена антенная решетка (рис. 4.16).

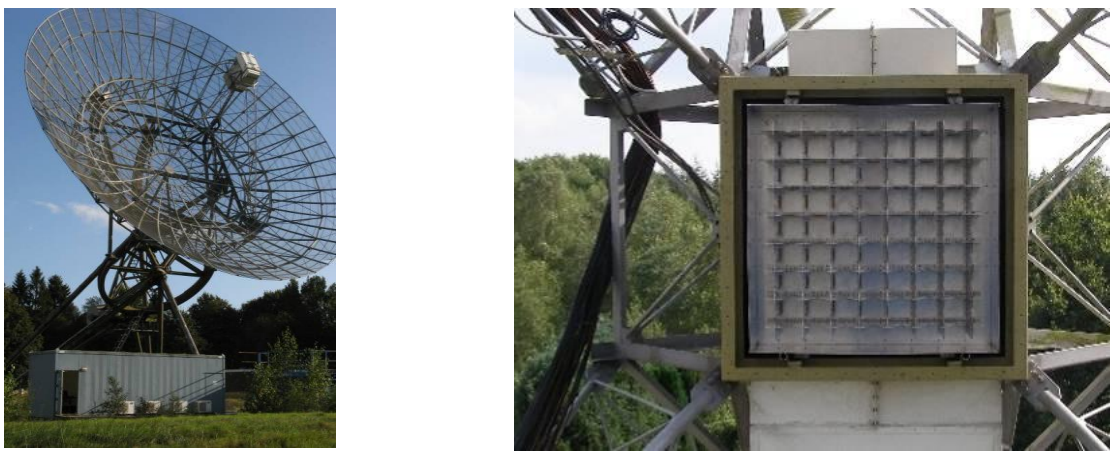

а)                                        б)

Рис. 4.16 Антенна Вестерборгского радиотелескопа: а) зеркальный отражатель; б) антенная решетка, установленная в фокусе зеркального отражателя

Ввиду финансовых ограничений на момент проведения измерений, было реализовано только 52 канала приема из 72 для каждой из ортогональных поляризаций. Под каналом приема имеется в виду цепочка «элемент антенной решетки — МШУ — коаксиальный кабель — приемник».

Измерение чувствительности в заданном направлении проводилось следующим образом.

По методике, описанной в подразделе 3.5, измерялся сигнальный вектор **e** и рассчитывались весовые коэффициенты **w**.

Выходной сигнал с каждого элемента решетки $e_n$ умножался на соответствующие весовые коэффициенты $w_n$, и вычислялось отношение сигнал/шум по выражению [31]

$$\left(\frac{S}{N}\right) = \mathbf{e}^H \mathbf{C}^{-1} \mathbf{e}. \tag{4.3}$$

По рассчитанному значению отношения сигнал/шум и спектральной плотности потока мощности источника $I$ по выражению (1.8) вычислялась чувствительность системы $S$ для соответствующего весовым коэффициентам луча.

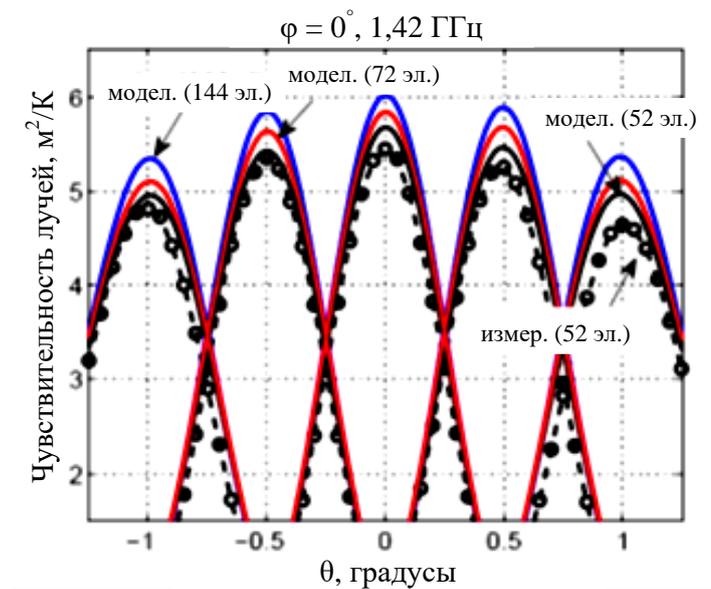

Рис. 4.17 Сечения лучей 16 … 19, результаты моделирования и измерения

На рис. 4.17 показаны сечения лучей 16 … 19 при $\varphi = 0°$ для схемы формирования лучей *MaxSNR*. Штриховой линией с круглыми маркерами показана чувствительность, вычисленная на основе измеренных данных (в соответствии с описанным способом). Сплошной линией показана чувствительность, полученная в результате моделирования при различном количестве активных элементов решетки: для 52 элементов (тех же, которые использовались при измерениях), для 72 элементов (все элементы решетки *одной поляризации*) и для 144 элементов (все элементы решетки).

Аналогичным способом были получены чувствительности (на основе измеренных данных и полученные путем моделирования) для направлений максимумов всех 37 лучей, а также направлений их пересечений. Результаты показаны на рис. 4.18 и 4.19. На рис. 4.18 чувствительность показана в поле обзора, а на рис. 4.19 — как зависимость от номера луча.

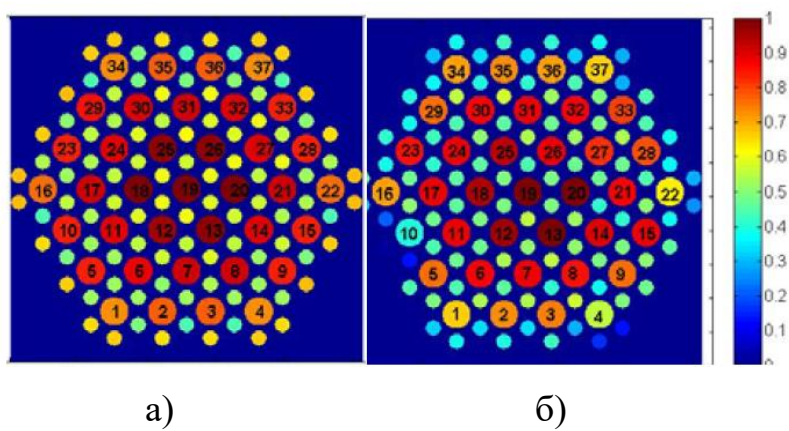

а)                                б)

Рис. 4.18 Нормированная чувствительность системы в направлении максимумов лучей (большие точки) и в направлениях их пересечения (маленькие точки): а) вычисленные значения; б) измеренные значения

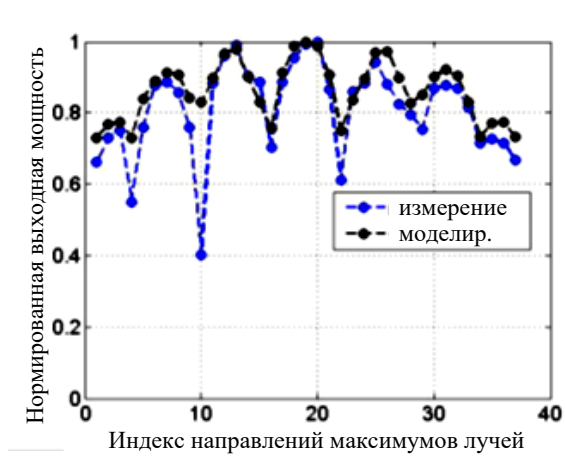

Рис. 4.19 Зависимость чувствительности системы в направлении максимумов лучей от номера луча в поле обзора

Из рис. 4.18 …4.19 видно, что для большинства лучей разница между моделированием и измерениями не превышает 10%. Исключение составляют лучи, находящиеся на краю поля обзора (например, лучи 4, 10, 22), то есть сканирующие на (1…1,5)°. Для этих лучей наблюдается значительно заниженная чувствительность при измерениях. Это можно объяснить наличием в системе металлического корпуса, в котором расположена решетка. Так как для этих лучей при сканировании на угол $(1\ldots1,5)°$ краевые элементы решетки наиболее возбуждены, то влияние корпуса при этом максимально. Также в реальной системе существуют внешние источники шума, которые не учитывались в модели. Например, стоящее рядом здание или радиопомехи (особенно GSM диапазона, т.к. они попадают в рабочий диапазон системы), отраженные от несущих конструкций антенны.

### 4.5. Расчет поляризационных свойств системы

Одной из целей данной работы является анализ поляризационных свойств системы, т.е. способности системы разделять поляризацию падающего поля на две ортогональные составляющие при приеме сигнала с любого направления в пределах поля обзора.

При проведения анализа по методу $MaxSNR$ были рассчитаны два набора весовых коэффициентов: $\mathbf{w}^{CO}$ и $\mathbf{w}^{XP}$ (или $\mathbf{w}^{p}$, где $p \in \{CO, XP\}$). Верхние индексы CO и XP обозначают поляризацию падающего поля $E^{CO}$ и $E^{XP}$, для оптимального приема которого рассчитывались весовые коэффициенты $\mathbf{w}$. При этом векторы поляризации являются ортогональными и соответствуют базисным векторам $Ludwig$-3 определения [104]. Другими словами, $\mathbf{w}^{p}$ — весовые коэффициенты, оптимальные (по критерию максимальной чувствительности) для приема $E^{p}$. Таким образом, имеются два формирователя: один для приема $E^{CO}$, другой — для приема $E^{XP}$.

Напряжения на выходе формирователей определялись как вектор

$$\mathbf{v}^{p} = \begin{bmatrix} v_1^p \\ v_2^p \end{bmatrix} = \begin{bmatrix} \left(\mathbf{w}^{CO}\right)^H \mathbf{e}^p \\ \left(\mathbf{w}^{XP}\right)^H \mathbf{e}^p \end{bmatrix}, \tag{4.4}$$

где $v_1^p$ и $v_2^p$ — напряжение на выходе первого и второго формирователей соответственно при приеме падающей волны с поляризацией $p$; $\mathbf{e}^p$ — вектор, элементы которого представляют собой напряжения на каждом элементе решетки при приеме $E^p$ (см. раздел 2).

В идеальном случае, при приеме, например $E^{CO}$, напряжение $v_1^{CO}$ должно быть пропорционально напряженности поля, в то время как $v_2^{CO}$ должно быть равно нулю. Однако, в реальной системе это не так по двум основным причинам: 1) из-за наличия кросс-поляризованной составляющей в характеристиках направленности антенных элементов решетки элементы, подключенные ко второму формирователю, имеют не нулевой отклик при приеме $E^{CO}$, и 2) из-за наличия взаимной связи между элементами решетки сигнал, принятый элементами решетки первого формирователя, проникает в каналы приема второго формирователя.

Для оценки поляризационной развязки каналов рассматриваемой системы был рассчитан коэффициент ортогональности векторов $\rho$ [10, 11]:

$$\rho = \frac{\left\langle \mathbf{v}^{CO}, \mathbf{v}^{XP} \right\rangle}{\sqrt{\left\langle \mathbf{v}^{CO}, \mathbf{v}^{CO} \right\rangle \left\langle \mathbf{v}^{XP}, \mathbf{v}^{XP} \right\rangle}}, \tag{4.5}$$

где оператор $\left\langle \mathbf{a}, \mathbf{b} \right\rangle$ обозначает эрмитово скалярное произведение двух векторов ($\left\langle \mathbf{a}, \mathbf{b} \right\rangle = \mathbf{a}^* \cdot \mathbf{b} = \mathbf{a}^H \mathbf{b}$).

Так как напряжение на выходе формирователей является функцией направления, с которого принимается сигнал, то и коэффициент ортогональности $\rho$ также является функцией направления. Чем ближе этот коэффициент к нулю, тем лучше система с точки зрения разделения поляризации.

На рис. 4.20 показаны зависимости коэффициента ортогональности от направления при формировании 37 лучей внутри поля обзора. При формировании лучей использовались либо 72 элемента решетки, *X*-ориентированные для первого формирователя и *Y*-ориентированные для второго формирователя (рис. 4.20а), либо все 144 элемента решетки для обоих формирователей (рис. 4.20б).

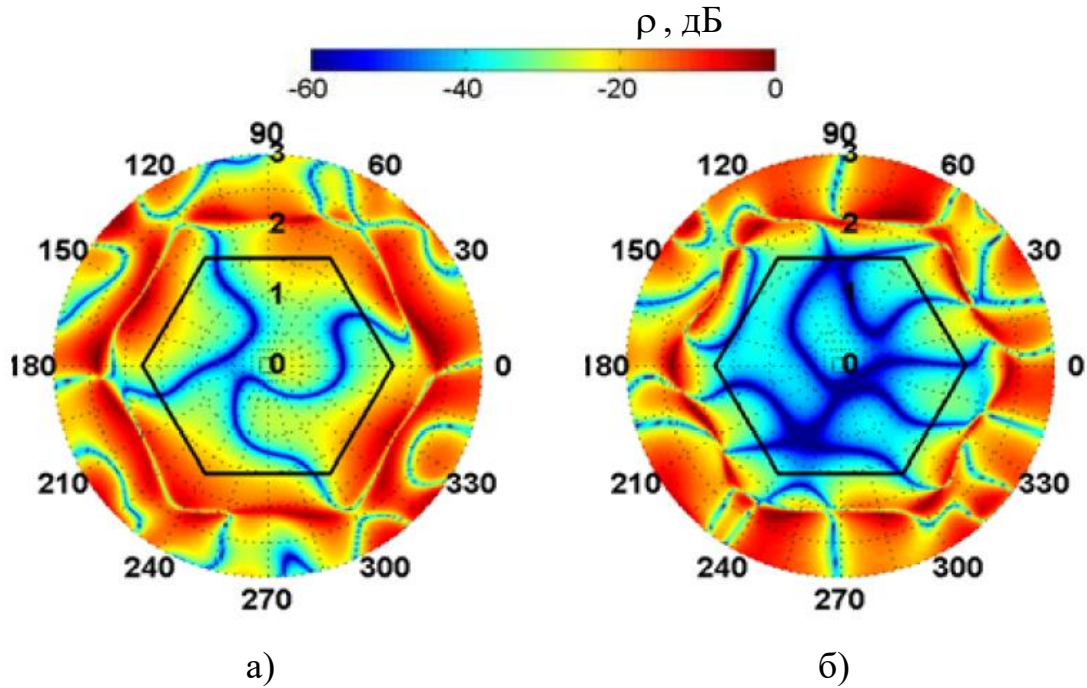

а)                                    б)

**Рис. 4.20** Зависимость коэффициента ортогональности ρ от направления:
а) при возбуждении 72-х одинаково ориентированных элементов решетки;
б) при возбуждении 144-х (всех) элементов решетки

Из рис. 4.20 видно, что для формирования лучей заданной поляризации использование элементов решетки, предназначенных для формирования кросс-поляризованных лучей, дает значительный выигрыш в разделении поляризации. Однако при этом возрастает объем вычислений (так как при формировании используется в 2 раза больше элементов решетки) и требуются большие вычислительные мощности.

**4.5. Выводы по разделу**

4.5.1. Рассчитаны оптимальные параметры $g_{const}$ для метода формирования лучей *LCMV* для нескольких частот. Показано, что наибольший выигрыш в равномерности чувствительности $\Delta S$ метод *LCMV* дает на нижней частоте рабочего диапазона (22 % для метода *MaxSNR* и 5 % для метода *LCMV* на частоте 1 ГГц). Также можно отметить уменьшение оптимального уровня усиления луча в направлении ограничения (параметр $g_{const}$) по отношению к усилению в направлении максимума с ростом частоты (см. табл. 4.1). Это можно объяснить тем, что ширина лучей уменьшается с повышением частоты, и, следовательно, в заданных направлениях ограничений (см. рис. 4.10), которые для всех частот одинаковы, уровень усиления падает.

4.5.2. Рассчитаны основные параметры зеркальной антенной системы с фокальной решеткой *APERTIF*, такие как:

— весовые коэффициенты для элементов решетки;

— первичные и вторичные ДН для некоторых лучей;

— коэффициенты эффективности системы (КИП, коэффициент перехвата, коэффициенты амплитудного и фазового распределений в апертуре зеркала);

— шумовая температура и ее составляющие;

— КНД системы;

— чувствительность системы для нескольких лучей в поле обзора.

Все перечисленные параметры приведены в сравнении для трех схем формирования лучей (*CFM*, *MaxSNR* и *LCMV*). Также для многих из параметров приведены их частотные зависимости.

4.5.3. Используя разработанную модель, проведено моделирование системы APERTIF при трех схемах формирования лучей. Показано, что:

— схема *CFM* не дает максимальной чувствительности системы, хотя КНД при ней максимален, и обладает существенной

неравномерностью в поле обзора (34 % на частоте 1,42 ГГц), поэтому целесообразно применять другие схемы формирования лучей;

— схема *LCMV* позволяет достичь компромисса между чувствительностью системы и ее неравномерностью внутри поля обзора; для рассмотренной системы уменьшение чувствительности на 10% позволило уменьшить ее неравномерность в поле обзора с 34 % до 24 % на частоте 1,42 ГГц и с 22 % до 5 % на частоте 1 ГГц.

— при использовании схемы *LCMV* чувствительность быстро падает при увеличении угла сканирования из-за искажения формы луча; если для MaxSNR критерия отношение чувствительности центрального луча с чувствительности краевого луча равна приблизительно 1,3, то для критерия LCMV это отношение равно 1,7;

4.5.4. Для второго прототипа *APERTIF* проведены измерения чувствительности пяти лучей в сечении $\varphi = 0°$ и результаты сравнены с промоделированными значениями чувствительности. Разница не превышает 4 % для центрального луча и 10% для краевого луча (см. рис. 4.17).

4.5.5. В результате проведения исследования системы как поляриметра, установлено, что при формировании лучей заданной поляризации всеми элементами решетки (как *X*-, так и *Y*-ориентированными), средняя в поле обзора величина коэффициент ортогональности лучей $\rho$ улучшается с –25 дБ до –33 дБ. Однако при этом возрастает объем вычислений (так как при формировании используется в 2 раза больше элементов решетки) и требуются большие вычислительные мощности.

# ВЫВОДЫ

В результате выполнения диссертационной работы решена актуальная научно-прикладная задача моделирования гибридных приемных антенных систем, состоящих из рефлектора, плотной антенной решетки, малошумящих усилителей и формирователя диаграммы направленности. Получены новые результаты, а именно проанализирована система в многолучевом режиме работы, в котором высокая чувствительность достигается в широком поле обзора, что позволяет существенно увеличить скорость обзора неба радиотелескопом.

В результате выполнения работы получены следующие основные научные и практичные результаты:

1. Разработана единая модель гибридной приемной антенной системы с фокальной решеткой и приемной частью, позволяющая рассчитывать важнейшие параметры радиотелескопов, такие как а) оптимальные по произвольному критерию весовые коэффициенты для элементов фокальной решетки; б) шумовая температура системы и ее составляющие (за счет приема шумов земли, потери из-за рассогласования элементов решетки с МШУ и взаимной связи между элементами решетки); в) КИП и его составляющие (эффективность амплитудного и фазового распределения поля в апертуре зеркала, коэффициент перехвата); г) эффективность поляризационной дискриминации.

2. Создано прикладное программное обеспечение для расчетов и моделирования характеристик гибридной приемной антенной системы в пакете MATLAB, написан интерфейс взаимодействия с программами моделирования *CAESAR* и *GRASP*9.

3. Предложен критерий оптимизации весовых коэффициентов для достижения равномерной чувствительности в многолучевом поле обзора при незначительном ее уменьшении.

4. Рассчитаны параметры системы *APERTIF*, даны рекомендации по улучшению характеристик антенной решетки.

5. Результаты проверены экспериментально как для отдельных элементов системы (антенный элемент Вивальди и его устройство питания) и небольшой решетки, так и для всей системы в целом (измерения выполнены в институте *ASTRON* на одной из антенн Вестерборгского радиотелескопа).

**Рекомендованные направления дальнейших исследований.**

Следующим важнейшим шагом в развитии многолучевых радиотелескопов (в том числе зеркальных систем с фокальными решетками) является разработка методики эффективной калибровки инструмента, которая заключается в корректировке изменений формы ДН всех лучей, их усиления, усиления и фазового сдвига в каналах приемника и т.д.

Также в данной работе лишь немного затрагивалась тема эффективности поляризационной дискриминации интерферометра. Сравнивались только две схемы возбуждения антенной решетки. Актуальным является также разработка оптимальных схем возбуждения с точки зрения наилучшего разделения поляризации падающего поля на две компоненты, в то время как чувствительность не должна уменьшаться значительно.